\documentclass[preprint,acmsmall,screen]{acmart}

\acmJournal{PACMPL}
\acmVolume{1}
\acmNumber{POPL}
\acmArticle{1}
\acmYear{2021}
\acmMonth{1}
\acmDOI{}
\startPage{1}
\setcopyright{none}

\newif\ifappendices
\appendicestrue

\usepackage{float}
\usepackage{mathpartir}
\usepackage{bm}
\usepackage{soul}
\usepackage{xspace}
\usepackage{mathtools}
\usepackage{hyperref}
\usepackage{breakurl}
\usepackage[scaled=0.75]{beramono}
\usepackage{pifont}
\usepackage{framed}
\usepackage{subdepth}
\usepackage{dashundergaps}
\usepackage{listings}
\usepackage{mleftright}
\usepackage{extarrows}
\usepackage{placeins}
\usepackage[a]{esvect}
\usepackage{colortbl}
\usepackage{tabularx}
\usepackage{multirow}
\usepackage{subcaption}
\usepackage{enumitem}
\usepackage{sfmath}
\usepackage{tikz}
\usetikzlibrary{calc}

\newcommand*{\derivationWidth}{0.6\textwidth}
\input{tex-common/proof-helpers}
\newcommand*{\proofContext}[1]{\def\currentprefix{proof:#1}}
\newcommand*{\locallabel}[1]{\label{\currentprefix:#1}}
\newcommand*{\localref}[1]{\ref{\currentprefix:#1}}

\newcolumntype{L}[1]{>{\raggedright\let\newline\\\arraybackslash\hspace{0pt}}p{#1}}
\newcolumntype{C}[1]{>{\centering\arraybackslash}p{#1}}

\makeatletter
\DeclareRobustCommand{\reloverrideleft}{\mathrel{\mathpalette\rel@override\vartriangleleft}}
\DeclareRobustCommand{\reloverrideright}{\mathrel{\mathpalette\rel@override\vartriangleright}}

\newcommand{\rel@override}[2]{%
  \vphantom{#2}%
  \ooalign{$\m@th#1-$\cr$\m@th#1#2$}%
}
\makeatother

\everymath{\displaystyle}

\allowdisplaybreaks

\def\arraystretch{1.1}
\hypersetup{ 
    colorlinks=false,
    pdfborder={0 0 0},
}

\definecolor{Yellow}{RGB}{255,255,0}
\definecolor{LightGreen}{RGB}{144,238,144}

\definecolor{GainsboroDarkened}{RGB}{211,211,211}
\definecolor{GainsboroFaded}{RGB}{232,232,232}

\def\mystrut{\vrule height6pt depth1.5pt width0pt}
\newcommand*{\codeSel}[1]{\fboxsep=1pt\fcolorbox{Yellow!90!black}{Yellow!100}{\mystrut#1}}
\newcommand*{\codeSelTwo}[1]{\fboxsep=0pt\fcolorbox{LightGreen!90!black}{LightGreen!100}{\mystrut#1}}
\newcommand*{\codeErase}[1]{\fboxsep=1pt\fcolorbox{GainsboroFaded!85!black}{GainsboroFaded!60}{\mystrut#1}}
\newcommand*{\codeEraseTwo}[1]{\fboxsep=0pt\fcolorbox{GainsboroDarkened!85!black}{GainsboroDarkened!60}{\mystrut#1}}

\newbool{anonymous}

\makeatletter
\if@ACM@anonymous
   \booltrue{anonymous}%
\else
   \boolfalse{anonymous}%
\fi
\makeatother

\newcommand*{\piCalculus}{$\pi$-calculus\xspace}

\newcommand*{\ttt}[1]{\texttt{#1}}
\newcommand*{\kw}[1]{{\text{\ttt{#1}}}} 

\newcommand*{\twoPrime}{{\prime\mkern-2.6mu\prime\mkern-2.2mu}}

\DeclareTextFontCommand{\textbfit}{%
  \fontseries\bfdefault 
  \itshape
}

\newcommand*{\after}{\circ}

\newcommand*{\concat}{\cdot}

\newcommand*{\eqdef}{\stackrel{\smash{\text{\tiny def}}}{=}}

\newcommand*{\join}{\sqcup}
\newcommand*{\meet}{\sqcap}
\newcommand*{\bigjoin}{\bigsqcup}

\newcommand*{\lowlight}[1]{\textcolor{darkgray}{#1}}
\newcommand*{\sub}[2]{#1_{\lowlight{#2}}}

\newcommand*{\appref}[1]{Appendix~\ref{app:#1}}

\newcommand*{\figref}[1]{Figure~\ref{fig:#1}}
\newcommand*{\figrefTwo}[2]{Figures \ref{fig:#1} and \ref{fig:#2}}

\newcommand*{\lemref}[1]{Lemma~\ref{lem:#1}}

\newcommand*{\secref}[1]{Section~\ref{sec:#1}}

\newcommand*{\Secref}[1]{\S\,\ref{sec:#1}}

\newcommand*{\thmref}[1]{Theorem~\ref{thm:#1}}

\newenvironment{nop}{}{}
\newenvironment{sdisplaymath}
   {\begin{nop}\small\begin{displaymath}}
   {\end{displaymath}\end{nop}\ignorespacesafterend}
\newenvironment{smathpar}
   {\begin{nop}\small\begin{mathpar}}
   {\end{mathpar}\end{nop}\ignorespacesafterend}
\newenvironment{salign}
   {\par\nobreak\small\noindent\csname align*\endcsname}
   {\csname endalign*\endcsname}

\newenvironment{mathfig}{\begin{sdisplaymath}}{\end{sdisplaymath}}
\newenvironment{syntaxfig}{\begin{mathfig}\begin{array}{@{}l@{\quad}r@{~~}c@{\quad}ll}}{\end{array}\end{mathfig}}

\makeatletter
\newbox\sf@box
  {\def\sf@one{#1}%
   \def\sf@two{#2}%
   \setbox\sf@box\hbox
      \bgroup}%
  { \egroup
   \ifx\@empty\sf@two\@empty\relax
     \def\sf@two{\@empty}
   \fi
   \ifx\@empty\sf@one\@empty\relax
      \subfloat[\sf@two]{\box\sf@box}%
   \else
      \subfloat[\sf@one][\sf@two]{\box\sf@box}%
   \fi}
\makeatother

\RequirePackage{xcolor}

\definecolor{highlightcolor}{rgb}{1.0,0.8,0.8}
\definecolor{shadecolor}{rgb}{0.9,0.9,0.9}
\definecolor{lightgray}{rgb}{0.8,0.8,0.8}
\newcommand*{\shadebox}[1]{\fcolorbox{lightgray}{shadecolor}{\raisebox{0pt}[0.60\baselineskip][0.05\baselineskip]{#1}}}

\newcommand*{\ruleName}[1]{\textnormal{\textsf{#1}}}

\makeatletter
\newcommand{\superimpose}[2]{%
  {\ooalign{$#1\@firstoftwo#2$\cr\hfil$#1\@secondoftwo#2$\hfil\cr}}}
\makeatother

\newenvironment{nscenter}
 {\parskip=0pt\par\nopagebreak\centering}
 {\par\noindent\ignorespacesafterend}

\newsavebox{\vardisplaymathbox}

\newcommand{\crossrule}{\noindent\textcolor{lightgray}{\cleaders\hbox{.}\hfill}}

\newcommand*{\OurLanguage}{%
   \ifanonymous%
      [xxxxx]%
   \else%
      Fluid%
   \fi%
   }

\newcommand*{\ProofInSupplementaryMaterial}{Included with supplementary materials.}
\newcommand*{\IncludedWithSupplementaryMaterial}{included with the supplementary materials}

\newcommand*{\filledtriangleright}{\blacktriangleright}

\newcommand*{\hole}{\square}
\newcommand*{\blackhole}{\blacksquare}
\newcommand\grayuline{\bgroup\markoverwith{\textcolor{gray}{\rule[-0.9ex]{2pt}{0.4pt}}}\ULon}
\newcommand*{\TT}{\mathsf{tt}}
\newcommand*{\FF}{\mathsf{ff}}

\renewcommand*{\ldots}{...} 
\renewcommand*{\ruleName}[1]{\textcolor{gray}{\textnormal{\textsf{#1}}}}

\renewcommand*{\secref}{\Secref}

\newcommand*{\mathSf}[1]{\textup{\textsf{#1}}}
\newcommand*{\Set}[1]{\mathSf{#1}}

\newcommand*{\Int}{\mathbb{Z}}

\newcommand*{\seqEmpty}{\varepsilon}

\newcommand*{\primOp}{\oplus}
\newcommand*{\primFwd}[2]{\fwdF{#1}{#2}}
\newcommand*{\primFwdBool}[2]{\upperAdj{\sub{#1}{#2}}} 
\newcommand*{\primBwd}[2]{\bwdF{#1}{#2}}
\newcommand*{\primBwdBool}[2]{\lowerAdj{\sub{#1}{#2}}} 
\newcommand*{\primGCBool}[2]{\sub{#1}{#2}} 

\newcommand*{\funConcatMap}{\exVar{\kw{concatMap}}}
\newcommand*{\funEnumFromTo}{\exVar{\kw{enumFromTo}}}

\newcommand*{\annot}[2]{#1_{#2}}

\newcommand*{\sym}[1]{\textbf{\kw{#1}}}
\newcommand*{\symCons}{\sym{:}}

\newcommand*{\matrixLBrack}{\langle}
\newcommand*{\matrixRBrack}{\rangle}

\newcommand*{\exApp}[2]{{#1}\,{#2}}
\newcommand*{\exAppPrim}[2]{{#1}({#2})}
\newcommand*{\exBinaryApp}[3]{{#1} \mathbin{#2} {#3}}
\newcommand*{\exCons}[2]{{#1}\,\symCons\,{#2}}
\newcommand*{\exClosure}[3]{\kw{cl}({#1},{#2},{#3})}
\newcommand*{\exFalse}{\kw{false}}

\newcommand*{\exIfThenElse}[3]{\kw{if}\;{#1}\;\kw{then}\;{#2}\;\kw{else}\;{#3}}
\newcommand*{\exInt}[1]{#1}
\newcommand*{\exList}[2]{\kw{[}\;#1 \ #2}
\newcommand*{\exListComp}[2]{\kw{[}\,{#1}\;\sym{$\vert$}\;{#2}\kw{]}}
\newcommand*{\exListEnum}[2]{\kw{[}\,{#1}\;\kw{..}\;{#2}\,\kw{]}}
\newcommand*{\exLambda}[1]{\uplambda{#1}}
\newcommand*{\exLet}[3]{\kw{let}\;{#1}\equal{#2}\;\kw{in}\;{#3}}

\newcommand*{\exLetRecMutual}[2]{\kw{let}\;{#1}\;\kw{in}\;{#2}}
\newcommand*{\exMatch}[2]{\kw{match}~{#1}~\kw{as}\;{#2}}
\newcommand*{\exNil}{\kw{[]}}
\newcommand*{\exOp}[1]{\kw{(}{#1}\kw{)}}
\newcommand*{\exPair}[2]{\kw{(}#1\comma#2\kw{)}}

\newcommand*{\exPrimOp}[2]{{#1}({#2})} 
\newcommand*{\exRec}[1]{\kw{(}#1\kw{)}}
\newcommand*{\exRecProj}[2]{{#1}\textbf{\kw{.}}{#2}}
\newcommand*{\exRecEmpty}{\kw{()}}
\newcommand*{\exStr}[1]{#1}
\newcommand*{\exTrue}{\kw{true}}
\newcommand*{\exVar}[1]{#1}
\newcommand*{\exVec}[3]{\matrixLBrack{#1}\;\sym{$\vert$}\;#2\;\kw{in}\;{#3}\matrixRBrack}
\newcommand*{\exVecVal}[2]{\matrixLBrack{#1}\;\sym{$\vert$}\;{#2}\matrixRBrack}
\newcommand*{\exVecLookup}[2]{{#1}\,\sym{!}\,{#2}}
\newcommand*{\exVecLen}[1]{\kw{len}\;{#1}}

\newcommand*{\exListEnd}{\kw{]}}
\newcommand*{\exListNext}[2]{\comma\;#1 \ #2}
\newcommand*{\annListEnd}[1]{\kw{]}_{#1}}
\newcommand*{\annListNext}[3]{\comma_{#3}\;#1 \ #2}

\newcommand*{\annClosure}[4]{\kw{cl}({#1},{#2},{#3},{#4})}
\newcommand*{\annCons}[3]{{#1}\,\symCons_{\hspace{-0.1em}#3}\,{#2}}
\newcommand*{\annFalse}[1]{\kw{false}_{#1}}
\newcommand*{\annList}[3]{\kw{[}_{#3}\;#1 \ #2}
\newcommand*{\annListComp}[3]{\kw{[}\,{#1}\;\sym{$\vert$}\;{#2}\kw{]}_{#3}}
\newcommand*{\annInt}[2]{#1_{#2}}

\newcommand*{\annNil}[1]{\kw{[]}_{#1}}
\newcommand*{\annPair}[3]{\exPair{#1}{#2}_{#3}}
\newcommand*{\annRec}[2]{\exRec{#1}_{#2}}
\newcommand*{\annStr}[2]{#1_{#2}}
\newcommand*{\annTrue}[1]{\kw{true}_{#1}}
\newcommand*{\annVec}[4]{\matrixLBrack{#1}\;\sym{$\vert$}\;#2\;\kw{in}\;{#3}\matrixRBrack_{#4}}
\newcommand*{\annVecVal}[3]{\exVecVal{#1}{#2}_{#3}}

\newcommand*{\clause}[2]{{#1}\equal{#2}}
\newcommand*{\clauseUncurried}[2]{{#1}\equal{#2}}

\newcommand*{\qualGuard}[1]{{#1}}
\newcommand*{\qualDeclaration}[2]{\kw{let}\;{#1}\equal{#2}}

\newcommand*{\qualGenerator}[2]{{#1}\;\kw{$\leftarrow$}\;{#2}}

\newcommand*{\pattVar}[1]{#1}
\newcommand*{\pattTrue}{\exTrue}
\newcommand*{\pattFalse}{\exFalse}
\newcommand*{\pattCons}[2]{\exCons{#1}{#2}}
\newcommand*{\pattNil}{\exNil}
\newcommand*{\pattList}[2]{\exList{#1}{#2}}
\newcommand*{\pattListNext}[2]{\exListNext{#1}{#2}}
\newcommand*{\pattListEnd}{\exListEnd}
\newcommand*{\pattPair}[2]{\exPair{#1}{#2}}
\newcommand*{\pattRec}[1]{\exRec{#1}}
\newcommand*{\pattRecEmpty}{\exRecEmpty}

\newcommand*{\elimmapsto}{{:}\;}
\newcommand*{\branchCons}[1]{(\symCons)\elimmapsto{#1}}
\newcommand*{\branchNil}[1]{\exNil\elimmapsto{#1}}
\newcommand*{\branchTrue}[1]{\exTrue\elimmapsto{#1}}
\newcommand*{\branchFalse}[1]{\exFalse\elimmapsto{#1}}

\newcommand*{\elimNil}[1]{\{\branchNil{#1}\}}
\newcommand*{\elimCons}[1]{\{\branchCons{#1}\}}
\newcommand*{\elimBool}[2]{\{\branchTrue{#1}, \branchFalse{#2}\}}
\newcommand*{\elimTrue}[1]{\{\branchTrue{#1}\}}
\newcommand*{\elimFalse}[1]{\{\branchFalse{#1}\}}
\newcommand*{\elimList}[2]{\{\branchNil{#1},\branchCons{#2}\}}
\newcommand*{\elimProd}[1]{\{\exPair{}{}\elimmapsto{#1}\}}
\newcommand*{\elimRec}[2]{\{\exRec{#1}\elimmapsto{#2}\}}
\newcommand*{\elimRecEmpty}[1]{\{\exRecEmpty\elimmapsto{#1}\}}
\newcommand*{\elimVar}[2]{{#1}\elimmapsto{#2}} 

\renewcommand*{\lowlight}[1]{\textcolor{gray}{#1}}

\newcommand*{\matchCons}[2]{\exCons{#1}{#2}}
\newcommand*{\matchFalse}{\exFalse}
\newcommand*{\matchNil}{\exNil}
\newcommand*{\matchPair}[2]{\exPair{#1}{#2}}
\newcommand*{\matchRec}[1]{\exRec{#1}}
\newcommand*{\matchRecEmpty}{\exRecEmpty}
\newcommand*{\matchTrue}{\exTrue}
\newcommand*{\matchVar}[1]{#1}

\newcommand*{\tyBool}[0]{\kw{Bool}}
\newcommand*{\tyFun}[2]{{#1}\rightarrow{#2}}
\newcommand*{\tyInt}[0]{\kw{Int}}
\newcommand*{\tyList}[1]{\kw{List}\;{#1}}

\newcommand*{\tyProd}[2]{{#1}\times{#2}}
\newcommand*{\tyRec}[1]{\kw{Rec}\;{\exRec{#1}}}
\newcommand*{\tyRecEmpty}{\kw{Rec}\;{\exRecEmpty}}

\newcommand*{\Below}[1]{\Sel{#1}{A}}

\newcommand*{\ValF}[1]{\Set{Val}\,{#1}}
\newcommand*{\TermF}[1]{\Set{Term}\,{#1}}
\newcommand*{\Sel}[2]{\SelF{#1}\,{\Ann{#2}}}
\newcommand*{\SelF}[1]{\Set{Sel}_{#1}}

\newcommand*{\Bool}{2}
\newcommand*{\Unit}{1}

\let\oldbigjoin\bigjoin

\renewcommand*{\bigjoin}[1]{{\textstyle{\oldbigjoin}}\hspace{-0.1em}#1}

\let\oldleq\leq
\let\oldgeq\geq
\newcommand*{\numleq}{\oldleq}
\newcommand*{\numgeq}{\oldgeq}
\newcommand*{\numgt}{>}

\renewcommand*{\leq}{\sqsubseteq}
\renewcommand*{\geq}{\sqsupseteq}
\newcommand{\eq}{\doteq}

\newcommand*{\envLookupS}{\in}
\newcommand*{\envLookupR}{\envLookupS}
\newcommand*{\envLookupBwdS}{\ni}
\newcommand*{\envLookupBwdR}[1]{\mathrel{\sub{\envLookupBwdS}{#1}}}
\newcommand*{\envLookupFwdR}[1]{\mathrel{\sub{\envLookupS}{#1}}}
\newcommand*{\envLookup}[3]{\bind{#2}{#3}\envLookupR{#1}}
\newcommand*{\envLookupFwdF}[4]{{#1}\envLookupFwdR{#2} \bind{#3}{#4}}
\newcommand*{\envLookupBwdF}[4]{{#1} \envLookupBwdR{#2} \bind{#3}{#4}}

\newcommand*{\envLookupBwd}[3]{{#1}\envLookupBwdR{#2}{#3}}
\newcommand*{\mapUpdate}[3]{{#1}\reloverrideleft\bind{#2}{#3}}

\newcommand*{\fwdF}[2]{\sub{\mathsf{#1{-}fwd}}{#2}} 
\newcommand*{\bwdF}[2]{\sub{\mathsf{#1{-}bwd}}{#2}} 

\newcommand*{\closeDefsS}{\twoheadrightarrow}
\newcommand*{\closeDefsR}{\closeDefsS}
\newcommand*{\closeDefsFwdR}{\mathrel{\rotatebox[origin=c]{45}{$\closeDefsS$}}}
\newcommand*{\closeDefsBwdR}{\mathrel{\rotatebox[origin=c]{-45}{$\closeDefsS$}}}

\newcommand*{\closeDefsFwdF}[1]{\sub{\closeDefsFwdR}{\hspace{-0.2em}#1}}
\newcommand*{\closeDefsBwdF}[1]{\sub{\closeDefsBwdR}{#1}}

\newcommand*{\evalS}{\Rightarrow}
\newcommand*{\evalFwdS}{\mathrel{\rotatebox[origin=c]{45}{$\evalS$}}}
\newcommand*{\evalBwdS}{\mathrel{\rotatebox[origin=c]{-45}{$\evalS$}}}
\newcommand*{\evalR}{\evalS}
\newcommand*{\evalBwdR}[1]{\sub{\evalBwdS}{#1}}
\newcommand*{\evalFwdR}[1]{\sub{\evalFwdS}{\hspace{-0.2em}#1}}
\newcommand*{\evalBwd}[5]{#1 \evalBwdR{#2} #3, #4, #5}
\newcommand*{\evalBwdLeq}[5]{#1 \evalBwdR{#2}\leq #3, #4, #5}
\newcommand*{\evalFwd}[5]{#1, #2, #3 \evalFwdR{#4} #5}
\newcommand*{\evalFwdEq}[5]{#1, #2, #3 \evalFwdR{#4}\eq #5}
\newcommand*{\evalFwdGeq}[5]{#1, #2, #3 \evalFwdR{#4}\geq #5}
\newcommand*{\evalFwdF}[1]{{\evalFwdR{#1}}}
\newcommand*{\evalBwdF}[1]{{\evalBwdR{#1}}}
\newcommand*{\evalGC}[1]{\mathrel{\sub{\evalR}{#1}}}

\newcommand*{\dual}[1]{{#1}^{\circ}}
\newcommand*{\lowerAdj}[1]{{#1}^*}
\newcommand*{\upperAdj}[1]{{#1}_*}
\newcommand*{\erase}[1]{\lfloor#1\rfloor}

\newcommand*{\matchS}{\rightsquigarrow}
\newcommand*{\matchFwdS}{\mathrel{\rotatebox[origin=c]{45}{$\matchS$}}}
\newcommand*{\matchBwdS}{\mathrel{\rotatebox[origin=c]{-45}{$\matchS$}}}
\newcommand*{\matchR}{\matchS}
\newcommand*{\matchFwdR}[1]{\sub{\matchFwdS}{\hspace{-0.2em}#1}}
\newcommand*{\matchBwdR}[1]{\sub{\matchBwdS}{#1}}
\newcommand*{\matchFwd}[6]{#1, #2 \matchFwdR{#3} #4, #5, #6}
\newcommand*{\matchFwdGeq}[6]{#1, #2 \matchFwdR{#3}\geq #4, #5, #6}
\newcommand*{\matchBwd}[6]{#1, #2, #3 \matchBwdR{#4} #5, #6}
\newcommand*{\matchBwdLeq}[6]{#1, #2, #3 \matchBwdR{#4}\leq #5, #6}
\newcommand*{\matchFwdF}[1]{{\matchFwdR{#1}}}
\newcommand*{\matchBwdF}[1]{{\matchBwdR{#1}}}

\newcommand*{\desugarS}{\twoheadrightarrow}
\newcommand*{\desugarR}{\desugarS}
\newcommand*{\desugarFwdS}{\mathrel{{\rotatebox[origin=c]{45}{$\desugarS$}}}}
\newcommand*{\desugarBwdS}{\mathrel{{\rotatebox[origin=c]{-45}{$\desugarS$}}}}
\newcommand*{\desugarFwdR}{\desugarFwdS}
\newcommand*{\desugarBwdR}[1]{\sub{\desugarBwdS}{#1}}
\newcommand*{\desugarFwdF}[1]{\sub{\desugarFwdR}{\hspace{-0.13em}#1}}
\newcommand*{\desugarBwdF}[1]{\desugarBwdR{#1}}
\newcommand*{\desugarGC}[1]{\sub{\desugarR}{#1}}

\newcommand*{\listRestFwdF}[1]{\desugarFwdF{#1}}
\newcommand*{\listRestBwdF}[1]{\desugarBwdF{#1}}

\newcommand*{\clauseS}{\rightharpoonup}
\newcommand*{\clauseBwdS}{\mathrel{{\rotatebox[origin=c]{-45}{$\clauseS$}}}}
\newcommand*{\clauseFwdR}{\mathrel{{\rotatebox[origin=c]{45}{$\clauseS$}}}}
\newcommand*{\clauseBwdR}[1]{\sub{\clauseBwdS}{#1}}
\newcommand*{\clauseN}{clause}
\newcommand*{\clauseFwdF}[1]{\fwdF{\clauseN}{#1}}
\newcommand*{\clauseBwdF}[1]{\bwdF{\clauseN}{#1}}

\newcommand*{\clausesBwdS}{\desugarBwdS}
\newcommand*{\clausesFwdR}{\desugarFwdR}
\newcommand*{\clausesBwdR}[1]{\desugarBwdR{#1}}
\newcommand*{\clausesN}{clauses}
\newcommand*{\clausesFwdF}[1]{\fwdF{\clausesN}{#1}}
\newcommand*{\clausesBwdF}[1]{\bwdF{\clausesN}{#1}}

\newcommand*{\totaliseFwdS}{\nearrow}
\newcommand*{\totaliseBwdS}{\searrow}
\newcommand*{\totaliseFwdR}[1]{\sub{\totaliseFwdS}{\hspace{-0.12em}#1}}
\newcommand*{\totaliseBwdR}[1]{\sub{\totaliseBwdS}{#1}}
\newcommand*{\totaliseFwd}[4]{{#1},{#2} \totaliseFwdR{#3} {#4}}
\newcommand*{\totaliseBwd}[4]{{#1} \totaliseBwdR{#2} {#3}, {#4}}
\newcommand*{\totaliseFwdGeq}[4]{{#1},{#2} \totaliseFwdR{#3}\geq {#4}}
\newcommand*{\totaliseN}{totalise}
\newcommand*{\totaliseFwdF}[1]{\fwdF{\totaliseN}{#1}}
\newcommand*{\totaliseBwdF}[1]{\bwdF{\totaliseN}{#1}}

\newcommand*{\bind}[2]{#1{:}\;#2}
\renewcommand*{\vec}[1]{\vv{#1}}
\newcommand*{\seq}[1]{\vv{#1}}
\newcommand*{\seqRange}[2]{\seqRangeOp{#1}{#2}{\concat}}

\newcommand*{\seqRangeOp}[3]{{#1} #3 .. #3 {#2}}

\newcommand*{\arity}[1]{\mathsf{arity}(#1)}
\newcommand*{\length}[1]{|#1|}

\newcommand*{\Sub}[2]{#1_{#2}}

\newcommand*{\trApp}[4]{\exApp{#1}{#2} \mathrel{\filledtriangleright} {\bind{#3}{#4}}}
\newcommand*{\trAppPrim}[5]{\exApp{\Sub{#1}{(#2,#3)}}{\Sub{#4}{#5}}}
\newcommand*{\trAppPrimNew}[3]{\exAppPrim{#1}{\vec{\Sub{#2}{#3}}}}

\newcommand*{\trCons}[2]{\exCons{#1}{#2}}
\newcommand*{\trFalse}[1]{\exFalse} 
\newcommand*{\trInt}[2]{\exInt{#1}} 
\newcommand*{\trLambda}[1]{\exLambda{#1}}

\newcommand*{\trLetRecMutual}[2]{\exLetRecMutual{#1}{#2}}

\newcommand*{\trNil}[1]{\exNil} 
\newcommand*{\trOp}[2]{\Sub{\exOp{#1}}{#2}}

\newcommand*{\trRec}[1]{\exRec{#1}}
\newcommand*{\trRecProj}[3]{\exRecProj{\Sub{#1}{#2}}{#3}}
\newcommand*{\trTrue}[1]{\exTrue} 
\newcommand*{\trVar}[2]{#1} 
\newcommand*{\trVec}[4]{\exVec{#1}{#2}{\Sub{#3}{#4}}}
\newcommand*{\trVecLookup}[4]{\exVecLookup{\Sub{#1}{#2}}{\Sub{#3}{#4}}}
\newcommand*{\trVecLen}[2]{\exVecLen{\Sub{#1}{#2}}}

\newcommand*{\explVal}[2]{{\textcolor{gray}{#1}}\mathrel{\textcolor{gray}{::}}{#2}}

\newcommand*{\comma}{\textbf{\kw{,}}}
\newcommand*{\equal}{\mathrel{\kw{=}}}

\newcommand*{\disjjoin}{\mathbin{\ooalign{$\join$\cr%
   \hfil\raise0.42ex\hbox{$\scriptscriptstyle+$}\hfil\cr}}}

\newcommand*{\Ann}[1]{\mathcal{#1}}
\newcommand*{\raw}[1]{\bm{#1}}
\newcommand*{\Lattice}[5]{\langle#1,#2,#3,#4,#5\rangle}
\newcommand*{\BoolLattice}[6]{\langle#1,#2,#3,#4,#5,#6\rangle}

\definecolor{verylightgray}{gray}{0.9}
\definecolor{lightgray}{gray}{0.5}
\definecolor{mediumgray}{gray}{0.45}
\newlength\lsthorizontalpadding
\newcommand*\lstnumberstyle{\ttfamily\scriptsize\textcolor{lightgray}}
\newlength\lstnumbersep
\newlength\lstnumberwidth
\lstdefinelanguage{Fluid}{%
   morekeywords={else,if,let,then,in}%
}
\lstnewenvironment{codecol}[1][]{\lstset{language=code,#1,moredelim=*[is][\textcolor{mediumgray}]{|}{|}}}{}

\setcitestyle{nosort}

\begin{abstract}
   We present new language-based dynamic analysis techniques for linking visualisations and other structured outputs to data in a fine-grained way, allowing a user to interactively explore how data attributes map to visual or other output elements by selecting (focusing on) substructures of interest. This can help both programmers and end-users understand how data sources and complex outputs are related, which can be a challenge even for someone with expert knowledge of the problem domain. Our approach builds on bidirectional program slicing techiques based on Galois connections, which provide desirable round-tripping properties.

   Unlike the prior work in program slicing, our approach allows selections to be negated. In a setting with negation, the bidirectional analysis has a De Morgan dual, which can be used to link different outputs generated from the same input. This offers a principled language-based foundation for a popular interactive visualisation feature called \emph{brushing and linking} where selections in one chart automatically select corresponding elements in another related chart. Although such view coordination features are valuable comprehension aids, they tend be to hard-coded into specific applications or libraries, or require programmer effort.
\end{abstract}

\begin{document}

\title{Linked visualisations via Galois dependencies}
\ifappendices
   \subtitle{\emph{Extended paper with additional supporting material}}
\fi
\author{Roly Perera}
\orcid{0000-0001-9249-9862}
\affiliation{%
   \institution{The Alan Turing Institute}
   \city{London}
   \country{UK}
}
\additionalaffiliation{%
   \institution{University of Bristol}
   \city{Bristol}
   \country{UK}
}
\email{rperera@turing.ac.uk}

\author{Minh Nguyen}
\orcid{0000-0003-3845-9928}
\email{min.nguyen@bristol.ac.uk}
\affiliation{%
   \institution{University of Bristol}
   \city{Bristol}
   \country{UK}
}

\author{Tomas Petricek}
\orcid{0000-0002-7242-2208}
\affiliation{%
   \institution{University of Kent}
   \city{Canterbury}
   \country{UK}
}
\additionalaffiliation{%
   \institution{The Alan Turing Institute}
   \city{London}
   \country{UK}
}
\email{tpetricek@kent.ac.uk}

\author{Meng Wang}
\orcid{0000-0001-7780-630X}
\email{meng.wang@bristol.ac.uk}
\affiliation{%
   \institution{University of Bristol}
   \city{Bristol}
   \country{UK}
}

\maketitle

\section{Introduction}
\label{sec:introduction}

Techniques for dynamic dependency analysis have been fruitful, with applications ranging from information-flow security~\cite{sabelfeld03} and optimisation~\cite{kildall73} to debugging and program comprehension~\cite{weiser81,delucia96}. There are, however, few methods suitable for fine-grained analysis of richly structured outputs, such as data visualisations and multidimensional arrays. Dataflow analyses \cite{reps95} tend to focus on analysing variables rather than parts of structured values. Where-provenance~\cite{buneman01} and related data provenance techniques are fine-grained, but are specific to relational query languages. Taint tracking \cite{newsome05} is also fine-grained, but works forwards from input to output. For many applications, it would be useful to be able to focus on a particular part of a structured output, and have an analysis isolate the input data pertinent only to that substructure.

This is a need that increasingly arises outside of traditional programming. Journalists and data scientists use programs to compute charts and other visual summaries from data, charts which must be interpreted by colleagues, policy makers and lay readers alike. Interpreting a chart correctly means understanding what the components of the visualisation actually \emph{represent}, i.e.~the mapping between data and visual elements. But this is a hard task, requiring time and expertise, even with access to the data and source code used to create the visualisation. In practice it is easy for innocent (but devastating) mistakes such as transposing two columns of data to go unnoticed~\cite{miller06}. Since visualisations are simply cases of programs that transform structured inputs (data tables) into structured outputs (charts and other graphics), general-purpose language-based techiques for fine-grained dependency tracking should be able to help with this, by making it possible to reveal these relationships automatically to an interested user.

\subsection{Linking structured outputs to structured inputs}
\label{sec:introduction:data-linking}

\begin{figure}
   \begin{subfigure}[b]{0.99\textwidth}
      \centering
      {\includegraphics[scale=0.55]{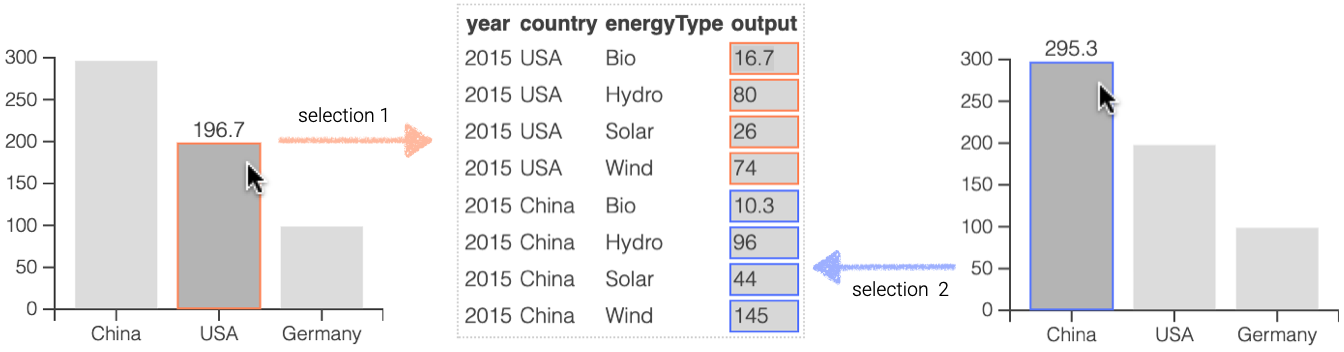}}
   \end{subfigure}\\
   \vspace{2mm}
   \begin{subfigure}{0.65\textwidth}
      \small
\begin{lstlisting}[language=Fluid]
let totalFor c rows =
   sum [ row.output | row $\leftarrow$ rows, row.country == c ];
let data2015 = [ row | row $\leftarrow$ data, row.year == 2015 ];
    countryData = [ { x: c, y: totalFor c data2015 }
                  | c $\leftarrow$ ["China", "USA", "Germany"] ]
in BarChart {
   caption: "Total output by country",
   data: countryData
}
\end{lstlisting}
   \end{subfigure}
   \caption{Fine-grained linking of outputs to inputs, focusing on data for USA (left) and China (right).}
   \label{fig:introduction:data-linking}
\end{figure}

First, interpreting a chart would be much easier if the user were able to explore the relationship between the various parts of the chart and the underlying data interactively, discovering the relevant relationships on a need-to-know basis. For example, selecting a particular bar in a bar chart could highlight the relevant data in a table, perhaps showing only the relevant rows, as illustrated in \figref{introduction:data-linking}. We could certainly do more and say something about the nature of the relationship (summation, in this case), but even just revealing the relevant data puts a reader in a much better position to fact-check or confirm their own understanding of what they are looking at. Indeed, visualisation designers sometimes create ``data-linked'' artefacts like these by hand, such as Nadieh Bremer's award-winning visualisation of population density growth in Asian cities~\cite{bremer15}, at the cost of significant programming effort. Libraries such as Altair \cite{vanderPlas18} alleviate some of this work, but require data transformations  to specified using a limited set of combinators provided (and understood) by the library.

What we would rather do instead is allow a data scientist to author analyses and visualisations using an expressive functional language like the one shown in \figref{introduction:data-linking}, with data linking provided automatically for the computed artefact, as baked-in transparency feature. At the core of this is a program analysis problem: we want to be able to focus on a particular chart element, and determine the inputs that contribute to it. Given the output selection, this is a matter of performing some kind of backwards analysis to identify the relevant data. As well providing a path to automation, framing data linking as a program analysis problem invites interesting questions that a hand-crafted solution is unlikely to properly address. For example, does the union of two output selections depend on the union of their respective dependencies? Do dependencies ``round-trip'', in that they identify sufficient resources to reconstruct the selected output? Are they minimal? These questions are important to establishing trust and a language-based approach offers a chance to address them.

\subsection{Linking structured outputs to other structured outputs}
\label{sec:introduction:vis-linking}

Second, authors often present distinct but related aspects of data in separate charts. In this situation a reader should be able to focus on (select) a visual element in one chart or other structured output and automatically see elements of a different chart which were computed using related inputs. For example in \figref{introduction:vis-linking} below, selecting the bar on the left should automatically highlight all the related visual elements on the right. This is a well-recognised use case called \emph{brushing and linking}~\cite{becker87}, which is supported by geospatial applications like GeoDa~\cite{anselin06} and charting libraries like Plotly, but tends to be baked into specific views, or require programmer effort and therefore anticipation in advance by the chart designer. Moreover these applications and libraries provide no direct access to the common data which explains why elements are related.

\begin{figure}
  \begin{subfigure}[b]{0.99\textwidth}
     \centering
     {\includegraphics[scale=0.58]{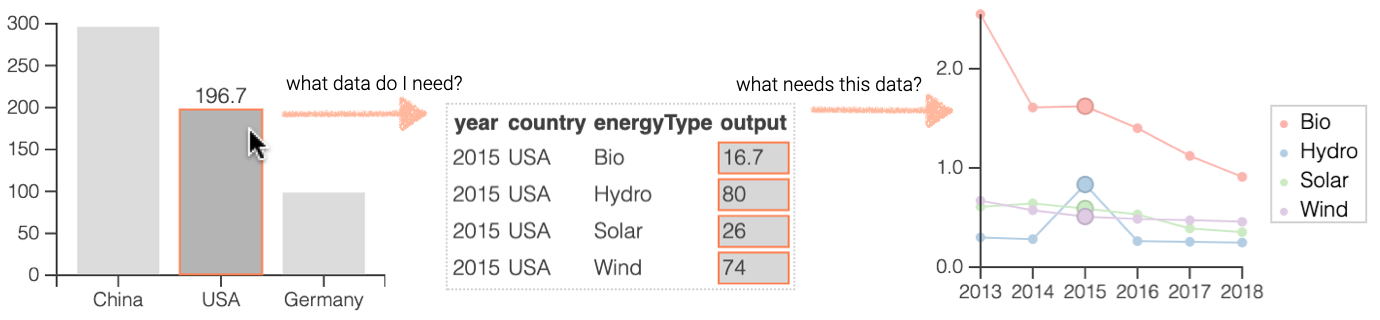}}
  \end{subfigure}\\[2mm]
  \begin{subfigure}{0.8\textwidth}
     \small
\begin{lstlisting}[language=Fluid]
let series type country = [
   { x: year, y: row.output }
   | year $\leftarrow$ [2013..2018], row $\leftarrow$ data,
     row.year == year, row.energyType == type, row.country == country
] in LineChart {
   caption: "Output of USA relative to China",
   plots: [
      LinePlot { name: type, data: plot }
      | type $\leftarrow$ ["Bio", "Hydro", "Solar", "Wind"],
        let plot = zipWith (fun p1 p2 $\rightarrow$ { x: p1.x, y: p1.y / p2.y })
                           (series type "USA") (series type "China")
   ]
}
\end{lstlisting}
  \end{subfigure}
 \caption{Linking visualisations via common data dependencies}
 \label{fig:introduction:vis-linking}
\end{figure}

Again, we would like to enable a more automated (and ubiquitous) version of brushing and linking, without imposing a burden on the programmer. They should be able to express visualisations and other data transformations using standard functional programming features such as those shown in \figref{introduction:vis-linking}, and have brushing and linking enabled automatically between computed artefacts which depend on common data. At the core of this requirement is a variant of our original program analysis problem: we want to select a part of the output and perform a backwards analysis to identify the required inputs, as before, but then also to perform a forwards analysis to identify dependent parts of the other output. Moreover, we would also like the brushing and linking feature to be able to provide a concise view of the data that explain why the two selections are linked. Note that the intuition behind the forwards analysis here is not the same as the one we appealed to in the context of round-tripping: there the (hypothetical) question was whether the selected data was \emph{sufficient} to reconstruct the selected output, whereas to identify related items in another view, we must determine those parts for which the selected data is \emph{necessary}. As before, a language-based approach offers the prospect of addressing these sorts of question in a robust way.

\subsection{Contributions}

To make progress towards these challenges, we present a bidirectional analysis which tracks fine-grained data dependencies between input and output selections, with round-tripping properties characterised by Galois connections. Selections have a complement, which we use to adapt the analysis to compute fine-grained dependencies between two outputs which depend on common inputs. Recent program slicing techniques~\cite{perera12a,perera13a,ricciotti17} allow the user to focus on the output by ``erasing'' parts deemed to be irrelevant; the erased parts, called \emph{holes}, are propagated backwards by a backwards analysis which identifies parts of the program and input which are no longer needed. Although these approaches also enjoy useful round-tripping properties characterised by Galois connections, they only allow focusing on \emph{prefixes} of a structured output, rather than arbitrary substructures, a notion which is not closed under complement. Our specific contributions are as follows:

\begin{itemize}
   \item[--] a new bidirectional dynamic dependency analysis which operates on selections of arbitrary parts of data values, for a core calculus with lists, records and mutual recursion, and a proof that the analysis is a Galois connection (\secref{data-dependencies});
   \item[--] a second bidirectional dependency analysis, derived from the first by De Morgan duality, which is also a Galois connection and which can be composed with the first analysis to link outputs to outputs, with an extended example based on matrix convolution  (\secref{de-morgan});
   \item[--] a richer surface language called \OurLanguage, implemented in PureScript, with familiar functional programming features such as piecewise definitions and list comprehensions, and a further Galois connection linking selections between the core and surface languages (\secref{surface-language}).
\end{itemize}

\noindent First we introduce the core calculus which provides the setting for the rest of the paper, and which will serve as the desugaring target for the surface syntax presented in \secref{surface-language}.

\section{Core language}
\label{sec:core-language}

The core calculus is a mostly standard call-by-value functional language with datatypes and records. The main distinguishing feature is the use of \emph{eliminators}, a trie-like pattern-matching construct that more familiar pattern-matching features like case expressions and piecewise function definitions easily desugar into. (A single core construct allows us to define the semantics of pattern-matching once and defer syntactic considerations to the surface language.) We give a big-step environment-based semantics, which is easier for the backward and forward dependency analyses in \secref{data-dependencies}, and introduce a compact (term-like) representation of derivation trees in the operational semantics, called \emph{traces}, which we will use to define the analyses over a fixed execution. Mutual recursion requires some care for the backwards analysis so we also treat that as a core language feature.

\subsection{Syntax and typing}
\label{sec:core-language:syntax-typing}

Although our implementation in untyped, types help describe the structure of the core language. \figref{core-language:syntax} introduces the types $A, B$ which include $\tyBool$, $\tyInt$ and function types $\tyFun{A}{B}$, but also lists $\tyList{A}$ and records $\tyRec{\vec{\bind{x}{A}}}$ which exemplify the two kinds of structured data of which are of interest: recursive datatypes with varying structure, and tabular data with a fixed shape. As usual the notation $\bind{x}{A}$ denotes the binding of $x$ to $A$ (understood formally as a pair); $\seq{\bind{x}{A}}$ denotes the sequence of bindings that results from zipping same-length sequences $\vec{x}$ and $\vec{A}$. In a record type $\tyRec{\vec{\bind{x}{A}}}$ the field names in $\vec{x}$ are required to be unique.

\begin{figure}
{\small
\begingroup
\renewcommand*{\arraystretch}{1}
\begin{minipage}[t]{0.5\textwidth}
\begin{tabularx}{\textwidth}{rL{2cm}L{3cm}}
&\textbfit{Type}&
\\
$A, B ::=$
&
$\tyBool$
&
Booleans
\\
&
$\tyInt$
&
integers
\\
&
$\tyRec{\vec{\bind{x}{A}}}$
&
records
\\
&
$\tyList{A}$
&
lists
\\
&
$\tyFun{A}{B}$
&
functions
\\[2mm]
$\Gamma, \Delta ::=$
&
$\seq{\bind{x}{A}}$
&
typing context
\\[2mm]
&\textbfit{Term}&
\\
$e ::=$
&
$\exTrue \mid \exFalse$
&
Boolean
\\
&
$\exInt{n}$
&
integer
\\
&
$\exVar{x}$
&
variable
\\
&
$\exAppPrim{\phi}{\vec{e}}$
&
primitive application
\\
&
$\exApp{e}{e'}$
&
application
\\
&
$\exNil$
&
nil
\\
&
$\exCons{u}{v}$
&
cons
\\
&
$\exRec{\vec{\bind{x}{e}}}$
&
record
\\
&
$\exRecProj{e}{x}$
&
record projection
\\
&
$\exLambda{\sigma}$
&
anonymous function
\\
&
$\exLetRecMutual{h}{e}$
&
recursive let
\end{tabularx}
\end{minipage}%
\begin{minipage}[t]{0.5\textwidth}
\begin{tabularx}{\textwidth}{rL{2.5cm}L{2.9cm}}
&\textbfit{Eliminator}&
\\
$\sigma, \tau ::=$
&
$\elimVar{x}{\kappa}$
&
variable
\\
&
$\elimBool{\kappa}{\kappa'}$
&
Boolean
\\
&
$\elimRec{\vec{x}}{\kappa}$
&
record
\\
&
$\elimList{\kappa}{\sigma}$
&
list
\\[2mm]
&\textbfit{Continuation}&
\\
$\kappa ::=$
&
$e$
&
term
\\
&
$\sigma$
&
eliminator
\\[2mm]
&\textbfit{Value}&
\\
$u, v ::=$
&
$\exTrue \mid \exFalse$
&
Boolean
\\
&
$\exInt{n}$
&
integer
\\
&
$\exNil$
&
nil
\\
&
$\exCons{u}{v}$
&
cons
\\
&
$\exRec{\vec{\bind{x}{v}}}$
&
record
\\
&
$\exClosure{\rho}{h}{\sigma}$
&
closure
\\[2mm]
$\rho ::=$
&
$\seq{\bind{x}{v}}$
&
environment
\\
$h ::=$
&
$\seq{\bind{x}{\sigma}}$
&
recursive functions
\\
\\[2mm]
\end{tabularx}
\end{minipage}
\endgroup
}
\caption{Syntax of core language}
\label{fig:core-language:syntax}
\end{figure}

The terms of the language are defined in \figref{core-language:syntax}. These include Boolean constants $\exTrue$ and $\exFalse$, integers $n$, variables $x$, and applications $\exApp{e}{e'}$. Primitives are not first-class; the expression $\exAppPrim{\phi}{\vec{e}}$ is the fully saturated application of $\phi$ to a sequence of arguments. (First-class and infix primitives are provided by desugarings in \secref{surface-language}). We also provide list constructors $\exNil$ and $\exCons{e}{e'}$, record construction $\exRec{\vec{\bind{x}{e}}}$ and record projection $\exRecProj{e}{x}$. The final two term forms, anonymous functions $\exLambda{\sigma}$ and recursive let-bindings $\exLetRecMutual{h}{e}$ where $h$ is of the form $\vec{\bind{x}{\sigma}}$, are explained below after we introduce the pattern-matching construct $\sigma$ (\emph{eliminator}). The typing rules for terms are given in \figref{core-language:typing}, and are intended only to help a reader understand the language; therefore the rules are simple and do not include features such as polymorphism. The main typing rules of interest are the ones which involve eliminators.

\subsection{Eliminators}
\label{sec:core-language:syntax-eliminator}

Eliminators $\sigma, \tau$ are also defined in \figref{core-language:syntax}, and are inspired by \citeauthor{connelly95}'s notion of generalised trie \cite{connelly95}, elaborated further by \citet{hinze00}. Eliminators specify how to match against a partial value of some type and select a continuation $\kappa$ for further execution, where $\kappa$ is either a term $e$ or another eliminator $\sigma$. The Boolean eliminator $\elimBool{\kappa}{\kappa'}$ selects either $\kappa$ or $\kappa'$ depending on whether a Boolean value is $\exTrue$ or $\exFalse$. The record eliminator $\elimRec{\seq{x}}{\kappa}$ matches a record with fields $\seq{x}$ and then returns $\kappa$ with the variables $\seq{x}$ bound to the components of the record. The list eliminator $\elimList{\kappa}{\sigma}$ maps the empty list to $\kappa$ and a non-empty list to another eliminator $\sigma$ which specifies how the head and tail of the list are to be matched. The variable eliminator $\elimVar{x}{\kappa}$ extends the usual notion of trie, matching any value, binding it to $x$, and then returning $\kappa$. Eliminators resemble the ``case trees'' commonly used as an intermediate form when compiling languages with pattern-matching \cite{graf20}, and can serve as an elaboration target for more advanced features such as the piecewise definitions described in \secref{surface-language}.

\begin{figure}
\small{\flushleft \shadebox{$\Gamma \vdash e: A$}%
\hfill \textbfit{$e$ has type $A$ under $\Gamma$}}
\begin{smathpar}
\inferrule*[right={$x : A \in \Gamma$}]
{
   \strut
}
{
   \Gamma \vdash \exVar{x}: A
}
\and
\inferrule*
{
   \strut
}
{
   \Gamma \vdash \exInt{n}: \tyInt
}
\and
\inferrule*
{
   \strut
}
{
   \Gamma \vdash \exTrue: \tyBool
}
\and
\inferrule*
{
   \strut
}
{
   \Gamma \vdash \exFalse: \tyBool
}
\and
\inferrule*
{
   \Gamma \vdash e_i: A_i
   \quad
   (\forall i \numleq \length{\vec{x}})
}
{
   \Gamma \vdash \exRec{\vec{\bind{x}{e}}}: \tyRec{\vec{\bind{x}{A}}}
}
\and
\inferrule*[
   right={$i \numleq \length{\vec{x}}$}
]
{
   \Gamma \vdash e: \tyRec{\vec{\bind{x}{A}}}
}
{
   \Gamma \vdash \exRecProj{e}{x_i}: A_i
}
\and
\inferrule*
{
   \Gamma \vdash \sigma : \tyFun{A}{B}
}
{
   \Gamma \vdash \exLambda{\sigma} : \tyFun{A}{B}
}
\and
\inferrule*[
   right={$\phi \in \Int^j \to \Int$}
]
{
   \Gamma \vdash e_i: \tyInt
   \quad
   (\forall i \numleq j)
}
{
   \Gamma \vdash \exAppPrim{\phi}{\vec{e}}: \tyInt
}
\and
\inferrule*
{
   \Gamma \vdash e: \tyFun{A}{B}
   \\
   \Gamma \vdash e': A
}
{
   \Gamma \vdash \exApp{e}{e'}: B
}
\and
\inferrule*
{
   \strut
}
{
   \Gamma \vdash \exNil: \tyList{A}
}
\and
\inferrule*
{
   \Gamma \vdash e: A
   \\
   \Gamma \vdash e': \tyList{A}
}
{
   \Gamma \vdash (\exCons{e}{e'}): \tyList{A}
}
\and
\inferrule*
{
   \Gamma \vdash h: \Delta
   \\
   \Gamma \concat \Delta \vdash e: A
}
{
   \Gamma \vdash \exLetRecMutual{h}{e}: A
}
\end{smathpar}
{\small \flushleft \shadebox{$\Gamma \vdash \sigma: \tyFun{A}{B}$}%
\hfill \textbfit{$\sigma$ has type $\tyFun{A}{B}$ under $\Gamma$}}
\begin{smathpar}
\inferrule*
{
   \Gamma \concat \bind{x}{A} \vdash \kappa: B
}
{
   \Gamma \vdash (\elimVar{x}{\kappa}): \tyFun{A}{B}
}
\and
\inferrule*
{
   \Gamma \vdash \kappa: A
   \\
   \Gamma \vdash \kappa': A
}
{
   \Gamma \vdash \elimBool{\kappa}{\kappa'}: \tyFun{\tyBool}{A}
}
\and
\inferrule*
{
   \Gamma \vdash \kappa: A
}
{
   \Gamma
   \vdash
   \elimRecEmpty{\kappa}: \tyFun{\tyRecEmpty}{A}
}
\and
\inferrule*
{
   \Gamma \vdash \elimRec{\vec{x}}{\sigma}: \tyFun{\tyRec{\vec{\bind{x}{A}}}}{\tyFun{A'}{B}}
}
{
   \Gamma
   \vdash
   \elimRec{\vec{x} \concat y}{\sigma}: \tyFun{\tyRec{\vec{\bind{x}{A}} \concat \bind{y}{A'}}}{B}
}
\and
\inferrule*
{
   \Gamma \vdash \kappa: B
   \\
   \Gamma \vdash \sigma: \tyFun{A}{\tyFun{\tyList{A}}{B}}
}
{
   \Gamma \vdash \elimList{\kappa}{\sigma}: \tyFun{\tyList{A}}{B}
}
\end{smathpar}
\\[2mm]
{\small \flushleft \shadebox{$\vdash v: A$}%
\hfill \textbfit{$v$ has type $A$}}
\begin{smathpar}
\inferrule*
{
   \strut
}
{
   \vdash \exInt{n}: \tyInt
}
\and
\inferrule*
{
   \strut
}
{
   \vdash \exTrue: \tyBool
}
\and
\inferrule*
{
   \strut
}
{
   \vdash \exFalse: \tyBool
}
\and
\inferrule*
{
   \vdash v_i: A_i
   \quad
   (\forall i \numleq \length{\vec{x}})
}
{
   \vdash \exRec{\vec{\bind{x}{v}}}: \tyRec{\vec{\bind{x}{A}}}
}
\and
\inferrule*
{
   \strut
}
{
   \vdash \exNil: \tyList{A}
}
\and
\inferrule*
{
   \vdash u: A
   \\
   \vdash v: \tyList{A}
}
{
   \vdash (\exCons{u}{v}): \tyList{A}
}
\and
\inferrule*
{
   \vdash \rho: \Gamma
   \\
   \Gamma \vdash h: \Delta
   \\
   \Gamma \concat \Delta \vdash \sigma: \tyFun{A}{B}
}
{
   \vdash \exClosure{\rho}{h}{\sigma}: \tyFun{A}{B}
}
\end{smathpar}
\\[2mm]
\begin{minipage}[t]{0.4\textwidth}%
{\small \flushleft \shadebox{$\vdash \rho: \Gamma$}%
\hfill \textbfit{$\rho$ has type $\Gamma$}}
\begin{smathpar}
\inferrule*
{
   \vdash v_i: A_i
   \quad
   (\forall i \numleq \length{\vec{x}})
}
{
   \vdash \vec{\bind{x}{v}}: \vec{\bind{x}{A}}
}
\end{smathpar}
\end{minipage}%
\hspace{1.5cm}
\begin{minipage}[t]{0.48\textwidth}%
{\small \flushleft \shadebox{$\Gamma \vdash h: \Delta$}%
\hfill \textbfit{$h$ has type $\Delta$ under $\Gamma$}}
\begin{smathpar}
   \inferrule*[right={
      \textnormal{$\Delta = \vec{\bind{x}{\tyFun{A}{B}}}$}
   }]
   {
      \Gamma \concat \Delta \vdash \sigma_i: \tyFun{A_i}{B_i}
      \quad
      (\forall i \numleq \length{\vec{x}})
   }
   {
      \Gamma \vdash \vec{\bind{x}{\sigma}}: \Delta
   }
\end{smathpar}
\end{minipage}
\caption{Typing rules for core language}
\label{fig:core-language:typing}
\end{figure}

The use of nested eliminators to match sub-values will become clearer if we consider the typing judgement $\Gamma \vdash \sigma: \tyFun{A}{B}$ for eliminators given in \figref{core-language:typing}. (Eliminators always have function type, so this should be read as a four-place relation, with $\to$ part of the notation.) The typing rule for variable eliminators reveals the connection between eliminators and functions: it converts a continuation $\kappa$ which can be assigned type $B$ under the assumption that $x$ is of type $A$ into an eliminator of type $\tyFun{A}{B}$. The typing rule for Boolean eliminators says that to make an eliminator of type $\tyFun{\tyBool}{A}$, we simply need two continuations $\kappa$ and $\kappa'$ of type $A$. The rule for the empty record states that to make an eliminator of type $\tyFun{\tyRec{}}{A}$, we simply need a continuation $\kappa$ of type $A$. The rule for non-empty records allows us to treat an eliminator of type $\tyFun{\tyRec{\vec{\bind{x}{A}}}}{\tyFun{A'}{B}}$ as an eliminator of type $\tyFun{\tyRec{\vec{\bind{x}{A}} \concat \bind{y}{A'}}}{B}$, representing the isomorphism between $\tyFun{\tyFun{A}{B}}{C}$ and $\tyFun{\tyProd{A}{B}}{C}$ but at the level of record types \cite{hinze00}. (Formalising eliminators precisely requires \citeauthor{bird98}-style nested datatypes \cite{bird98} and polymorphic recursion, but these details need not concern us here.)

The typing rule for list eliminators $\elimList{\kappa}{\sigma}$ combines some of the flavour of record and Boolean eliminators. To make an eliminator of type $\tyFun{\tyList{A}}{B}$, we need a continuation of type $B$ for the empty case, and in the non-empty case, an eliminator of type $\tyFun{A}{\tyFun{\tyList{A}}{B}}$ which can be treated as an eliminator of type $\tyFun{\tyProd{A}{\tyList{A}}}{B}$ in order to process the head and tail.

\subsubsection{Eliminators as functions}

We can now revisit the term forms $\exLambda{\sigma}$ and $\exLetRecMutual{h}{e}$. If $\sigma$ is an eliminator of type $\tyFun{A}{B}$, then $\exLambda{\sigma}$ is an anonymous function of the same type. If $h$ is of the form $\vec{\bind{x}{\sigma}}$, then $\exLetRecMutual{h}{e}$ introduces a sequence of mutually recursive functions which are in scope in $e$. The typing rule for $\exLetRecMutual{h}{e}$ uses an auxiliary typing judgement $\Gamma \vdash h : \Delta$ which assigns to every $x$ in $\Delta$ a function type $\tyFun{A}{B}$ as long as the $\sigma$ to which $x$ is bound in $h$ has that type.

\subsubsection{Values}
Values $v, u$, and environments $\rho$ are also defined in \figref{core-language:syntax}, and are standard for call-by-value. (Environments are more convenient than substitution for tracking variable usage.) To support mutual recursion, the closure form  $\exClosure{\rho}{h}{\sigma}$ captures the (possibly empty) sequence $h$ of functions with which the function was mutually defined, in addition to the ambient environment $\rho$. For the typing judgements $\vdash \rho: \Gamma$ and $\vdash v: A$ for environments and values (\figref{core-language:typing}), only the closure case is worth noting, which delegates to the typing rules for recursive definitions and eliminators.

\subsubsection{Evaluation}
\label{sec:core-language:eval}

\begin{figure}
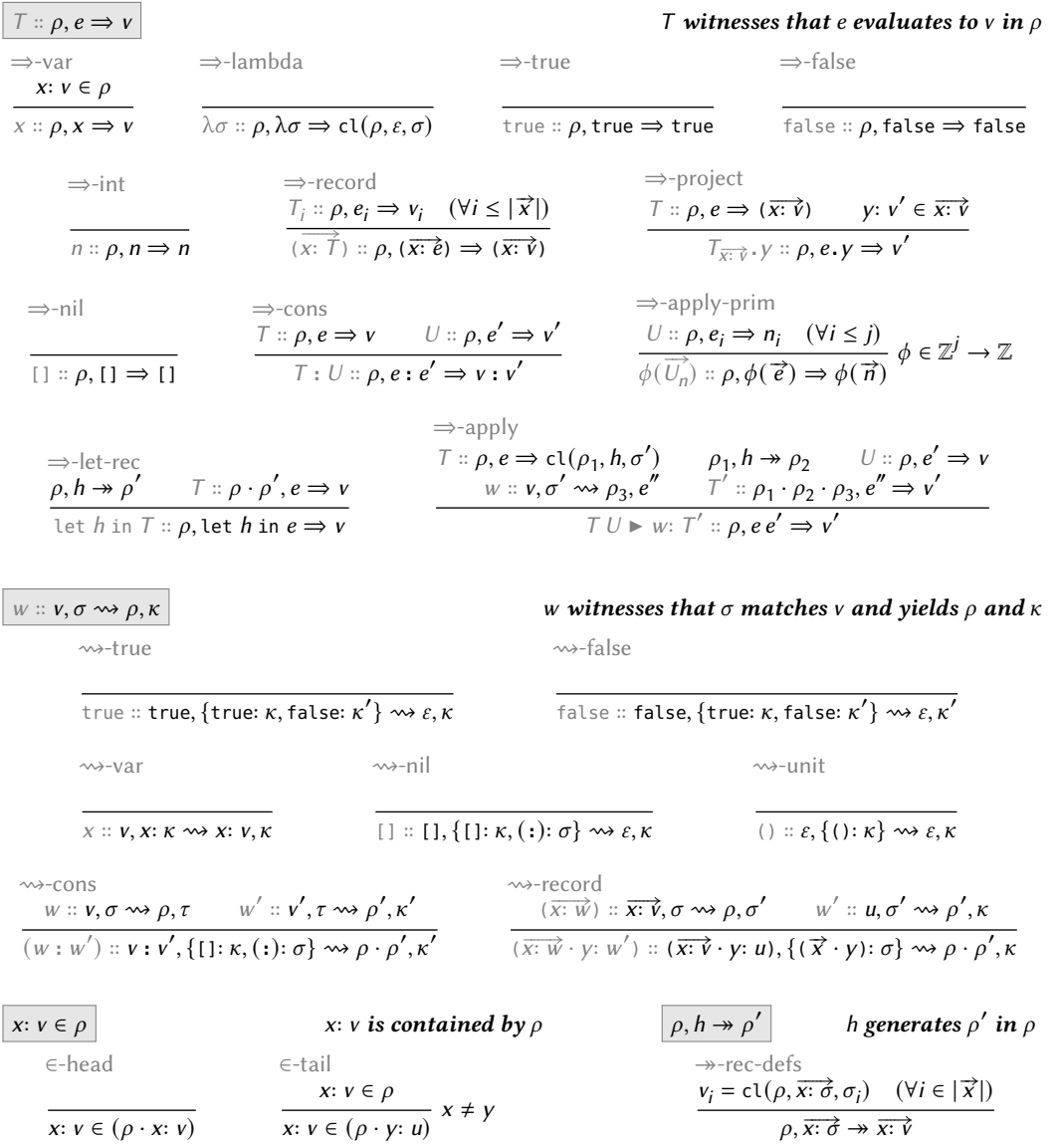

   \input{fig/core-language/eval}\vspace{1mm}
   \input{fig/core-language/match}
   \input{fig/core-language/eval-aux}
   \caption{Operational semantics}
   \label{fig:core-language:semantics}
\end{figure}

\figref{core-language:semantics} gives the operational semantics of the core language. In \secref{data-dependencies} we will define forward and backward analyses over a single execution; in anticipation of that use case, we treat the operational semantics as an inductive data type, following the ``proved transitions'' approach of \citeauthor{boudol89} for reversible CCS~\cite{boudol89}. The inhabitants of this data type are derivation trees explaining how a result was computed, and the analyses will be defined by structural recursion over these trees. Expressed in terms of inference rules, these trees can become quite cumbersome, so we introduce an equivalent but more term-like syntax for them, called a \emph{trace} (\figref{core-language:syntax-trace}), similar to the approach taken by \citeauthor{perera16d} for $\pi$-calculus~\cite{perera16d}.

\begin{figure}[H]
{\small
\begingroup
\renewcommand*{\arraystretch}{1}
\begin{minipage}[t]{0.5\textwidth}
\begin{tabularx}{\textwidth}{rL{2cm}L{3cm}}
&\textbfit{Trace}&
\\
$T, U ::=$
&
$\trVar{x}{\rho}$
&
variable
\\
&
$\trTrue{\rho} \mid \trFalse{\rho}$
&
Boolean
\\
&
$\trInt{n}{\rho}$
&
integer
\\
&
$\trRec{\vec{\bind{x}{T}}}$
&
record
\\
&
$\trRecProj{T}{\vec{\bind{x}{v}}}{y}$
&
record projection
\\
&
$\trNil{\rho}$
&
nil
\\
&
$\trCons{T}{U}$
&
cons
\\
&
$\trLambda{\sigma}$
&
anonymous function
\\
&
$\trApp{T}{U}{w}{T'}$
&
application
\end{tabularx}
\end{minipage}%
\begin{minipage}[t]{0.5\textwidth}
\begin{tabularx}{\textwidth}{rL{2.5cm}L{3cm}}
\\
&
$\trAppPrimNew{\phi}{U}{\exInt{n}}$
&
primitive application
\\
&
$\trLetRecMutual{h}{T}$
&
recursive let
\\[2mm]
&\textbfit{Match}&
\\
$w ::=$
&
$\matchVar{x}$
&
variable
\\
&
$\matchTrue \mid \matchFalse$
&
Boolean
\\
&
$\matchRec{\vec{\bind{x}{w}}}$
&
record
\\
&
$\matchNil$
&
nil
\\
&
$\matchCons{w}{w'}$
&
cons
\end{tabularx}
\end{minipage}
\endgroup
}
\caption{Syntax of traces and matches}
\label{fig:core-language:syntax-trace}
\end{figure}

The judgement $T :: \rho, e \evalR v$ defined at the top of \figref{core-language:semantics} states that term $e$ under environment $\rho$ evaluates to value $v$, and that $T$ is a proof term that witness that fact. (In the figure the traces appear in grey, to reinforce the idea that they are not part of the definition of $\evalR$ but rather a notation for its inhabitants.) The rules for Booleans, integers and lists are standard and have unsurprising trace forms. For variables, we give an explicit inductive definition of the environment lookup relation $\envLookupR$ at the bottom of the figure, again so that later we can perform analysis over a proof that an environment contains a binding. The lambda rule is standard except that we specify $\seqEmpty$ for the sequence of definitions being simultaneously defined, since a lambda is not recursive. For record construction, the trace form contains a subtrace $T_i$ for each field, and for record projection, which also uses the lookup relation $\envLookupR$, the trace form $\trRecProj{T}{\vec{\bind{x}{v}}}{y}$ records both the record $\vec{\bind{x}{v}}$ and the field name $y$ that was selected.

The rule for (mutually) recursive functions $\exLetRecMutual{h}{e}$, where $h$ is a sequence $\vec{\bind{x}{\sigma}}$ of function definitions, makes use of the auxiliary relation $\rho, h \closeDefsR \rho'$ at the bottom of \figref{core-language:semantics} which turns $h$ into an environment $\rho'$ binding each function name $x_i$ to a closure $\exClosure{\rho}{h}{\sigma_i}$ capturing $\rho$ and a copy of $h$. For primitive applications, the trace records the values of the arguments which were passed to the operation $\phi$. The rule for application $\exApp{e}{e'}$ is slightly non-standard, because it must deal with both mutual recursion and pattern-matching. First we unpack the recursive definitions $h$ from the closure $\exClosure{\rho_1}{h}{\sigma}$ computed by $e$, and again use the auxiliary relation $\closeDefsR$ to promote this into an environment $\rho_2$ of closures. We then use the relation $\matchR$ explained below to match $v$ against the eliminator $\sigma$, obtaining the branch $e^\twoPrime$ of the function to be executed and parameter bindings $\rho_3$. In addition to subtraces $T$ and $U$ for the function and argument, the application trace $\trApp{T}{U}{w}{T'}$ also has subtraces $w$ for the pattern-match and $T'$ for the selected branch.

\subsubsection{Pattern matching}
\label{sec:core-language:pattern-match}

The judgement $w :: v, \sigma \matchR \rho, \kappa$ also defined in \figref{core-language:semantics} states that eliminator $\sigma$ can match $v$ and produce environment $\rho$ and continuation $\kappa$, with $\rho$ containing the variable bindings that arose during the match. \emph{Matches} $w$ are a compact notation for proof terms for the $\matchR$ relation, analogous to traces for the $\evalR$ relation, and again appear in grey in the figure.

Variable eliminators $\elimVar{x}{\kappa}$ match any value, returning the singleton environment $\bind{x}{v}$ and continuation $\kappa$. Boolean eliminators match any Boolean value, returning the appropriate branch and empty environment $\seqEmpty$. List eliminators $\elimList{\kappa}{\sigma}$ match any list. The nil case is analogous to the handling of Booleans; the cons case depends on the fact that the nested eliminator $\sigma$ for the cons branch has the curried type $\tyFun{A}{\tyFun{\tyList{A}}{B}}$. First, we recursively match the head $v$ of type $A$ using $\sigma$, obtaining bindings $\rho$ and eliminator $\tau: \tyFun{\tyList{A}}{B}$ as the continuation. Then the tail $v'$ is matched using $\tau$ to yield additional bindings $\rho'$ and final continuation $\kappa'$. Record matching is similar: the empty record case resembles the nil case, and the non-empty case relies on the nested eliminator having curried type $\tyFun{\tyRec{\vec{\bind{x}{A}}}}{\tyFun{A'}{B}}$. The initial part $\vec{\bind{x}{v}}$ of the record is matched using $\sigma$, returning another eliminator $\sigma'$ of type $\tyFun{A'}{B}$. Then the last field $\bind{y}{u}$ is matched using $\sigma'$ to yield final continuation $\kappa$.

\section{A bidirectional dynamic dependency analysis}
\label{sec:data-dependencies}

We now extend the core language from \secref{core-language} with a bidirectional mechanism for tracking data dependencies. \secref{data-dependencies:lattices-of-selections} establishes a way of selecting (parts of) values, such as the height of a bar in a bar chart. \secref{data-dependencies:analyses:fwd} defines a forward analysis function $\evalFwdF{T}$ which specifies how selections on programs and environments (collectively: \emph{input selections}) becomes selection on outputs; selections represent \emph{availability}, with the computed output selection indicating the data available to the downstream computation. \secref{data-dependencies:analyses:bwd} defines a backward dependency function $\evalBwdF{T}$ specifying how output selections are mapped back to inputs; then selections represent demands, with the computed input selection identifying the data needed from the upstream computation. Both functions are monotonic; this will become important in \secref{data-dependencies:galois-connections}, where show that $\evalBwdF{T}$ and $\evalFwdF{T}$ form a Galois connection, establishing the round-tripping properties sketched intuitively in \secref{introduction:data-linking} which are important for trusting the results of the analyses.

\subsection{Lattices of selections}
\label{sec:data-dependencies:lattices-of-selections}

Our approach to representing selections is shown in \figref{core-syntax-selection}. The basic idea is to parameterise the type $\Set{Val}$ of values by a (bounded) lattice $\Ann{A}$ of \emph{selection states} $\alpha$. We add selection states to Booleans, integers, records and lists. For our present purposes, we are only interested in dependencies between first-order data, so closures are not (directly) selectable, although they have selectable parts; closures also capture the \emph{ambient availability} (explained in \secref{data-dependencies:forward-eval} below), which is also a selection state $\alpha$. We parameterise the type $\Set{Term}$ of terms in the same way, but only add selection states to the term constructors corresponding to selectable values, allowing us to trace data dependencies back to expressions that appear in the source code. We return to this in \secref{surface-language}.

\begin{figure}[H]
{\small
\begingroup
\renewcommand*{\arraystretch}{1}
\begin{minipage}[t]{0.5\textwidth}
   \begin{tabularx}{\textwidth}{rL{2.5cm}L{3cm}}
   &\textbfit{Selectable term}&
   \\
   $e \in \TermF{\Ann{A}} ::=$
   &
   \ldots
   \\
   &
   $\hole$
   &
   hole
   \\
   &
   $\annTrue{\alpha} \mid \annFalse{\alpha}$
   &
   Boolean
   \\
   &
   $\annInt{n}{\alpha}$
   &
   integer
   \\
   &
   $\annRec{\vec{\bind{x}{e}}}{\alpha}$
   &
   record
   \\
   &
   $\annNil{\alpha} \mid \annCons{e}{e'}{\alpha}$
   &
   list
   \\[2mm]
   $\alpha, \beta \in \Ann{A}$
   &&
   \text{selection state}
   \end{tabularx}
\end{minipage}%
\begin{minipage}[t]{0.5\textwidth}
   \begin{tabularx}{\textwidth}{rL{2.5cm}L{3cm}}
   &\textbfit{Selectable value}&
   \\
   $u, v \in \ValF{\Ann{A}}::=$
   &
   $\hole$
   &
   hole
   \\
   &
   $\annTrue{\alpha} \mid \annFalse{\alpha}$
   &
   Boolean
   \\
   &
   $\annInt{n}{\alpha}$
   &
   integer
   \\
   &
   $\annRec{\vec{\bind{x}{v}}}{\alpha}$
   &
   record
   \\
   &
   $\annNil{\alpha} \mid \annCons{u}{v}{\alpha}$
   &
   list
   \\
   &
   $\annClosure{\rho}{h}{\alpha}{\sigma}$
   &
   closure
   \\
   \\[2mm]
   \end{tabularx}
\end{minipage}
\endgroup
}
\caption{Selection states, selectable terms and selectable values}
\label{fig:core-syntax-selection}
\end{figure}

The fact that $\Ann{A}$ is a bounded lattice means it has top and bottom elements $\top$ and $\bot$ representing fully selected/unselected, plus meet and join operations $\meet$ and $\join$ for combining selection information, with $\top$ and $\bot$ as their respective units. Recall the \kw{data} field of the bar chart computed in \figref{introduction:data-linking}, which contained a list of records; the record corresponding to China was $\exRec{\bind{\kw{x}}{\kw{"China"}} \concat \bind{\kw{y}}{\kw{295.3}}}$.
The two-point lattice $\Bool \eqdef \Lattice{\Bool}{\TT}{\FF}{\wedge}{\vee}$ can represent the selection of the field $\kw{y}$ as $\annRec{\bind{\kw{x}}{\annStr{\kw{"China"}}{\FF}} \concat \bind{\kw{y}}{\annInt{\kw{295.3}}{\TT}}}{\FF}$, indicating that the number $\exInt{\kw{295.3}}$ is selected, but neither the string $\exStr{\kw{"China"}}$, nor the record itself is selected. Because lattices are closed under component-wise products, we often write $(\alpha, \beta) \leq (\alpha', \beta')$ to mean that $\alpha \leq \alpha'$ and $\beta \leq \beta'$. This also suggests more interesting lattices of selections, such as vectors of Booleans to represent multiple selections simultaneously, which might be visualised using different colours (as in \figref{introduction:data-linking}).

\begin{figure}
   \input{fig/core-language/slicing/leq-value}
   \caption{Partial order on values}
   \label{fig:data-dependencies:leq}
\end{figure}

\subsubsection{Selections of a value}
\label{sec:data-dependencies:selections}

The analyses which follow will be defined with respect to a fixed computation, and so we will often need to talk about the selections of a given value. To make this notion precise, consider that the raw (unselectable) syntax described in \secref{core-language} can be recovered from a selectable term via an erasure operation $\erase{v}: \ValF{\Ann{A}} \to \ValF{\Unit}$ which forgets the selection information, where $\Unit$ is the trivial one-point lattice. We refer to $\erase{v}$ as the \emph{shape} of $v$. Allowing $\raw{u}, \raw{v}$ from now on to range over raw values, and reserving $u, v$ for selectable values, we can then define:

\begin{definition}[Selections of $\raw{v}$]
   Define $\Sel{\raw{v}}{A}$ to be the set of all values $v \in \ValF{\Ann{A}}$ with shape $\raw{v}$, i.e.
   ~that erase to $\raw{v}$.
\end{definition}

Since its elements have a fixed shape, $\SelF{\raw{v}}$ is a ``representable'' functor: it is isomorphic to a function space whose domain is the set of selection positions in $\raw{v}$. In particular, the pointwise comparison of any $v, v'$ in $\Sel{\raw{v}}{A}$ using the partial order $\leq$ of $\Ann{A}$ is well defined, as is the pointwise application (zip) of a binary operation~\cite{gibbons17}. It should be clear that if $\Ann{A}$ is a lattice, then $\Sel{\raw{v}}{A}$ is also a lattice, with $\top_{\raw{v}}$, $\bot_{\raw{v}}$, $\meet_{\raw{v}}$, and $\join_{\raw{v}}$ defined pointwise. For example, if $u$ and $u'$ have the same shape and $v$ and $v'$ have the same shape, the join of the lists $(\annCons{u}{v}{\alpha})$ and $(\annCons{u'}{v'}{\alpha'})$ is defined and equal to $\annCons{(u \join u')}{(v \join v')}{\alpha \join \alpha'}$. Similarly, the top element of $\Sel{\raw{v}}{A}$ is the selection of $\raw{v}$ which has $\top$ at every selection position. (We omit the $\raw{v}$ indices from the lattice operations if it is clear which lattice is being referred to.) The notion of the ``selections'' of $\raw{v}$ extends to the other syntactic forms.

\subsubsection{Environment selections and hole equivalence}

The notion of a ``selection'' of $\raw{v}$ also extends pointwise to environments, so that $\Sel{\raw{\rho}}{A}$ means the set of selectable environments $\rho'$ of shape $\raw{\rho}$, where the variables in $\rho'$ are bound to selections of the corresponding variables in $\raw{\rho}$. One challenge arises from the pointwise use of $\join$ to combine environment selections. Since environments contain other environments recursively through closures, environment join is a very expensive operation if treated naively. One implementation option is to employ an efficient representation of values which are fully unselected, which is the case for the majority of bindings in an environment.

We therefore enrich the set of selectable values $\ValF{\Ann{A}}$ with a distinguished element \emph{hole}, written $\hole$, which is an alternative notation for $\bot_{\raw{v}}$ for any $\raw{v}$, i.e.~the selectable value of shape $\raw{v}$ which has $\bot$ at every selection position. The equivalence of $\hole$ to any such bottom element is established explicitly by the partial order defined in \figref{data-dependencies:leq}: the first rule always allows $\hole$ on the left-hand side of $\leq$, and other rules allow $\hole$ on the right-hand side of $\leq$ as long as all the selections that appear on the left-hand side are $\bot$. (The rules for recursive definitions $h$, eliminators $\sigma$ and terms $e$ are analogous and are omitted.) If we write $\eq$ for the equivalence relation induced by $\leq$ on selectable values, which we call \emph{hole-equivalence}, it should be clear that $\hole \join v \eq v$ and $\hole \meet v \eq \hole$. This means the join of two selections $v, v'$ of $\raw{v}$ can be implemented efficiently, whenever one selection is $\hole$, by simply discarding $\hole$ and returning the other selection without further processing.

\begin{definition}[Hole equivalence]
   Define $\eq$ as the intersection of $\leq$ and $\geq$.
\end{definition}

Because $\hole$ is equivalent to $\bot_{\raw{v}}$ for any $\raw{v}$, all such bottom elements are hole-equivalent. For example, the selectable value $\annCons{\hole}{\hole}{\top}$ is hole-equivalent to $\annCons{5_{\bot}}{\hole}{\top}$, but also to $\annCons{6_{\bot}}{{\exNil_{\bot}}}{\top}$, and so the last two selections, even though they have different shapes, are hole-equivalent by transitivity. In practice we only use the hole ordering to compare selections with the same shape.

\subsection{Forward data dependency}
\label{sec:data-dependencies:analyses:fwd}

We now define the core bidirectional data dependency analyses for a fixed computation $T :: \raw{\rho}, \raw{e} \evalR \raw{v}$, where $T$ is a trace. In practise one would first evaluate a program to obtain its trace $T$, and then run multiple forward or backward analyses over $T$ with appropriate lattices. We start with the forward dependency function $\evalFwdF{T}$ which ``replays'' evaluation, turing input availability into output availability, with $T$ guiding the analysis whenever holes in the input selection would mean the analysis would otherwise get stuck. $\evalFwdF{T}$ uses the auxiliary function $\matchFwdF{w}$ for forward-analysing a pattern-match, which introduces the key idea of a selection as the data available to a downstream computation.

\begin{figure}
{\small \flushleft \shadebox{$\matchFwd{v}{\sigma}{w}{\rho}{\kappa}{\alpha}$}%
\hfill \textbfit{$v$ and $\sigma$ forward-match along $w$ to $\rho$ and $\kappa$, with ambient availability $\alpha$}}
\begin{smathpar}
   \inferrule*[
      lab={\ruleName{$\matchFwdS$-hole-1}}
   ]
   {
      \hole \eq v
      \\
      \matchFwd{v}{\sigma}{w}{\rho}{\kappa}{\alpha}
   }
   {
      \matchFwd{\hole}{\sigma}{w}{\rho}{\kappa}{\alpha}
   }
   \and
   \inferrule*[
      lab={\ruleName{$\matchFwdS$-hole-2}}
   ]
   {
      \hole \eq \sigma
      \\
      \matchFwd{v}{\sigma}{w}{\rho}{\kappa}{\alpha}
   }
   {
      \matchFwd{v}{\hole}{w}{\rho}{\kappa}{\alpha}
   }
   \and
   \inferrule*[
      lab={\ruleName{$\matchFwdS$-var}}
   ]
   {
      \strut
   }
   {
      \matchFwd{v}{\elimVar{x}{\kappa}}{\matchVar{x}}{\bind{x}{v}}{\kappa}{\TT}
   }
   \and
   \inferrule*[
      lab={\ruleName{$\matchFwdS$-true}}
   ]
   {
      \strut
   }
   {
      \matchFwd{\annTrue{\alpha}}
               {\elimBool{\kappa}{\kappa'}}
               {\matchTrue}
               {\seqEmpty}{\kappa}{\alpha}
   }
   \and
   \inferrule*[
      lab={\ruleName{$\matchFwdS$-false}}
   ]
   {
      \strut
   }
   {
      \matchFwd{\annFalse{\alpha}}
               {\elimBool{\kappa}{\kappa'}}
               {\matchFalse}
               {\seqEmpty}{\kappa'}{\alpha}
   }
   \and
   \inferrule*[
      lab={\ruleName{$\matchFwdS$-unit}}
   ]
   {
      \strut
   }
   {
      \matchFwd{\annot{\exRecEmpty}{\alpha}}
               {\elimRecEmpty{\kappa}}
               {\matchRecEmpty}
               {\seqEmpty}
               {\kappa}
               {\alpha}
   }
   \and
   \inferrule*[
      lab={\ruleName{$\matchFwdS$-record}}
   ]
   {
      \matchFwd{\annRec{\vec{\bind{x}{v}}}{\top}}
               {\elimRec{\vec{x}}{\sigma}}
               {\matchRec{\vec{\bind{x}{w}}}}
               {\rho}
               {\sigma'}
               {\beta}
      \\
      \matchFwd{u}{\sigma'}{w}{\rho'}{\kappa}{\beta'}
   }
   {
      \matchFwd{\annRec{\vec{\bind{x}{v}} \concat \bind{y}{u}}{\alpha}}
               {\elimRec{\vec{x} \concat y}{\sigma}}
               {\matchRec{\vec{\bind{x}{w}} \concat \bind{y}{w'}}}
               {\rho \concat \rho'}
               {\kappa}
               {\alpha \meet \beta \meet \beta'}
   }
   \and
   \inferrule*[
      lab={\ruleName{$\matchFwdS$-nil}}
   ]
   {
      \strut
   }
   {
      \matchFwd{\annNil{\alpha}}
               {\elimList{\kappa}{\sigma'}}
               {\matchNil}
               {\seqEmpty}{\kappa}{\alpha}
   }
   \and
   \inferrule*[
      lab={\ruleName{$\matchFwdS$-cons}}
   ]
   {
      \matchFwd{v}{\sigma}{w}{\rho}{\tau}{\beta}
      \\
      \matchFwd{v'}{\tau}{w'}{\rho'}{\kappa}{\beta'}
   }
   {
      \matchFwd{\annCons{v}{v'}{\alpha}}
               {\elimList{\kappa}{\sigma}}
               {\matchCons{w}{w'}}
               {\rho \concat \rho'}{\kappa}{\alpha \meet \beta \meet \beta'}
   }
\end{smathpar}

\vspace{2mm}
{\small \flushleft \shadebox{$\evalFwd{\rho}{e}{\alpha}{T}{v}$}%
\hfill \textbfit{$\rho$ and $e$, with ambient availability $\alpha$, forward-evaluate along $T$ to $v$}}
\begin{smathpar}
   \mprset{center}
   \inferrule*[
      lab={\ruleName{$\evalFwdS$-hole}}
   ]
   {
      \hole \eq e
      \\
      \evalFwd{\rho}{e}{\alpha}{T}{v}
   }
   {
      \evalFwd{\rho}{\hole}{\alpha}{T}{v}
   }
   \and
   \inferrule*[
      lab={\ruleName{$\evalFwdS$-var}}
   ]
   {
      \envLookup{\rho}{x}{v}
   }
   {
      \evalFwd{\rho}{\exVar{x}}{\alpha}{\trVar{x}{\rho}}{v}
   }
   \and
   \inferrule*[lab={\ruleName{$\evalFwdS$-lambda}}]
   {
      \strut
   }
   {
      \evalFwd{\rho}
              {\exLambda{\sigma}}
              {\alpha}
              {\trLambda{\sigma'}}
              {\annClosure{\rho}{\seqEmpty}{\alpha}{\sigma}}
   }
   \and
   \inferrule*[lab={\ruleName{$\evalFwdS$-int}}]
   {
      \strut
   }
   {
      \evalFwd{\rho}
              {\annInt{n}{\alpha'}}
              {\alpha}
              {\trInt{n}{\rho}}
              {\annInt{n}{\alpha \meet \alpha'}}
   }
   \and
   \inferrule*[lab={\ruleName{$\evalFwdS$-record}}]
   {
      \evalFwd{\rho}{e_i}{\alpha}{T_i}{v_i}
      \quad
      (\forall i \numleq \length{\vec{x}})
   }
   {
      \evalFwd{\rho}
              {\annRec{\vec{\bind{x}{e}}}{\alpha'}}
              {\alpha}
              {\trRec{\vec{\bind{x}{T}}}}
              {\annRec{\vec{\bind{x}{v}}}{\alpha \meet \alpha'}}
   }
   \and
   \inferrule*[
      lab={\ruleName{$\evalFwdS$-project}}
   ]
   {
      \evalFwdEq{\rho}{e}{\alpha}{T}{\annRec{\vec{\bind{x}{u}}}{\beta}}
      \\
      \envLookup{\vec{\bind{x}{u}}}{y}{v'}
   }
   {
      \evalFwd{\rho}
              {\exRecProj{e}{y}}
              {\alpha}
              {\trRecProj{T}{\vec{\bind{x}{v}}}{y}}
              {v'}
   }
   \and
   \inferrule*[lab={\ruleName{$\evalFwdS$-nil}}]
   {
      \strut
   }
   {
      \evalFwd{\rho}
              {\annNil{\alpha'}}
              {\alpha}
              {\trNil{\rho}}
              {\annNil{\alpha \meet \alpha'}}
   }
   \and
   \inferrule*[
      lab={\ruleName{$\evalFwdS$-cons}},
   ]
   {
      \evalFwd{\rho}{e}{\alpha}{T}{v}
      \\
      \evalFwd{\rho}{e'}{\alpha}{U}{v'}
   }
   {
      \evalFwd{\rho}
              {\annCons{e}{e'}{\alpha'}}
              {\alpha}
              {\trCons{T}{U}}
              {\annCons{v}{v'}{\alpha \meet \alpha'}}
   }
   \and
   \inferrule*[
      lab={\ruleName{$\evalFwdS$-apply-prim}},
      right={$\primFwdBool{\phi}{\vec{n}}(\vec{\beta}) = \alpha'$}
   ]
   {
      \evalFwdEq{\rho}{e_i}{\alpha}{U_i}{\annInt{{n_i}}{\beta_i}}
      \quad
      (\forall i \numleq \length{\vec{n}})
   }
   {
      \evalFwd{\rho}
              {\exAppPrim{\phi}{\vec{e}}}
              {\alpha}
              {\trAppPrimNew{\phi}{U}{n}}
              {\annInt{\exAppPrim{\phi}{\vec{n}}}{\alpha'}}
   }
   \and
   \inferrule*[
      lab={\ruleName{$\evalFwdS$-apply}},
      width={3.4in}
   ]
   {
      \evalFwdEq{\rho}{e}{\alpha}{T}{\annClosure{\rho_1}{h}{\beta}{\sigma}}
      \\
      \rho_1, h, \beta \closeDefsFwdR \rho_2
      \\
      \evalFwd{\rho}{e'}{\alpha}{U}{v}
      \\
      \matchFwd{v}{\sigma}{w}{\rho_3}{e^\twoPrime}{\beta'}
      \\
      \evalFwd{\rho_1 \concat \rho_2 \concat \rho_3}{e^\twoPrime}{\beta \meet \beta'}{T'}{v'}
   }
   {
      \evalFwd{\rho}{\exApp{e}{e'}}{\alpha}{\trApp{T}{U}{w}{T'}}{v'}
   }
   \and
   \inferrule*[lab={
      \ruleName{$\evalFwdS$-let-rec}}
   ]
   {
      \rho, h', \alpha \closeDefsFwdR \rho'
      \\
      \evalFwd{\rho \concat \rho'}{e}{\alpha}{T}{v}
   }
   {
      \evalFwd{\rho}{\exLetRecMutual{h'}{e}}{\alpha}{\trLetRecMutual{h}{T}}{v}
   }
\end{smathpar}
\vspace{2mm}
{\small \flushleft \shadebox{$\rho, h, \alpha \closeDefsFwdR \rho'$}%
\hfill \textbfit{$h$ forward-generates to $\rho'$ in $\rho$ and $\alpha$}}
\begin{smathpar}
   \inferrule*[
      lab={\ruleName{$\closeDefsFwdR$-rec-defs}}
   ]
   {
      v_i = \annClosure{\rho}{\vec{\bind{x}{\sigma}}}{\alpha}{\sigma_i}
      \quad
      (\forall i \in \length{\vec{x}})
   }
   {
      \rho, \vec{\bind{x}{\sigma}}, \alpha
      \closeDefsFwdR
      \vec{\bind{x}{v}}
   }
\end{smathpar}
\caption{Forward data dependency (Boolean cases for $\evalFwdR{T}$ omitted)}
\label{fig:data-dependencies:fwd}
\end{figure}

\subsubsection{Forward match}
\label{sec:data-dependencies:analyses:fwd:pattern-match}

\figref{data-dependencies:fwd} defines a family of \emph{forward-match} functions $\matchFwdF{w}$ of type $\Sel{\raw{v}, \raw{\sigma}}{A} \to \Sel{\raw{\rho}, \raw{\kappa}}{A} \times \Ann{A}$ for any $w :: \raw{v}, \raw{\sigma} \matchR \raw{\rho}, \raw{\kappa}$. (The definition is presented in a relational style for readability, but should be understood as a total function defined by structural recursion on $w$, which appears in grey to emphasise the connection to \figref{core-language:semantics}.) Forward-match replays the match witnessed by $w$, transferring the selections on the relevant parts of $v \in \Sel{\raw{v}}{A}$ to the output environment $\rho \in \Sel{\raw{\rho}}{A}$, and from the relevant part of $\sigma \in \Sel{\raw{\sigma}}{A}$ to the chosen continuation $\kappa \in \Sel{\raw{\kappa}}{A}$.

$\matchFwdF{w}$ also returns the \emph{meet} of the selection states associated with the part of $v$ consumed by $\sigma$, which we call the \emph{availability} of $v$ (in the context of $\sigma$), since it represents the extent to which the demand implied by $\sigma$ was met. A variable match consumes nothing of $v$ and so the availability is simply $\top$, the unit for meet. A Boolean match consumes either $\annot{\exTrue}{\alpha}$ or $\annot{\exFalse}{\alpha}$, with availability $\alpha$; empty list and record matches are similar. When we match a cons, we return the meet of the $\alpha$ on the cons node itself with the availabilities $\beta$ and $\beta'$ computed for $v$ and $v'$. Non-empty records are similar, but to process the initial part of the record, we supply the neutral selection state $\top$ on the subrecord in order to use the definition recursively. (Subrecords are not first-class, but exist only as intermediate artefacts of the interpreter.)

One might imagine dispensing with the need for $w$ by simply defining $\matchFwdF{w}$ by case analysis on $v$ and $\sigma$. However, it is then unclear how to proceed in the event that $v$ is a hole. It would be legitimate to produce $\hole$ as the output continuation and $\bot$ as the output selection state, but for the output environment to be well-typed, it must provide variable bindings corresponding to those introduced in the baseline computation where $\raw{v}$ was matched against $\raw{\sigma}$. If $\matchFwdS$ is defined with respect to a known $w$, this can be achieved via an additional rule \ruleName{$\matchFwdS$-hole-1} that defines the behaviour at hole to be the same as the behaviour at any $\eq$-equivalent value in $\Sel{\raw{v}}{A}$. The \ruleName{$\matchFwdS$-hole-2} rule makes a similar provision for $\sigma$, which may also be a hole. Operationally, these rules can be interpreted as ``expanding'' enough of the holes in $v$ or $\sigma$ to make another rule of the definition match. There will be exactly one non-hole rule that matches, corresponding to the execution path originally taken, and although there may be multiple such expansions, the following property implies that the result of $\matchFwdF{w}$ will be unique up to $\eq$.

\begin{lemma}[Monotonicity of $\matchFwdF{w}$]
   Suppose $w :: \raw{v}, \raw{\sigma} \matchR \raw{\rho}, \raw{\kappa}$, with $v, \sigma \matchFwdR{w} \rho, \kappa, \alpha$ and $v', \sigma' \matchFwdR{w} \rho', \kappa', \alpha'$. If $(v, \sigma) \leq (v', \sigma)$ then $(\rho, \kappa, \alpha) \leq (\rho', \kappa', \alpha')$.
\end{lemma}

\subsubsection{Forward evaluation}
\label{sec:data-dependencies:forward-eval}

\figref{data-dependencies:fwd} also defines a family of \emph{forward-evaluation} functions $\evalFwdF{T}$ of type $\Sel{\raw{\rho}, \raw{e}}{A} \times \Ann{A} \to \Sel{\raw{v}}{A}$ for any $T :: \raw{\rho}, \raw{e} \evalR \raw{v}$. Like forward-match, forward-evaluation is presented in a relational style, but should be understood as a total function defined by structural recursion on $T$. Forward evaluation replays $T$, using the input selection $(\rho, e) \in \Sel{\raw{\rho}, \raw{e}}{A}$ to determine an output selection $v \in \Sel{\raw{v}}{A}$. The rules mirror those of the evaluation relation $\evalR$, although there is an additional selection state input $\alpha$ called the \emph{ambient availability}. We explain this with reference to the application rule, which is the only point where a new ambient availability is assigned.

\paragraph{Function application} The rule assumes the application $\exApp{e}{e'}$ already has an ambient availability $\alpha$; at the outermost level this will usually be $\top$. The rule passes $\alpha$ down when recursively forward-evaluating $e$ and $e'$, but computes a new selection state $\beta \meet \beta'$ when transferring control to the function, combining ambient availability $\beta$ captured by the closure and $\beta'$ obtained by forward-matching the argument $v$ with the eliminator $\sigma$ of the closure, representing the availability of those parts of $v$ demanded by the function. The ambient availability is used to upper-bound the availability of any selectable values constructed in the dynamic context of that function, establishing a dependency between the resources consumed by functions and resources they produce. The auxiliary function $\closeDefsFwdF{\rho,h}: \Sel{\raw{\rho},\raw{h}}{A} \times \Ann{A} \to \Sel{\raw{\rho}'}{A}$ for any $\smash{\raw{\rho}, \raw{h} \closeDefsR \raw{\rho'}}$ is given at the bottom of \figref{data-dependencies:fwd} and follows $\closeDefsR$, but captures the ambient availability into each closure.

\paragraph{Primitive application} Primitive operations are the other source of input-output dependencies beyond user-defined functions. Since a primitive operation is opaque, these dependencies cannot be derived from its execution, but must be specified by the primitive operation directly. More specifically, $\phi \in \tyInt^{i} \to \tyInt$ is required to provide a forward-dependency function $\primFwdBool{\phi}{\vec{n}}: \Ann{A}^i \to \Ann{A}$ for every $\vec{n} \in \tyInt^i$ which specifies how to turn an input selection $\vec{\alpha} \in \Ann{A}^i$ for $\vec{n}$ into an output selection $\alpha'$ on $\exAppPrim{\phi}{\vec{n}}$. There is one such function per possible input $\vec{n}$, since the dynamic dependencies for a primitive operation with an annihilator, such as multiplication, depend on the values passed to the operation. Primitives are free to implement forward-dependency however they want, except that \secref{data-dependencies:analyses:bwd:eval} will also require $\phi$ to provide a backward-dependency function for any input $\vec{n}$, and \secref{data-dependencies:galois-connections} will require these to be related in a certain way for the consistency of the whole system to be guaranteed.

\paragraph{Other rules} All other rules pass the ambient availability into any subcomputations unchanged. Variable lookup disregards the ambient $\alpha$, simply preserving the selection on the returned value. The lambda rule captures it in the closure, along with the environment; the letrec rule passes it on to $\closeDefsFwdR$ so it is captured by recursive closures as well. Record projection is more interesting, disregarding not only the ambient $\alpha$ but also the availability $\beta$ of the record itself. Containers are considered to be independent of the values they contain: here, $v_i$ has its own internal availability which is preserved by projection, but there is no implied dependency of the field on the record from which it was projected. Record construction also reflects this principle, preserving the field selections into the resulting record selection unchanged. But it also sets the availability of the record value to be the meet of $\alpha$ and $\alpha'$, reflecting the dependency of the container on the constructing expression and also on any resources consumed by the ambient function. The rules for nil, cons and integers are similar.

\paragraph{Hole case} A hole rule is also needed, because if $e$ is $\hole$, forward evaluation must continue; in particular, subsequent steps may extract non-$\hole$ values from $\rho$ and result in non-$\hole$ outputs. The rule is similar to the rules for $\matchFwdF{w}$ and again can be understood operationally as using the information in $T$ to expand $\hole$ sufficiently for another rule to apply, with a result which is unique up to $\eq$. Environments have no special $\hole$ form. However, application and record projection must accommodate the case where the selection on the closure or record being eliminated is represented by $\hole$. In these rules $\evalFwdR{T}\eq$ is used to denote the relational composition of $\evalFwdR{T}$ and $\eq$.

\begin{lemma}[Monotonicity of $\evalFwdF{T}$]
   Suppose $T :: \raw{\rho}, \raw{e} \evalR \raw{v}$ with $\rho, e, \alpha \evalFwdR{T} v$ and $\rho', e', \alpha' \evalFwdR{T} v'$. If $(\rho, e, \alpha) \leq (\rho', e', \alpha')$ then $v \leq v'$.
\end{lemma}

\subsection{Backward data dependency}
\label{sec:data-dependencies:analyses:bwd}

The backward dependency function $\evalBwdF{T}$ ``rewinds'' evaluation, turning output demand into input demand, with $T$ guiding the analysis backward. We start with the auxiliary function $\matchBwdF{w}$ which is used for backward-analysing a pattern-match.

\subsubsection{Backward match}
\label{sec:data-dependencies:analyses:bwd:pattern-match}

\figref{data-dependencies:bwd} defines a family of \emph{backward-match} functions $\matchBwdF{w}$ of type $\Sel{\raw{\rho}, \raw{\kappa}}{A} \times \Ann{A} \to \Sel{\raw{v}, \raw{\sigma}}{A}$ for any $w :: \raw{v}, \raw{\sigma} \matchR \raw{\rho}, \raw{\kappa}$. Backward-match rewinds the match witnessed by $w$, turning demand on the environment and continuation into demand on the value and eliminator that were originally matched. The additional input $\alpha$ represents the downstream demand placed on any resources that were constructed in the context of this match; $\matchBwdF{w}$ transfers this to the matched portion of $\raw{v}$, establishing a backwards link between resources produced and resources consumed in a given function context.

\begin{figure}
   {\small \flushleft \shadebox{$\matchBwd{\rho}{\kappa}{\alpha}{w}{v}{\sigma}$}%
   \hfill \textbfit{$\rho$ and $\kappa$, with ambient demand $\alpha$, backward-match along $w$ to $v$ and $\sigma$}}
   \begin{smathpar}
      \inferrule*[lab={\ruleName{$\matchBwdS$-true}}]
      {
         \strut
      }
      {
         \matchBwd{\seqEmpty}{\kappa}{\alpha}{\matchTrue}{\annTrue{\alpha}}{\elimBool{\kappa}{\hole}}
      }
      \and
      \inferrule*[lab={\ruleName{$\matchBwdS$-false}}]
      {
         \strut
      }
      {
         \matchBwd{\seqEmpty}{\kappa}{\alpha}{\matchFalse}{\annFalse{\alpha}}{\elimBool{\hole}{\kappa}}
      }
      \inferrule*[lab={\ruleName{$\matchBwdS$-var}}]
      {
         \strut
      }
      {
         \matchBwd{\bind{x}{v}}{\kappa}{\alpha}{\matchVar{x}}{v}{\elimVar{x}{\kappa}}
      }
      \and
      \inferrule*[lab={\ruleName{$\matchBwdS$-unit}}]
      {
         \strut
      }
      {
         \matchBwd{\seqEmpty}
                  {\kappa}
                  {\alpha}
                  {\matchRecEmpty}
                  {\annot{\exRecEmpty}{\alpha}}
                  {\elimRecEmpty{\kappa}}
      }
      \and
      \inferrule*[lab={\ruleName{$\matchBwdS$-nil}}]
      {
         \strut
      }
      {
         \matchBwd{\seqEmpty}{\kappa}{\alpha}{\matchNil}{\annNil{\alpha}}{\elimList{\kappa}{\hole}}
      }
      \and
      \inferrule*[lab={\ruleName{$\matchBwdS$-record}}]
      {
         \matchBwd{\rho'}{\kappa}{\alpha}{w'}{u}{\sigma}
         \\
         \matchBwd{\rho}
                  {\sigma}
                  {\alpha}
                  {\matchRec{\vec{\bind{x}{w}}}}
                  {\annRec{\vec{\bind{x}{v}}}{\beta}}
                  {\tau}
      }
      {
         \matchBwd{\rho \concat \rho'}
                  {\kappa}
                  {\alpha}
                  {\matchRec{\vec{\bind{x}{w}} \concat \bind{y}{w'}}}
                  {\annRec{\vec{\bind{x}{v}} \concat \bind{y}{u}}{\alpha}}
                  {\elimRec{\vec{x} \concat y}{\tau}}
      }
      \and
      \inferrule*[lab={\ruleName{$\matchBwdS$-cons}}]
      {
         \matchBwd{\rho'}{\kappa}{\alpha}{w'}{v'}{\sigma}
         \\
         \matchBwd{\rho}{\sigma}{\alpha}{w}{v}{\tau}
      }
      {
         \matchBwd{\rho \concat \rho'}
                  {\kappa}
                  {\alpha}
                  {\matchCons{w}{w'}}
                  {\annCons{v}{v'}{\alpha}}
                  {\elimList{\hole}{\tau}}
      }
   \end{smathpar}
   \vspace{1mm}

   {\small \flushleft \shadebox{$\evalBwd{v}{T}{\rho}{e}{\alpha}$}%
   \hfill \textbfit{$v$ backward-evaluates along $T$ to $\rho$ and $e$, with ambient demand $\alpha$}}
   \begin{smathpar}
      \inferrule*[
         lab={\ruleName{$\evalBwdS$-eq}}
      ]
      {
         \evalBwd{v}{T}{\rho}{e}{\alpha}
      }
      {
         \evalBwd{v \eq}{T}{\rho}{e}{\alpha}
      }
      \and
      \inferrule*[
         lab={\ruleName{$\evalBwdS$-var}}
      ]
      {
         \envLookupBwd{\rho'}{\rho}{\bind{x}{v}}
      }
      {
         \evalBwd{v}{\trVar{x}{\rho}}{\rho'}{\exVar{x}}{\bot}
      }
      \and
      \inferrule*[
         lab={\ruleName{$\evalBwdS$-lambda}}
      ]
      {
         \strut
      }
      {
         \evalBwd{\annClosure{\rho}{\seqEmpty}{\alpha}{\sigma}}
                 {\trLambda{\sigma'}}
                 {\rho}
                 {\exLambda{\sigma}}
                 {\alpha}
      }
      \and
      \inferrule*[
         lab={\ruleName{$\evalBwdS$-int}}
      ]
      {
         \strut
      }
      {
         \evalBwd{\annInt{n}{\alpha}}
                 {\trInt{n}{\rho}}
                 {\hole_{\raw{\rho}}}
                 {\annInt{n}{\alpha}}
                 {\alpha}
      }
      \and
      \inferrule*[lab={\ruleName{$\evalBwdS$-record}}]
      {
         \evalBwd{v_i}{T_i}{\rho_i}{e_i}{\alpha_i'}
         \quad
         (\forall i \numleq \length{\vec{x}})
      }
      {
         \evalBwd{\annRec{\vec{\bind{x}{v}}}{\alpha}}
                 {\trRec{\vec{\bind{x}{T}}}}
                 {\bigjoin\vec{\rho}}
                 {\annRec{\vec{\bind{x}{e}}}{\alpha}}
                 {\alpha \join \bigjoin\vec{\alpha}'}
      }
      \and
      \inferrule*[lab={\ruleName{$\evalBwdS$-project}}]
      {
         \envLookupBwd{\vec{\bind{x}{u}}}{\vec{\bind{x}{v}}}{\bind{y}{v'}}
         \\
         \evalBwd{\annRec{\vec{\bind{x}{u}}}{\bot}}
                 {T}
                 {\rho}
                 {e}
                 {\alpha}
      }
      {
         \evalBwd{v'}
                 {\trRecProj{T}{\vec{\bind{x}{v}}}{y}}
                 {\rho}
                 {\exRecProj{e}{y}}
                 {\alpha}
      }
      \and
      \inferrule*[
         lab={\ruleName{$\evalBwdS$-nil}}
      ]
      {
         \strut
      }
      {
         \evalBwd{\annNil{\alpha}}
                 {\trNil{\rho}}
                 {\hole_{\raw{\rho}}}
                 {\annNil{\alpha}}
                 {\alpha}
      }
      \and
      \inferrule*[
         lab={\ruleName{$\evalBwdS$-cons}}
      ]
      {
         \evalBwd{v}{T}{\rho}{e}{\alpha}
         \\
         \evalBwd{v'}{U}{\rho'}{e'}{\alpha'}
      }
      {
         \evalBwd{\annCons{v}{v'}{\beta}}
                 {\trCons{T}{U}}
                 {\rho \join \rho'}
                 {\annCons{e}{e'}{\beta}}
                 {\beta \join \alpha \join \alpha'}
      }
      \and
      \inferrule*[
         lab={\ruleName{$\evalBwdS$-let-rec}}
      ]
      {
         \evalBwd{v}{T}{\rho \concat \rho_1}{e}{\alpha}
         \\
         \rho_1 \closeDefsBwdR \rho', h', \alpha'
      }
      {
         \evalBwd{v}{\trLetRecMutual{h}{T}}{\rho \join \rho'}{\exLetRecMutual{h'}{e}}{\alpha \join \alpha'}
      }
      \and
      \inferrule*[
         lab={\ruleName{$\evalBwdS$-apply-prim}},
         right={$\primBwdBool{\phi}{\vec{n}}(\alpha') = \vec{\alpha}$}
      ]
      {
         \evalBwd{\annInt{{n_i}}{\alpha_i}}{U_i}{\rho_i}{e_i}{\beta_i}
         \quad
         (\forall i \in \length{\vec{n}})
      }
      {
         \evalBwd{\annInt{m}{\alpha'}}
                 {\trAppPrimNew{\phi}{U}{n}}
                 {\bigjoin\vec{\rho}}
                 {\exAppPrim{\phi}{\vec{e}}}
                 {\bigjoin\vec{\beta}}
      }
      \and
      \inferrule*[
         lab={\ruleName{$\evalBwdS$-apply}},
         width={3.3in}
      ]
      {
         \evalBwd{v}{T'}{\rho_1 \concat \rho_2 \concat \rho_3}{e}{\beta}
         \\
         \matchBwd{\rho_3}{e}{\beta}{w}{v'}{\sigma}
         \\
         \evalBwd{v'}{U}{\rho}{e_2}{\alpha}
         \\
         \rho_2 \closeDefsBwdR \rho_1', h, \beta'
         \\
         \evalBwd{\annClosure{\rho_1 \join \rho_1'}{h}{\beta \join \beta'}{\sigma}}{T}{\rho'}{e_1}{\alpha'}
      }
      {
         \evalBwd{v}{\trApp{T}{U}{w}{T'}}{\rho \join \rho'}{\exApp{e_1}{e_2}}{\alpha \join \alpha'}
      }
   \end{smathpar}
   \vspace{1mm}

\begin{minipage}[t]{0.48\textwidth}%
   {\small\flushleft \shadebox{$\envLookupBwd{\rho'}{\rho}{\bind{x}{v}}$}%
   \hfill \textbfit{$\rho'$ contains $\bind{x}{v}$}}
   \begin{smathpar}
      \inferrule*[
         lab={\ruleName{$\envLookupBwdS$-head}}
      ]
      {
         \strut
      }
      {
         \envLookupBwd{(\hole_{\raw{\rho}} \concat \bind{x}{u})}
                      {\rho \concat \bind{x}{v}}
                      {\bind{x}{u}}
      }
      \and
      \inferrule*[
         lab={\ruleName{$\envLookupBwdS$-tail}},
      ]
      {
         \envLookupBwd{\rho'}{\rho}{\bind{x}{u}}
         \\
         x \neq y
      }
      {
         \envLookupBwd{(\rho' \concat \bind{y}{\hole})}
                      {\rho \concat \bind{y}{v}}
                      {\bind{x}{u}}
      }
   \end{smathpar}
\end{minipage}%
\hfill
\begin{minipage}[t]{0.47\textwidth}%
   {\small\flushleft\shadebox{$\rho \closeDefsBwdR \rho', h, \alpha$}%
   \hfill \textbfit{$\rho$ backward-generates to $\rho'$, $h$, $\alpha$}}
   \begin{smathpar}
      \inferrule*[
         lab={\ruleName{$\closeDefsBwdR$-rec-defs}}
      ]
      {
         v_i = \annClosure{\rho_i}{h_i}{\alpha_i}{\sigma_i}
         \quad
         (\forall i \in \length{\vec{x}})
      }
      {
         \vec{\bind{x}{v}}
         \closeDefsBwdR
         \bigjoin\vec{\rho}, \vec{\bind{x}{\sigma}} \join {\bigjoin\vec{h}}, \bigjoin{\vec{\alpha}}
      }
   \end{smathpar}
\end{minipage}

   \caption{Backward data dependency (Boolean cases for $\evalBwdR{T}$ omitted)}
   \label{fig:data-dependencies:bwd}
   \end{figure}

In the variable case, no proper part of $\raw{v}$ was matched, so $\alpha$ is disregarded. The rule need only ensure that the demand $v$ in the singleton environment $\bind{x}{v}$ is propagated backward. If a Boolean constant was matched, $\alpha$ becomes the demand on that constant, and $\kappa$, capturing the demand on the continuation, is used to construct the demand on the original eliminator, with $\hole$ used to represent the absence of demand on the non-taken branch. (Using $\hole$ for this means matches $w$ need only retain information about taken branches.) The nil case is similar.

For a cons match $\matchCons{w}{w'}$, we split the environment into $\rho$ and $\rho'$ (there is a unique well-typed decomposition) and then backward-match $w$ and $w'$ recursively to obtain $v$ and $v'$, representing the demand on the head and tail of the list. These are combined into the demand on the entire list, using $\alpha$ as the demand on the cons node itself. $\sigma$ represents the demand on the interim eliminator used to match the tail, and $\tau$ the demand on the eliminator used to match the head, which are then combined into a demand on the eliminator used to match the whole list, with $\hole$ again used to represent the absence of demand on the nil branch. Records are similar, except there is only a single branch, and the selection state $\beta$ computed for the initial part of the record is an artefact of processing records recursively, and is disregarded.

\begin{lemma}[Monotonicity of $\matchBwdF{w}$]
   Suppose $w :: \raw{v}, \raw{\sigma} \matchR \raw{\rho}, \raw{\kappa}$, with $\rho, \kappa, \alpha \matchBwdR{w} v, \sigma$ and $\rho', \kappa', \alpha' \matchBwdR{w} v', \sigma'$. If $(\rho, \kappa, \alpha) \leq (\rho', \kappa', \alpha')$ then $(v, \sigma) \leq (v', \sigma)$.
\end{lemma}

\subsubsection{Backward evaluation}
\label{sec:data-dependencies:analyses:bwd:eval}

\figref{data-dependencies:bwd} also defines a family of \emph{backward-evaluation} functions $\evalBwdF{T}$ of type $\Sel{\raw{v}}{A} \to \Sel{\raw{\rho}, \raw{e}}{A} \times \Ann{A}$ for any $T :: \raw{\rho}, \raw{e} \evalR \raw{v}$. Backward evaluation rewinds $T$, using the output selection $v \in \Sel{\raw{v}}{A}$ to determine an input selection $(\rho, e) \in \Sel{\raw{\rho}, \raw{e}}{A}$. The rules resemble those of the evaluation relation $\evalR$ with inputs and outputs flipped, and with an additional output $\alpha$ called the \emph{ambient demand}. The general pattern is that each backward rule takes the join of the demand attached to any partial values constructed at that step, and the ambient demand associated with any subcomputations, and passes it upwards as the new ambient demand. The output environment is constructed similarly, by joining the demand flowing back through the environment copies used to evaluate subcomputations. Demand is also attached to the source expression when it is the expression form responsible for the construction of a demanded value.

\paragraph{Function application} The application rule is where the ambient demand is used and the function context changes, so we start here. The rule essentially runs the forward evaluation rule in reverse, using the trace $T'$ to backward-evaluate the function body. The ambient demand $\beta$ associated with $T'$ is the join of the demand on any resources constructed directly by that function invocation, and is transferred to the matched part of the function argument by the backward-match function $\matchBwdF{w}$. The ambient demand passed upwards into the enclosing function context is $\alpha \join \alpha'$, representing the resources needed along $T$ and $U$. The auxiliary function $\smash{\closeDefsBwdF{\rho, h}}: \smash{\Sel{\raw{\rho'}}{A} \to \Sel{\raw{\rho}, \raw{h}}{A}} \times \Ann{A}$ for any $\raw{\rho}, \raw{h} \closeDefsR \raw{\rho'}$ defined at the bottom of \figref{data-dependencies:bwd} is used to turn $\rho_2$, capturing the demand flowing back through any recursive uses of the function and any others with which it was mutually defined, into information that can be merged back into the demand on the closure. The function $\closeDefsBwdF{\rho,h}$ is also used in the letrec rule, which otherwise follows the generic pattern described above.

\paragraph{Primitive application} Each primitive operation $\phi: \tyInt^{i} \to \tyInt$ must provide a backward-dependency  function $\primBwdBool{\phi}{\vec{n}}: \Ann{A} \to \Ann{A}^i$ for every $\vec{n} \in \tyInt^i$ which specifies how to turn the output selection $\alpha'$ on $\exAppPrim{\phi}{\vec{n}}$ into an input selection $\vec{a} \in \Ann{A}^i$ on $\vec{n}$. The rule for primitive application uses this information to pair each argument $n_i$ with its demand $\alpha_i$ and then backwards-evaluate the argument. The ambient demand passed upward is the join of those arising from these subcomputations, and is unrelated to the execution of the primitive itself, similar to a function application. Here $\bigjoin{\vec{\beta}}$ means the fold of $\join$ (with unit $\bot$) over the sequence of selection states $\seqRange{\beta_1}{\beta_{\length{\vec{x}}}'}$. Environment demands $\vec{\rho} = \seqRange{\rho_1}{\rho_{\length{\vec{n}}}}$ are joined (pointwise) in a similar fashion.

\paragraph{Other rules} In the variable case, no partial values were constructed during evaluation and there are no subcomputations, so the ambient demand is $\bot$, the unit for $\join$. The returned environment selection demands $v$ for the variable $x$ and $\hole$ for all other variables, using the family of \emph{backwards lookup} functions $\envLookupBwdF{-}{\rho}{x}{-}$ of type $\Sel{\raw{v}}{A} \to \Sel{\raw{\rho}}{A}$ for any $\envLookup{\raw{\rho}}{x}{\raw{v}}$ also defined in \figref{data-dependencies:bwd}. (The output of the function is on the left in the relational notation.) For atomic values such as integers and nil, the ambient demand is simply the demand $\alpha$ associated with the constructed value, which is also attached to the corresponding expression, and the environment demand has $\hole$ for every variable in the original environment $\raw{\rho}$, written $\hole_{\raw{\rho}}$.

For closures, the ambient demand is unpacked along with the other components, preserving any internal selections on $\rho$ and $\sigma$. Composite values such as records and cons cells follow the general pattern; thus for records, the ambient demands $\alpha'_i$ for the subcomputations are joined with the $\alpha$ on the record itself to produce the ambient demand passed upward. Record projection never demands the record constructor itself, but simply promotes the field demand into a record demand, using using $\envLookupBwdR{\vec{\bind{x}{\raw{v}}}}$ to demand fields other that $y$ with $\hole$.

\paragraph{Hole rule} The hole rule, as elsewhere, ensures that the function is defined when $v$ is $\hole$, and it is easy to show that $\evalBwdF{T}$  preserves $\leq$, and thus $\eq$.

\begin{lemma}[Monotonicity of $\evalBwdF{T}$]
   Suppose $T :: \raw{\rho}, \raw{e} \evalR \raw{v}$ with $v \evalBwdR{T} \rho, e, \alpha $ and $v' \evalBwdR{T} \rho', e', \alpha' $. If $v \leq v'$ then $(\rho, e, \alpha) \leq (\rho', e', \alpha')$.
\end{lemma}

\subsection{Round-tripping properties of $\evalFwdF{T}$ and $\evalBwdF{T}$}
\label{sec:data-dependencies:galois-connections}

We now establish more formally the round-tripping properties, alluded at the beginning of the section, that relate $\evalFwdF{T}$ to $\evalBwdF{T}$. For the analyses to be coherent, we expect $\evalFwdF{T}(\evalBwdF{T}(v))$ to produce a value selection $v' \geq v$, and $\evalBwdF{T}(\evalFwdF{T}(\rho,e))$ to produce an input selection $(\rho',e') \leq (\rho,e)$. Pairs of (monotonic) functions $f: X \to Y$ and $g: Y \to X$ that are related in this way are called \emph{Galois connections}. Galois connections generalise isomorphisms: $f$ and $g$ are not quite mutual inverses, but are the nearest to an inverse each can get to the other. We will present a visual example of some of these round-tripping properties in \secref{de-morgan:example}; here we establish the relevant theorems.

\begin{definition}[Galois connection]
   Suppose $X$ and $Y$ are sets equipped with partial orders $\numleq_X$ and $\numleq_Y$. Then monotonic functions $f: X \to Y$ and $g: Y \to X$ form a \emph{Galois connection} $(g, f): Y \to X$ iff $f(g(y)) \numgeq_Y y$ and $g(f(x)) \numleq_X x$.
\end{definition}

\noindent Galois connections are also adjoint functors between poset categories, with right and left adjoints $f$ and $g$  usually called the \emph{upper} and \emph{lower} adjoints, because $f$ approximates an inverse of $g$ from above, and $g$ an inverse of $f$ from below. Galois connections compose component-wise, so it is useful to think of them as having a type $X \to Y$, with the direction (by convention) given by the lower adjoint. If $f: X \to Y$ is a Galois connection, we will write $\lowerAdj{f}$ and $\upperAdj{f}$ for the lower and upper adjoints respectively; an important property we will return to is that $\lowerAdj{f}$ preserves joins and $\upperAdj{f}$ preserves meets. We now show that $\evalBwdF{T}$ and $\evalFwdF{T}$ form a Galois connection $\evalGC{T}$ for any $\Ann{A}$ (\thmref{core-language:eval:gc}) by first establishing that the relevant auxiliary functions also form Galois connections.

\begin{theorem}[Galois connection for pattern-matching]
   \label{thm:core-language:match:gc}
      Suppose $w :: \raw{v}, \raw{\sigma} \matchR \raw{\rho}, \raw{\kappa}$.  Then $(\matchBwdF{w}, \matchFwdF{w}): \Sel{\raw{\rho},\raw{\kappa}}{A} \to \Sel{\raw{v}, \raw{\sigma}}{A}$ is a Galois connection.
\end{theorem}

\begin{proof}
\ifappendices See \appref{proofs:match:gc}. \else \ProofInSupplementaryMaterial \fi
\end{proof}

\begin{lemma}[Galois connection for environment lookup]
\label{lem:core-language:env-get-put}Suppose $\envLookup{\raw{\rho}}{x}{\raw{v}}$. Then $(\envLookupBwdF{-}{\raw{\rho}}{x}{},\envLookupFwdF{}{\raw{\rho}}{x}{-}): \Sel{\raw{v}}{A} \to \Sel{\raw{\rho}}{A}$ is a Galois connection.
\end{lemma}

\begin{proof}
   \ifappendices See \appref{proofs:lookup:gc}. \else \ProofInSupplementaryMaterial \fi
   \end{proof}

   \begin{theorem}[Galois connection for recursive bindings]
\label{thm:core-language:closeDefs:gc}
   Suppose $\raw{\rho}, \raw{h} \closeDefsR \raw{\rho}'$. Then $({\closeDefsBwdF{\raw{\rho},\raw{h}}, \closeDefsFwdF{\raw{\rho},\raw{h}}}): \Sel{\raw{\rho}'}{A} \to \Sel{\raw{\rho},\raw{h}}{A}$ is a Galois connection.
\end{theorem}

\begin{proof}
   \ifappendices See \appref{proofs:closeDefs:gc}. \else \ProofInSupplementaryMaterial \fi
   \end{proof}

We assume (rather than prove) that the forward and backwards dependency functions $\primGCBool{\phi}{\vec{n}} = (\primFwdBool{\phi}{\vec{n}}, \primBwdBool{\phi}{\vec{n}})$ provided for every primitive operation $\phi: \tyInt^i \to \tyInt$ form a Galois connection of type $\Ann{A} \to \Ann{A}^i$. Under this assumption the following holds.

\begin{theorem}[Galois connection for evaluation]
\label{thm:core-language:eval:gc}
   Suppose $T :: \raw{\rho}, \raw{e} \evalR \raw{v}$.  Then $(\evalBwdF{T}, \evalFwdF{T}): \Sel{\raw{\rho}, \raw{e}}{A} \to \Sel{\raw{v}}{A}$ is a Galois connection.
\end{theorem}

\begin{proof}
   \ifappendices See \appref{proofs:eval:gc}. \else \ProofInSupplementaryMaterial \fi
\end{proof}

Establishing that $(\evalBwdF{T}, \evalFwdF{T})$ is an adjoint pair might seem rather weak as a correctess property: it merely ensures that the two analyses are related in a sensible way, not that they actually capture any useful information. This is a familiar problem from other approximate analyses like type systems and model checking, where properties like soundness or completeness are essential but do not by themselves guarantee utility. One could certainly define versions of $\evalBwdF{T}$ and $\evalFwdF{T}$ that are too coarse grained to be useful yet still satisfy \thmref{core-language:eval:gc}. However Galois connections do at least require that every tightening or tweak to the forward analysis is paired with a corresponding adjustment to the backward analysis, and vice-versa. In \secref{conclusion} we consider how other ideas from provenance and program slicing might be adapted to provide additional correctness criteria.

\newpage
\section{De Morgan dependencies for brushing and linking}
\label{sec:de-morgan}

\secref{data-dependencies} addresses the first kind of question we motivated in the introduction (\secref{introduction:data-linking}). In particular $\evalBwdF{T}$ can answer questions like: ``what data is needed to compute this bar in a bar chart?'', and indeed we were able to use our implementation to generate \figref{introduction:data-linking}. The second problem we set ourselves was how to link selections between \emph{cognate} outputs, i.e.~outputs computed from the same data~(\secref{introduction:vis-linking}). This is called ``brushing and linking'' \cite{becker87} in data visualisation, and has been extensive studied as an interaction paradigm, but with little emphasis on techniques for automation. Intuitively, the problem has a bidirectional flavour: one must consider how dependencies flow backward from a selection in one output to a selection $v$ in the common data, and then forward from the selected data $v$ to a corresponding selection in the other output. A natural question then is whether the analysis established in \secref{data-dependencies} can supply the information required to support an automated solution.

An immediate problem is that the flavour of the forward dependency required here differs from that provided by the forward analysis $\evalFwdF{T}$ defined in \secref{data-dependencies:analyses:fwd}. That was able to answer the question: what can we compute given only the data selected in $v$? But to identify the related data in another output, we must determine not what the input selection $v$ is sufficient for, but what it is necessary for: those parts of the other output that depend on $v$. In fact the question can be formulated as a kind of dual: what would we \emph{not} be able to compute if the data selected in $v$ were \emph{unavailable}?

\subsection{De Morgan duality}
\label{sec:de-morgan:de-morgan-duality}

Why $\evalFwdF{T}$ is unsuitable as a forward dependency relation for linking cognate outputs can also be understood in terms of compositionality. Suppose $\Ann{V}_1$ and $\Ann{V}_2$ are the lattices of selections for two views computed from a shared input source, and $\Ann{D}$ is the lattice of selections for the shared input. Using the procedure given in \secref{data-dependencies}, we can obtain two Galois connections $f: \Ann{V}_1 \to \Ann{D}$ and $g: \Ann{V}_2 \to \Ann{D}$ as shown in \figref{example:de-morgan:non-composable} below; we ask the reader to ignore \figref{example:de-morgan:composable} for the moment.

\begin{figure}[H]
   \begin{subfigure}[t]{0.43\textwidth}
      \begin{nscenter}
         \begin{tikzpicture}[node distance=1.5cm, auto]
            \node (AA) [node distance=2cm] {
               $\Ann{V}_1$
            };
            \node (BB) [right of=AA] {
               $\Ann{D}$
            };
            \node (CC) [right of=BB] {
               $\Ann{V}_2$
            };
            \node (A) [below of=AA] {
               $\Ann{V}_1$
            };
            \node (B) [right of=A] {
               $\Ann{D}$
            };
            \node (C) [right of=B] {
               $\Ann{V}_2$
            };
            \draw[->] (AA) to node {$f$} (BB);
            \draw[->] (CC) to node [swap] {$g$} (BB);

            \draw[->, bend right] (A) to node [swap] {$\lowerAdj{f}$} (B);
            \draw[->, bend left] (B) to node {$\upperAdj{g}$} (C);
            \draw[->, bend left] (C) to node {$\lowerAdj{g}$} (B);
            \draw[->, bend right] (B) to node [swap] {$\upperAdj{f}$} (A);

            \draw[red,thick,dotted,line cap=round,dash pattern=on 0pt off 0.2em] ($(BB.north west)+(0.1,0.3)$)  rectangle ($(C.south east)+(-0.1,-0.5)$);
         \end{tikzpicture}
      \end{nscenter}
      \caption{Non-composable Galois connections}
      \label{fig:example:de-morgan:non-composable}
   \end{subfigure}
   \begin{subfigure}[t]{0.55\textwidth}
      \begin{nscenter}
         \begin{tikzpicture}[node distance=1.5cm, auto]
            \node (AA) [node distance=2cm] {
               $\Ann{V}_1$
            };
            \node (BB) [right of=AA] {
               $\Ann{D}$
            };
            \node (CC) [right of=BB] {
               $\Ann{V}_2$
            };
            \node (A) [below of=AA] {
               $\Ann{V}_1$
            };
            \node (A1) [right of=A] {
               $\Ann{D}$
            };
            \node (B) [right of=A1] {
               $\Ann{D}$
            };
            \node (C) [right of=B] {
               $\Ann{V}_2$
            };
            \node (C1) [right of=C] {
               $\Ann{V}_2$
            };
            \draw[->] (AA) to node {$f$} (BB);
            \draw[->] (BB) to node {$\dual{g}$} (CC);

            \draw[->, bend right] (A) to node [swap] {$\lowerAdj{f}$} (A1);
            \draw[->, bend right] (A1) to node [swap] {$\neg_{\Ann{D}}$} (B);
            \draw[->, bend right] (B) to node [swap] {$\upperAdj{g}$} (C);
            \draw[->, bend right] (C) to node [swap] {$\neg_{\Ann{V}_2}$} (C1);
            \draw[->, bend right] (C1) to node [swap] {$\neg_{\Ann{V}_2}$} (C);
            \draw[->, bend right] (C) to node [swap] {$\lowerAdj{g}$} (B);
            \draw[->, bend right] (B) to node [swap] {$\neg_{\Ann{D}}$} (A1);
            \draw[->, bend right] (A1) to node [swap] {$\upperAdj{f}$} (A);

            \draw[red,thick,dotted,line cap=round,dash pattern=on 0pt off 0.2em] ($(BB.north west)+(0.1,0.3)$)  rectangle ($(C1.south east)+(-0.1,-0.5)$);
         \end{tikzpicture}
   \end{nscenter}
   \caption{Composing via De Morgan duality}
   \label{fig:example:de-morgan:composable}
\end{subfigure}
\caption{Dualising $g: \Ann{V}_2 \to \Ann{D}$ for composition with $f: \Ann{V}_1 \to \Ann{D}$}
\end{figure}

\vspace{-1mm}

\noindent Unfortunately, $f$ and $g$ are not composable, as their types makes clear. While the upper adjoint $\upperAdj{g}: \Ann{D} \to \Ann{V}_2$ has the correct type to compose with the lower adjoint $\lowerAdj{f}: \Ann{V}_1 \to \Ann{D}$, the result is not a Galois connection: $\upperAdj{g}$ preserves meets, whereas $\lowerAdj{f}$ preserves joins. However, it turns out that if selections are closed under complement, we can derive an analysis of what is \emph{necessary} for a given input selection from an analysis of what it is \emph{sufficient} for. The effect is to invert $g$, yielding a Galois connection $\dual{g}$ with a type that allows it to compose with $f$. Then the composite $\dual{g} \after f$ is a Galois connection linking $\Ann{V}_1$ to $\Ann{V}_2$ via $\Ann{D}$, as shown in \figref{example:de-morgan:composable}, offering a general mechanism for brushing and linking, with nice round-tripping properties. We now unpack this in more detail.

First we shift settings from the lattices used in \secref{data-dependencies} to Boolean lattices (or Boolean algebras) $\Ann{A}= \BoolLattice{\Ann{A}}{\top}{\bot}{\meet}{\join}{\neg}$, which are lattices equipped with an involution $\neg: \Ann{A} \to \Ann{A}$ called \emph{complement}. Boolean algebras satisfy complementation laws $x \meet \neg x = \bot$ and $x \join \neg x = \top$ and De Morgan laws $\neg x \meet \neg y = \neg(x \join y)$ and $\neg x \join \neg y = \neg(x \meet y)$. If $\Ann{A}$ is a Boolean algebra, then $\Sel{\raw{v}}{A}$ is also a Boolean algebra, with the Boolean operations, and in particular $\neg_{\raw{v}}: \Sel{\raw{v}}{A} \to \Sel{\raw{v}}{A}$, defined pointwise. An additional distinguished selectable value $\blackhole$ serves as the negation of $\hole$. The two-point lattice $\Bool$ we used to illustrate \secref{data-dependencies} is also a Boolean algebra $\BoolLattice{\Bool}{\TT}{\FF}{\wedge}{\vee}{\neg}$ with $\neg$ corresponding to logical negation.

\begin{figure}
   \begin{subfigure}[t]{3.3in}
      \small
\begin{lstlisting}[language=Fluid]
let zero n = const n;
    wrap n n_max = ((n - 1) `mod` n_max) + 1;
    extend n = min (max n 1);
    nth2 i j xss = nth (j - 1) (nth (i - 1) xss);

let convolve image kernel method =
    let ((m, n), (i, j)) = (dims image, dims kernel);
        (half_i, half_j) = (i `quot` 2, j `quot` 2);
        area = i * j
    in  $\langle$ let weightedSum = sum [
           image!(x, y) * kernel!(i' + 1, j' + 1)
           | (i', j') $\leftarrow$ range (0, 0) (i - 1, j - 1),
             let x = method (m' + i' - half_i) m,
             let y = method (n' + j' - half_j) n,
             x $\numgeq$ 1, x $\numleq$ m, y $\numgeq$ 1, y $\numleq$ n
         ] in weightedSum `quot` area
         | (m', n') in (m, n) $\rangle$;
\end{lstlisting}
   \end{subfigure}
   \begin{subfigure}[t]{2.1in}
      \small
\begin{lstlisting}[language=Fluid]
let emboss = [[-2, -1, 0],
              [-1,  1, 1],
              [ 0,  1, 2]];
    filter = $\langle$ nth2 i j emboss
             | (i, j) in (3, 3) $\rangle$;
    image' = [[15, 13,  6, 9, 16],
              [12,  5, 15, 4, 13],
              [14,  9, 20, 8,  1],
              [ 4, 10,  3, 7, 19],
              [ 3, 11, 15, 2,  9]];
    image = $\langle$ nth2 i j image'
            | (i, j) in (5, 5) $\rangle$
in convolve image filter zero
\end{lstlisting}
   \end{subfigure}
   \vspace{-2mm}
   \caption{Matrix convolution example, with methods \kw{zero}, \kw{wrap} and \kw{extend} for dealing with boundaries}
   \vspace{-1mm}
   \label{fig:example:convolve}
\end{figure}

It is an easy consequence of the complementation and De Morgan laws that any meet-preserving operation $g: \Ann{A} \to \Ann{B}$ on Boolean algebras has a join-preserving De Morgan dual $\dual{g}: \Ann{A} \to \Ann{B}$ given by $\neg_{\Ann{B}} \after g\after \neg_{\Ann{A}}$, and any join-preserving operation $h$ has a meet-preserving De Morgan dual $\dual{h}$ defined similarly. Moreover if $h$ is the lower adjoint of $g$, then $\dual{g}$ is the lower adjoint of $\dual{h}$. Thus Galois connections on Boolean algebras also admit a (contravariant) notion of De Morgan duality, defined component-wise.

\begin{definition}[De Morgan dual of a Galois connection]
   Suppose $\Ann{A}$ and $\Ann{B}$ are Boolean algebras and $f: \Ann{A} \to \Ann{B}$ is a Galois connection $(\lowerAdj{f},\upperAdj{f})$. Define the \emph{De Morgan dual} $\dual{f}$ of $f$ to be the Galois connection $(\dual{\upperAdj{f}}, \dual{\lowerAdj{f}}): \Ann{B} \to \Ann{A}$.
\end{definition}

\noindent Dualising a Galois connection flips the direction of the arrow by swapping the roles of the upper and lower adjoints. So while $f: \Ann{A} \to \Ann{B}$ and $g: \Ann{C} \to \Ann{B}$ are not composable, $f$ and $\dual{g}: \Ann{B} \to \Ann{C}$ are, and the composition is achieved by transforming $\upperAdj{g}$ from something which determines what we can compute with $v$ into something which determines what we cannot compute without $v$. This offers a principled basis for an automated brushing and linking feature between cognate computations $T$ and $U$. When the user selects part of the output of $T$, we can use $\evalBwdF{T}$ to compute the needed data $v$, and then use $\dual{\evalFwdF{U}}$ to compute the parts of the output of $U$ that depend on $v$. This is the approach implemented in \OurLanguage, and we used this to generate \figref{introduction:vis-linking} in \secref{introduction:vis-linking}.

\subsection{Example: matrix convolution}
\label{sec:de-morgan:example}

We now illustrate the $\dual{\evalGC{T}}$ Galois connection, contrasting it with $\evalGC{T}$, using an example which computes the convolution of a $5 \times 5$ matrix with a $3 \times 3$ kernel. Convolution has an intuitive dependency structure and the values involved have an easy visual presentation, making it useful for conveying the flavour of the four distinct (but connected) dependency relations that arise in the framework. The source code for the example is given in \figref{example:convolve}, and shows the \kw{convolve} function, plus \kw{zero}, \kw{wrap} and \kw{extend} which provide different methods for handling the boundaries of the input matrix. The angle-bracket notation is used to construct matrices, which were omitted from \secref{core-language}. (The formal treatment is similar to records.)

\begin{figure}
   \begin{subfigure}{0.53\textwidth}
      {\includegraphics[scale=0.5]{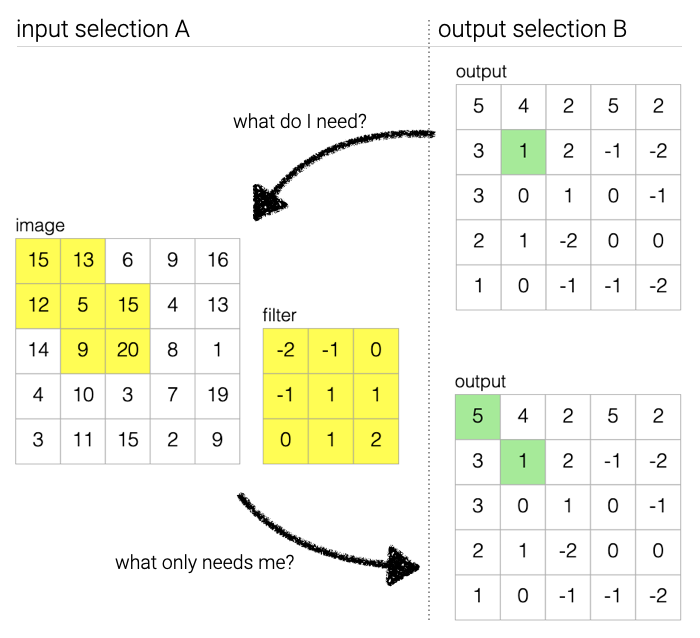}}
      \vspace{2mm}
      \caption{Galois dependency $\evalGC{T}$}
      \label{fig:example:convolve-viz:galois-dependency}
   \end{subfigure}
   \begin{subfigure}{0.46\textwidth}
      {\includegraphics[scale=0.5]{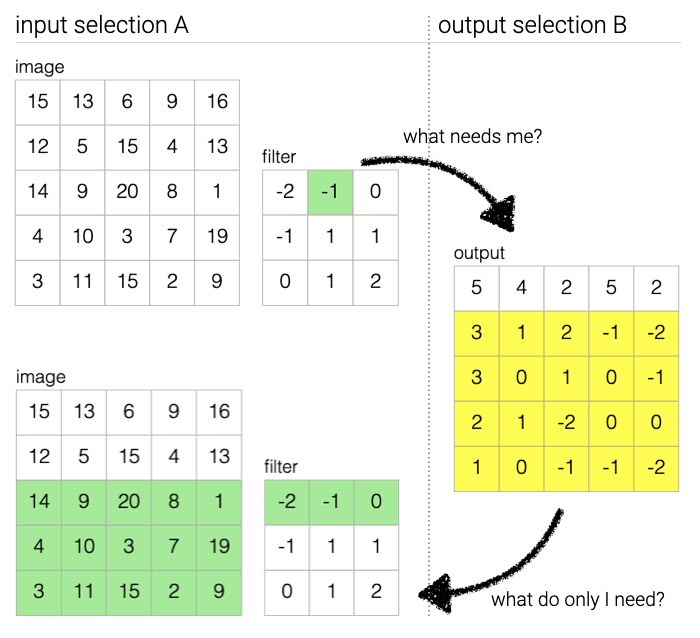}}
      \vspace{2mm}
      \caption{De Morgan dual $\dual{\evalGC{T}}$}
      \label{fig:example:convolve-viz:de-morgan-dual}
   \end{subfigure}
   \caption{Upper and lower pairs are dual; left and right pairs are adjoint}
   \label{fig:example:convolve-viz}
\end{figure}

\OurLanguage{} was used to generate the diagrams in \figref{example:convolve-viz}, which show the four dependency relations and two of their four possible round-trips. \figref{example:convolve-viz:galois-dependency} shows the $\evalGC{T}$ Galois connection defined in \secref{data-dependencies:galois-connections}. In the upper figure, the user selects (in green) the output cell at position $(2,2)$ (counting rows downwards from $1$). This induces a demand (via the lower adjoint $\evalBwdF{T}$) on the input matrix \kw{image} and the kernel \kw{filter}, revealing (in yellow) that the entire kernel was needed to compute the value $1$, but only some of the input matrix. In particular the elements at $(1,3)$ and $(3,1)$ in \kw{image} were not needed, because of zeros present in \kw{filter}. If we then ``round-trip'' that input selection, computing the corresponding availability on the output using the upper adjoint $\evalFwdF{T}$, the green selection grows: it turns out that the data needed to make $(2,2)$ available are sufficient to make $(1,1)$ available as well.

\figref{example:convolve-viz:de-morgan-dual} shows the De Morgan dual $\dual{\evalGC{T}}$. In the upper part of the figure, the user selects (green) kernel cell $(1, 2)$ to see the output cells that depend on it. This is computed using the De Morgan dual of $\evalFwdF{T}$. First we negate the input selection, marking $(1, 2)$ as unavailable, and all other inputs as available. Then we forward-analyse with $\evalFwdF{T}$ to determine that with this data selection, we can only compute the top row of the output. (If it seems odd that we can compute even the top row, notice that the example uses the method \kw{zero} for dealing with boundaries; \kw{wrap} or \kw{extend} would give a different behaviour.) Then we negate that top row selection to produce the (yellow) output selection shown in the figure. These are exactly the output cells which depend on kernel cell ($1, 2)$ in the sense that they cannot be computed if that input is unavailable.

We can then round-trip this output selection using the De Morgan dual of $\evalBwdF{T}$. We first negate the yellow output selection (selecting the top row of the output again), and then use $\evalBwdF{T}$ to determine the needed inputs, which turn out to be the top two rows of \kw{image}, and the top row of \kw{filter}. Negating again produces the green output selection shown in the lower figure. Thus the backwards De Morgan dual computes the inputs that would \emph{not} be needed if the selected outputs were not needed: more economically, the inputs that are \emph{only} needed for the selected output. Here the round-trip reveals that if kernel cell $(1, 2)$ is unavailable, then the entire top row of the kernel might as well have been unavailable too, and similarly for the bottom 3 rows of the input.

\subsection{Relationship to Galois slicing}
\label{sec:de-morgan:galois-slicing}

\begin{figure}
   \small
   \begin{centering}
      \begin{subfigure}{0.45\textwidth}
         {
\begin{lstlisting}[language=Fluid,escapeinside={(*@}{@*)}]
let (x, y) = (2, 3);
    sum = x + y in
if sum == 0
    then (0, 0)
    else (x / sum, y / sum)
(*@$\Rightarrow$@*) (0.4, 0.6)
\end{lstlisting}}
      \caption{Original program}
      \label{fig:example:diff-slicing:original}
      \end{subfigure}
      \begin{subfigure}{0.45\textwidth}
         {
\begin{lstlisting}[language=Fluid,escapeinside={(*@}{@*)}]
let (x, y) = (2, 3);
    sum = x + y in
if sum == 0
    then (*@$\hole$@*)
    else (x / sum, (*@$\hole$@*))
(*@$\Rightarrow$@*) (0.4, (*@$\hole$@*))
\end{lstlisting}}
      \caption{Backward slice for \kw{(0.4, $\hole$)}}
      \label{fig:example:diff-slicing:subtree}
      \end{subfigure}
      \\
      \begin{subfigure}{0.45\textwidth}
         {
\begin{lstlisting}[language=Fluid,escapeinside={(*@}{@*)}]
let (x, y) = (2, 3);
    sum = x + y in
if sum == 0
    then (*@$\hole$@*)
    else ((*@$\hole$@*), (*@$\hole$@*))
(*@$\Rightarrow$@*) ((*@$\hole$@*), (*@$\hole$@*))
\end{lstlisting}}
      \caption{Backward slice for spine \kw{($\hole$, $\hole$)}}
      \label{fig:example:diff-slicing:spine}
      \end{subfigure}
      \begin{subfigure}{0.45\textwidth}
         {
\begin{lstlisting}[language=Fluid,escapeinside={(*@}{@*)}]
let (x, y) = (2, 3);
    sum = x + y in
if sum == 0
   then (0, 0)
   else ((*@\codeSel{x / sum}@*), y / sum)
(*@$\Rightarrow$@*) ((*@\codeSelTwo{0.4}@*), 0.6)
\end{lstlisting}}
      \caption{Differential backward slice for \kw{(\codeSelTwo{0.4}, $\hole$)}}
      \label{fig:example:diff-slicing:differential}
      \end{subfigure}
   \end{centering}
   \caption{Differential Galois slicing selects input (yellow) needed \emph{only} for selected output (green)}
   \label{fig:example:diff-slicing}
\end{figure}

The De Morgan dual puts us in a better position to consider the relationship between the present system and earlier work on \emph{Galois slicing}, a program slicing technique that has been explored for pure functional programs~\cite{perera12a}, functional programs with effects~\cite{ricciotti17}, and \piCalculus~\cite{perera16d}. We consider other related work in \secref{conclusion:other-related-work}.

Galois slicing operates on lattices of \emph{slices}, which are programs (or values) where parts deemed irrelevant are replaced by a hole $\hole$. (If we think of the notion of selection defined in \secref{data-dependencies:selections} as picking out a subset of the paths in a term, then slices resembles selections which are prefix-closed, meaning that if a given path in a term is selected, then so are all of its prefixes.) For a fixed computation, a meet-preserving \emph{forward-slicing} function is defined which takes input slices to output slices, discarding parts which cannot be computed because the needed input is not present, plus a join-preserving \emph{backward-slicing} function taking output slices to input slices, retaining the parts needed for the output slice. For example \figref{example:diff-slicing:original} shows a computation with output \lstinline{(0.4, 0.6)}, and \figref{example:diff-slicing:subtree} gives the backward slice for output slice \kw{(0.4, $\hole$)}. Forward and backward slicing, for a given computation, form a Galois connection, giving the analyses the nice round-tripping properties we motivated in \secref{data-dependencies:galois-connections}.

Unfortunately, the notion of slice does not lend itself to computing dependencies where the needed input or output is a proper part of a value, such as a component of a tuple. \emph{Differential} slicing \cite{perera12a} improves on this by using Galois slicing to compute a pair of input slices $(e,e'$) for a pair of output slices $(v,v')$ where $v \leq v'$. By monotonicity, $e \leq e'$. This can be used to compute a (differential) slice for an arbitrary subtree, by setting $v$ to be the ``spine'' of the original output up to the location of the subtree, and $v'$ to be $v$ with the subtree of interest plugged back in. Here we could focus on the value \kw{0.4} in the output by computing the backward slice for $\kw{($\hole$, $\hole$)}$ (\figref{example:diff-slicing:spine}) and then comparing it with the backward slice for \kw{(0.4, $\hole$)}, generating a differential slice where the parts that are different are highlighted (\figref{example:diff-slicing:differential}). But although it supports a notion of selection which is closer to what we need, the differential slice highlights only the program parts that are needed \emph{exclusively} by the selected output. Here, \lstinline{2} and \lstinline{3} are needed to compute the spine as well (in order to decide which conditional branch to execute), so they are excluded from the differential slice. In fact differential slicing is similar to the dual of $\evalBwdF{T}$, and as such underapproximates the dependency information needed for data linking.

\newpage
\section{Galois connections for desugaring}
\label{sec:surface-language}

Elaborating a richer surface language into a simpler core is a common pattern with well known benefits. It can, however, make it harder to express certain information to the programmer in terms of the surface language. We face this problem with the analysis in \secref{data-dependencies}, which links outputs not only to inputs, but also to expressions responsible for introducing data. We could use this information in an IDE to link structured outputs to relevant code fragments, but only if we are able to map term selections back to the surface program. We now sketch a bidirectional desugaring procedure which addresses this, and which composes with the Galois dependency analysis defined in \secref{data-dependencies}.

\begin{figure}
{\small
\begingroup
\renewcommand*{\arraystretch}{1}
\begin{minipage}[t]{0.5\textwidth}
\begin{tabularx}{\textwidth}{rL{2.3cm}L{3cm}}
&\textbfit{Identifier}&
\\
$x, y ::=$
&
$\ldots$
&
\\
&
$\primOp$
&
operator name
\\[2mm]
&\textbfit{Surface term}&
\\
$s, t ::=$
&
$\ldots$
&
\\
&
$\exOp{\primOp}$
&
first-class operator
\\
&
$\exBinaryApp{s}{\primOp}{s'}$
&
binary application
\\
&
$\exLetRecMutual{\vec{g}}{s}$
&
recursive functions
\\
&
$\exIfThenElse{s}{s}{s}$
&
if
\\
&
$\exMatch{s}{\vec{\clauseUncurried{p}{s}}}$
&
match
\\
&
$\exLet{p}{s}{s}$
&
structured let
\\
&
$\annList{s}{r}{\alpha}$
&
non-empty list
\\
&
$\exListEnum{s}{s}$
&
list enum
\\
&
$\annListComp{s}{\vec{q}}{\alpha}$
&
list comprehension
\\[2mm]
&\textbfit{List rest term}&
\\
$r ::=$
&
$\annListEnd{\alpha}$
&
end
\\
&
$\annListNext{s}{r}{\alpha}$
&
cons
\\
\\
\end{tabularx}
\end{minipage}%
\begin{minipage}[t]{0.5\textwidth}
\begin{tabularx}{\textwidth}{rL{2.8cm}L{2.9cm}}
&\textbfit{Recursive function}&
\\
$g ::=$
&
$\bind{x}{\vec{c}}$
&
\\[2mm]
&\textbfit{Clause}&
\\
$c ::=$
&
$\clause{\vec{p}}{s}$
&
\\[2mm]
&\textbfit{Pattern}&
\\
$p ::=$
&
$\pattVar{x}$
&
variable
\\
&
$\pattRec{\vec{\bind{x}{p}}}$
&
record
\\
&
$\pattNil$
&
nil
\\
&
$\pattCons{p}{p}$
&
cons
\\
&
$\pattList{p}{o}$
&
non-empty list
\\[2mm]
&\textbfit{List rest pattern}&
\\
$o ::=$
&
$\pattListEnd$
&
end
\\
&
$\pattListNext{p}{o}$
&
cons
\\[2mm]
&\textbfit{Qualifier}&
\\
$q ::=$
&
$\qualGuard{s}$
&
guard
\\
&
$\qualDeclaration{p}{s}$
&
declaration
\\
&
$\qualGenerator{p}{s}$
&
generator
\end{tabularx}
\end{minipage}
\endgroup
}
\vspace{-2mm}
\caption{(Selectable) syntax for surface language}
\vspace{-2mm}
\label{fig:surface-language:syntax}
\end{figure}

\subsection{Surface language syntax}

The surface language \OurLanguage{} extends the core syntax with list notation $\kw{[} s \comma \ldots \comma s' \kw{]}$, Haskell 98-style list comprehensions \cite{peytonJones03}, list enumerations, first-class primitives and piecewise function definitions and pattern-matching, as shown in \figref{surface-language:syntax}. Typing rules are \ifappendices given in \appref{surface-language}, \figrefTwo{surface-language:typing-term}{surface-language:typing-pattern}\else \IncludedWithSupplementaryMaterial\fi. We attach selection states $\alpha$ to (selectable) surface terms $s, t$ that desugar to selectable core terms, and let $\raw{s}$, $\raw{t}$ range over ``raw'' surface terms, which are isomorphic to the selectable terms where the type of selection states is the unit lattice $\Unit$.

\figref{surface-language:example-1} shows how the end-to-end mapping would appear to a user. (For illustrative purposes the library function \kw{map} and some raw data are included in the same source file.) On the left, the user selects a cons cell (green) in the output; by backwards evaluating and then backwards desugaring, we are able to highlight the list comprehension, the cons in the second clause of \kw{map}, and both occurrences of the constant \kw{"Hydro"}. These last two are highlighted because the selected cons cell was constructed by eliminating a Boolean that was in turn constructed by the primitive \kw{==} operator, which consumed the two strings. The user might then conjecture that the two occurrences of \kw{"Geo"} were somehow responsible for the inclusion of the third cons cell in the output; they can confirm this by making the green selection on the right. (Highlighting \kw{==} too would clearly be helpful here; we discuss this possibility in \secref{conclusion:other-related-work}.) The grey selection is included to contrast the cons highlighting with the data demanded by the list elements themselves, which is quite different.

\begin{figure}
   \begin{subfigure}{0.48\textwidth}
      \small
\begin{lstlisting}[language=Fluid,escapeinside={(*@}{@*)}]
let map f [] = [];
    map f (x : xs) = f x (*@\codeSel{:}@*) map f xs;
let data = [
   { energyType: "Bio", output: 6.2 },
   { energyType: (*@\codeSel{"Hydro"}@*), output: 260 },
   { energyType: "Solar", output: 19.9 },
   { energyType: "Wind", output: 91 },
   { energyType: "Geo", output: 14.4 }
];
let xs = (*@\codeSel{[}@*) row.output
   | type $\leftarrow$ [(*@\codeSel{"Hydro"}@*), "Solar", "Geo"],
     row $\leftarrow$ data, row.energyType == type
(*@\codeSel{]}@*) in
map (fun x $\rightarrow$ floor (x / sum xs * 100)) xs
$\Rightarrow$ (88 (*@\codeSelTwo{:}@*) (6 : (4 : [])))
\end{lstlisting}
   \end{subfigure}
   \hfill
   \begin{subfigure}{0.48\textwidth}
      \small
\begin{lstlisting}[language=Fluid,escapeinside={(*@}{@*)}]
let map f [] = [];
    map f (x : xs) = f x (*@\codeSel{:}@*) map f xs;
let data = [
   { energyType: "Bio", output: 6.2 },
   { energyType: "Hydro", output: (*@\codeErase{260}@*) },
   { energyType: "Solar", output: (*@\codeErase{19.9}@*) },
   { energyType: "Wind", output: 91 },
   { energyType: (*@\codeSel{"Geo"}@*), output: (*@\codeErase{14.4}@*) }
];
let xs = (*@\codeSel{[}@*) row.output
   | type $\leftarrow$ ["Hydro", "Solar", (*@\codeSel{"Geo"}@*)],
     row $\leftarrow$ data, row.energyType == type
(*@\codeSel{]}@*) in
map (fun x $\rightarrow$ floor (x / sum xs * (*@\codeErase{100}@*))) xs
$\Rightarrow$ (88 : ((*@\codeEraseTwo{6}@*) : (4 (*@\codeSelTwo{:}@*) [])))
\end{lstlisting}
   \end{subfigure}
   \vspace{-3mm}
   \caption{Source selections (yellow) resulting from selecting different list cells (green)}
   \vspace{-2mm}
\label{fig:surface-language:example-1}
\end{figure}

\subsection{Forward desugaring}

To define the forward evaluation function $\evalFwdF{T}$ in \secref{data-dependencies:analyses:fwd}, we performed a regular evaluation using $\evalR$ to obtain a trace $T$, and then defined $\evalFwdF{T}$ by recursion over $T$, with $T$ guiding the analysis in the presence of $\hole$. There are no holes in the surface language, so we can take a simpler approach, defining a single \emph{forward desugaring} relation $\desugarFwdR$, and then showing that for every raw surface term $\raw{t} \desugarFwdR \raw{e}$, there is a monotonic function $\desugarFwdF{t}: \Sel{\raw{t}}{A} \to \Sel{\raw{e}}{A}$, which is simply $\desugarFwdR$ domain-restricted to $\Sel{\raw{t}}{A}$. The full definition of $\desugarFwdR$ is \ifappendices given in \appref{surface-language} \else \IncludedWithSupplementaryMaterial \fi; \figref{surface-language:desugar} gives a representative selection of the rules.

\begin{figure}
   \input{fig/surface-language/slicing/desugar-short.tex}
   \caption{Forwards and backwards desugaring (selected rules only)}
   \label{fig:surface-language:desugar}
\end{figure}

The definition follows a similar pattern to $\evalFwdF{T}$. At each step, we take the meet of the availability on any parts of $s$ being consumed at that step, and use that as the availability of any parts of $e$ being generated at that step. Thus the rules for list notation simply transfer the selection state $\alpha$ on the opening and closing brackets $\annot{\kw{[}}{\alpha}$ and $\annot{\kw{]}}{\alpha}$ to the corresponding cons and nil of the resulting list, and those on intervening delimiters $\annot{\comma}{\alpha}$ to the corresponding cons. List comprehensions $\annListComp{s}{\vec{q}}{\alpha}$ have a rule for each kind of qualifier $q$ at the head of $\vec{q}$, plus a rule for when $\vec{q}$ is $\seqEmpty$. The general pattern is to push the $\alpha$ on the comprehension itself through recursively, so it ends up on all core terms generated during its elaboration: in particular the \kw{false} branch when $q$ is a guard, and the singleton list when $\vec{q}$ is empty. Auxiliary relations $\clauseFwdR$ and $\totaliseFwdR{p}$ (\ifappendices defined in \appref{surface-language}, \figrefTwo{surface-language:clause-fwd}{surface-language:totalise-fwd}\else \IncludedWithSupplementaryMaterial\fi) transfer availability on guards and generators onto the eliminators they elaborate into.

\subsection{Backward desugaring}

The backwards analysis is then defined as a family of \textit{backward desugaring} functions $\desugarBwdR{\raw{t}}: \Sel{\raw{e}}{A} \to \Sel{\raw{t}}{A}$ for any $\raw{t} \desugarFwdR \raw{e}$, with the raw surface term $\raw{t}$ guiding the analysis backwards. (The role of $\raw{t}$ in disambiguating the backwards rules should be clear if you consider that $e$ typically matches multiple rules but only one for a given $\raw{t}$.) \figref{surface-language:desugar} gives some representative rules; the full definition is {\ifappendices given in \appref{surface-language} \else \IncludedWithSupplementaryMaterial.\fi} To reverse a desugaring step, we take the join of the demand on any parts of $e$ which were constructed at this step, and use that as the demand on the parts of $s$ which were consumed at this step, turning demand on the core term into (minimal) demand on the surface term. Thus the effect of the list comprehension rules and auxiliary judgements is to set the demand on the comprehension itself to be the join of the demand of all the syntax generated during the elaboration of the comprehension, using auxiliary judgments $\totaliseBwdR{p}$ and $\clauseBwdR{p}$ \ifappendices (\appref{surface-language}, \figrefTwo{surface-language:totalise-bwd}{surface-language:clause-bwd}) \fi to transfer demand from eliminators back onto the guards and generators.

\subsection{Round-tripping properties and compositionality}

It is easy to verify that $\desugarFwdF{t}$ and $\desugarBwdF{t}$ are monotonic. Moreover they form an adjoint pair.

\begin{theorem}[Galois connection for desugaring]
  \label{thm:surface-language:desugar:gc}
     Suppose $t \desugarFwdR e$. Then ${\desugarGC{t}} \eqdef (\desugarBwdF{t}, \desugarFwdF{t}): \Sel{\raw{e}}{A} \to \Sel{\raw{t}}{A}$ is a Galois connection.
\end{theorem}

\begin{proof}
   \ifappendices See \appref{proofs-surface:desugar:gc}. \else \ProofInSupplementaryMaterial \fi
\end{proof}

\noindent The $\desugarGC{\raw{t}}$ Galois connection readily composes with $\evalGC{T}$ to produce surface-language selections like the ones shown in \figref{surface-language:example-1}. This is useful, although somewhat monolithic. In future work we will investigate techniques for backwards desugaring at arbitrary steps in the computation (perhaps by interleaving desugaring with execution), and presenting selections on intermediate values (such as lists) in the surface language, even though they were not obtained via desugaring.

\section{Conclusion}
\label{sec:conclusion}

Our research was motived by the goal of making computational outputs which are automatically able to reveal how they relate to data in a fine-grained way. A casual reader who wants to understand or fact-check a chart, or a scientist evaluating another's work, should be able to do so by interacting directly with an output. Recent work by \citeauthor{walny19} suggests that developers would also benefit from such a feature while implementing visualisations, for example to check whether a quantity is represented by diameter or area in a bubble chart \cite{walny19}.

Galois connections provide an appealing setting for this problem because of their elegant round-tripping properties. However,  existing dynamic analysis techniques based on Galois connections do not lend themselves to richly structured outputs like visualisations and matrices. We presented an approach that allows focusing on arbitrary substructures, which also means data selections can be inverted. This enables linking not just of outputs to data, but of outputs to other outputs, providing a mathematical basis for a widely used (but so far ad hoc) feature in data visualisation. We implemented our approach in \OurLanguage, a realistic high-level functional programming language. %
\ifanonymous%
   A link to our GitHub repository is included in the non-anonymised version of the paper.
\else%
   Our implementation can be found at \url{https://github.com/explorable-viz/fluid}.%
\fi%

\subsection{Other related work and future directions}
\label{sec:conclusion:other-related-work}

We close by considering some limitations and opportunities in the context of other related work. Galois slicing \cite{perera12a,ricciotti17,perera16d} was considered in \secref{de-morgan:galois-slicing}.

\emph{Executable slicing.} Executable slices \cite{hall95} are programs with some parts removed, but which are still executable. Our approach computes data selections, not executable slices, but such a notion has obvious relevance in data science, since ``explaining'' a part of a result should (arguably) entail being able to recompute it. \emph{Expression provenance} \cite{acar12} explains how primitive values are computed using only primitive operations; however, this still omits crucial information, and does not obviously generalise to structured outputs. Work on executable slicing in term rewriting \cite{field98} could perhaps be adapted to structured values in our setting.

\emph{Dynamic program analysis.} Dynamic analysis techniques like dataflow analysis \cite{chen88,reps95} and taint tracking \cite{reps95} tend to focus on variables, rather than parts of structured values, and lack round-tripping properties; Galois dependencies have a clear advantage here. A limitation of dynamic techniques which is shared by our approach is that they can usually only reveal \emph{that} certain dependencies arise, not \emph{why}, which requires analysing path conditions \cite{hammer06}. In a data science setting this would clearly be valuable too, and it would be interesting to see if the benefits of the Galois framework can be extended to techniques for computing dynamic path conditions.

\emph{Brushing and linking.} Brushing and linking has been extensive studied in the data visualisation community~\cite{mcdonald82,becker87}, but although \citet{roberts06} argued it should be ubiquitous, no automated method of implementation has been proposed to date. Geospatial applications like GeoDa \cite{anselin06} hard-code view coordination features into specific views, and libraries like d3.js and Plotly support ad hoc linking mechanisms, with varying degrees of programmer effort required. No existing approach provides automation or round-tripping guarantees, or is able to provide data selections explaining why visual selections are linked.

\emph{Data provenance in data visualisation.} A recent vision paper by \citet{psallidas18} is the only work we are aware of that proposes that brushing and linking, and related view cooordination features like cross-filtering, can be understood in terms of data provenance. In a relational (query processing) setting, where the relevant notion of provenance is lineage, they propose backward-analysing to data, and then forward-analysing to another view (but again, without the round-tripping features of Galois connections). However, theirs is primarily a concept paper, proposing a research programme, rather than solving a specific problem.

\makeatletter%
\if@ACM@anonymous%
\else%
\paragraph{Acknowledgements}

Perera was supported by The UKRI Strategic Priorities Fund under EPSRC Grant EP/T001569/1, particularly the \emph{Tools, Practices and Systems} theme within that grant, and by The Alan Turing Institute under EPSRC grant EP/N510129/1.
\fi%
\makeatother%

\clearpage
\pagebreak

\ifappendices
   \clearpage
   \appendix
   \section{Surface language: typing and forwards and backwards desugaring}
\label{app:surface-language}

\begin{figure}

\flushleft \shadebox{$\Gamma \vdash \vec{g}: \Delta$}
\begin{smathpar}
   \inferrule*[right={
      \textnormal{$\Delta = \vec{\bind{x}{\tyFun{A}{B}}}$}
   }]
   {
      \Gamma \concat \Delta \vdash \vec{c_i}: \tyFun{A_i}{B_i}
      \quad
      (\forall i \numleq j)
   }
   {
      \Gamma \vdash \seqRange{\bind{x_1}{\vec{c_1}}}{\bind{x_j}{\vec{c_j}}}: \Delta
   }
\end{smathpar}

\vspace{5pt}
\flushleft \shadebox{$\Gamma \vdash s: A$}
\begin{smathpar}
   \inferrule*[right={$\primOp: A \in \Gamma$}]
   {
      \strut
   }
   {
      \Gamma \vdash \exOp{\primOp}: A
   }
   \and
   \inferrule*[right={$\primOp: \tyFun{\tyProd{\tyInt}{\tyInt}}{\tyInt} \in \Gamma$}]
   {
      \Gamma \vdash s: \tyInt
      \\
      \Gamma \vdash s': \tyInt
   }
   {
      \Gamma \vdash \exBinaryApp{s}{\primOp}{s'}: \tyInt
   }
   \and
   \inferrule*
   {
      \Gamma \vdash \vec{g}: \Delta
      \\
      \Gamma \concat \Delta \vdash s: A
   }
   {
      \Gamma \vdash \exLetRecMutual{\vec{g}}{s}: A
   }
   \and
   \inferrule*
   {
      \Gamma \vdash s: \tyBool
      \\
      \Gamma \vdash s_1: A
      \\
      \Gamma \vdash s_2: A
   }
   {
      \Gamma \vdash \exIfThenElse{s}{s_1}{s_2}: A
   }
   \and
   \inferrule*
   {
      \Gamma \vdash s: A
      \\
      \Gamma \vdash \vec{c}: \tyFun{A}{B}
   }
   {
      \Gamma \vdash \exMatch{s}{\vec{c}}: B
   }
   \and
   \inferrule*
   {
      p: A \dashv \Gamma'
      \\
      \Gamma \vdash s: A
      \\
      \Gamma \concat \Gamma' \vdash s': B
   }
   {
      \Gamma \vdash \exLet{p}{s}{s'}: B
   }
   \and
   \inferrule*
   {
      \Gamma \vdash s : A
      \\
      \Gamma \vdash r : \tyList{A}
   }
   {
      \Gamma \vdash \exList{s}{r} : \tyList{A}
   }
   \and
   \inferrule*
   {
      \Gamma \vdash s: \tyInt
      \\
      \Gamma \vdash s': \tyInt
   }
   {
      \Gamma \vdash \exListEnum{s}{s'}: \tyList{\tyInt}
   }
   \and
   \inferrule*
   {
      \Gamma \vdash \vec{q} \dashv \Delta
      \\
      \Gamma \concat \Delta \vdash s: A
   }
   {
      \Gamma \vdash \exListComp{s}{\vec{q}}: \tyList{A}
   }
   \end{smathpar}

   \vspace{5pt}
   \flushleft \shadebox{$\Gamma \vdash r: \tyList{A}$}
   \begin{smathpar}
   \inferrule*
   {
      \strut
   }
   {
      \Gamma \vdash \exListEnd : \tyList{A}
   }
   \and
   \inferrule*
   {
      \Gamma \vdash s : A
      \\
      \Gamma \vdash r : \tyList{A}
   }
   {
      \Gamma \vdash \exListNext{s}{r} : \tyList{A}
   }
   \end{smathpar}

   \vspace{0.1em}

   \flushleft \shadebox{$\Gamma \vdash \vec{q} \dashv \Delta $}
   \begin{smathpar}
   \inferrule*
   {
      \strut
   }
   {
      \Gamma \vdash \seqEmpty \dashv \seqEmpty
   }
   \and
   \inferrule*
   {
      \Gamma \vdash s: \tyBool
   }
   {
      \Gamma \vdash \qualGuard{s} \dashv \seqEmpty
   }
   \and
   \inferrule*
   {
      \Gamma \vdash s: \tyList{A}
      \\
      p: A \dashv \Delta
   }
   {
      \Gamma \vdash \qualGenerator{p}{s} \dashv \Delta
   }
   \and
   \inferrule*
   {
      \Gamma \vdash s: A
      \\
      p: A \dashv \Delta
   }
   {
      \Gamma \vdash \qualDeclaration{p}{s} \dashv \Delta
   }
   \and
   \inferrule*[
      right={$\vec{q} \neq \seqEmpty$}
   ]
   {
      \Gamma \vdash q \dashv \Gamma'
      \\
      \Gamma \concat \Gamma' \vdash \vec{q} \dashv \Delta
   }
   {
      \Gamma \vdash q \concat \vec{q} \dashv \Delta
   }
\end{smathpar}
\caption{Typing rules for surface terms (additional forms only) and qualifiers}
\label{fig:surface-language:typing-term}
\end{figure}

\begin{figure}


\flushleft \shadebox{$p: A \dashv \Gamma $}
\begin{smathpar}
\inferrule*
{
   \strut
}
{
   \pattVar{x}: A \dashv \seqEmpty, \exVar{x}: A
}
\and
\inferrule*[
   right={$\length{\vec{x}} = j$}
]
{
   p_i: A_i \dashv \Gamma_i
   \quad
   (\forall i \numleq j)
}
{
   \pattRec{\vec{\bind{x}{p}}}: \tyRec{\vec{\bind{x}{A}}} \dashv \seqRange{\Gamma_1}{\Gamma_j}
}
\and
\inferrule*
{
   \strut
}
{
   \pattNil: \tyList{A} \dashv \seqEmpty
}
\and
\inferrule*
{
   p: A \dashv \Gamma
   \\
   p': \tyList{A} \dashv \Gamma'
}
{
   \pattCons{p}{p'}: \tyList{A} \dashv \Gamma \concat \Gamma'
}
\and
\inferrule*
{
   p: A \dashv \Gamma
   \\
   o: \tyList{A} \dashv \Gamma'
}
{
   \pattList{p}{o} : \tyList{A} \dashv \Gamma \concat \Gamma'
}
\end{smathpar}
\vspace{3mm}

\flushleft \shadebox{$o: \tyList{A} \dashv \Gamma $}
\begin{smathpar}
\inferrule*
{
   \strut
}
{
   \pattListEnd : \tyList{A} \dashv \seqEmpty
}
\and
\inferrule*
{
   p : A \dashv \Gamma
   \\
   o : \tyList{A} \dashv \Gamma'
}
{
   (\pattListNext{p}{o}) : \tyList{A} \dashv \Gamma \concat \Gamma'
}
\end{smathpar}
\vspace{3mm}

\flushleft \shadebox{$\Gamma \vdash \vec{c}: \tyFun{A}{B}$}
\begin{smathpar}
\inferrule*
{
   p: A \dashv \Gamma'
   \\
   \Gamma \cdot \Gamma' \vdash s: B
}
{
   \Gamma \vdash \clauseUncurried{p}{s} : \tyFun{A}{B}
}
\and
\inferrule*[
   right={$\vec{p} \neq \seqEmpty$}
]
{
   p : A \dashv \Gamma'
   \\
   \Gamma \cdot \Gamma' \vdash \clause{\vec{p}}{s} : B
}
{
   \Gamma \vdash \clauseUncurried{(p \cdot \vec{p})}{s} : \tyFun{A}{B}
}
\and
\inferrule*[
]
{
   \Gamma \vdash c_i: \tyFun{A}{B}
   \quad
   (\forall i \numleq j)
}
{
   \Gamma \vdash \seqRange{c_1}{c_{j + 1}}: \tyFun{A}{B}
}
\end{smathpar}

\caption{Typing rules for patterns and clauses}
\label{fig:surface-language:typing-pattern}
\end{figure}

\begin{figure}
\begin{syntaxfig}
   \mbox{Eliminator}
   &
   \sigma
   &
   ::=
   &
   \ldots
   \\
   &&&
   \elimTrue{\kappa}
   &
   \text{true}
   \\
   &&&
   \elimFalse{\kappa}
   &
   \text{false}
   \\
   &&&
   \elimNil{\kappa}
   &
   \text{nil}
   \\
   &&&
   \elimCons{\sigma}
   &
   \text{cons}
\end{syntaxfig}
\vspace{3mm}

\flushleft\shadebox{$\kappa \disjjoin \kappa'$}
\begin{salign}
   \exLambda{\sigma} \disjjoin \exLambda{\sigma'}
   &=
   \exLambda{(\sigma \disjjoin \sigma')}
   \\[3mm]
   (\elimVar{x}{\sigma}) \disjjoin (\elimVar{x}{\sigma'})
   &=
   \elimVar{x}{\sigma \disjjoin \sigma'}
   \\
   \elimTrue{\sigma} \disjjoin \elimTrue{\sigma'}
   &=
   \elimTrue{\sigma \disjjoin \sigma'}
   \\
   \elimFalse{\tau} \disjjoin \elimFalse{\tau'}
   &=
   \elimFalse{\tau \disjjoin \tau'}
   \\
   \elimTrue{\kappa} \disjjoin \elimFalse{\kappa'}
   &=
   \elimBool{\kappa}{\kappa'}
   \\
   \elimTrue{\sigma} \disjjoin \elimBool{\sigma'}{\kappa}
   &=
   \elimBool{\sigma \disjjoin \sigma'}{\kappa}
   \\
   \elimFalse{\tau} \disjjoin \elimBool{\kappa}{\tau'}
   &=
   \elimBool{\kappa}{\tau \disjjoin \tau'}
   \\
   \elimBool{\sigma}{\tau} \disjjoin \elimBool{\sigma'}{\tau'}
   &=
   \elimBool{\sigma \disjjoin \sigma'}{\tau \disjjoin \tau'}
   \\
   \elimNil{\sigma} \disjjoin \elimNil{\sigma'}
   &=
   \elimNil{\sigma \disjjoin \sigma'}
   \\
   \elimCons{\tau} \disjjoin \elimCons{\tau'}
   &=
   \elimCons{\tau \disjjoin \tau'}
   \\
   \elimNil{\kappa} \disjjoin \elimCons{\kappa'}
   &=
   \elimList{\kappa}{\kappa'}
   \\
   \elimNil{\sigma} \disjjoin \elimList{\sigma'}{\kappa}
   &=
   \elimBool{\sigma \disjjoin \sigma'}{\kappa}
   \\
   \elimCons{\tau} \disjjoin \elimList{\kappa}{\tau'}
   &=
   \elimList{\kappa}{\tau \disjjoin \tau'}
   \\
   \elimList{\sigma}{\tau} \disjjoin \elimList{\sigma'}{\tau'}
   &=
   \elimList{\sigma \disjjoin \sigma'}{\tau \disjjoin \tau'}
   \\
   \elimRec{\vec{x}}{\sigma} \disjjoin \elimRec{\vec{x}}{\sigma'}
   &=
   \elimRec{\vec{x}}{\sigma \disjjoin \sigma'}
\end{salign}
\caption{Partial eliminators and disjoint join of partial continuations}
\label{fig:surface-language:disjoin-join-elim}
\end{figure}

\begin{definition}[Disjoint join of partial continuations]
   Define $\disjjoin$ to be the smallest partial symmetric function satisfying the equations in \figref{surface-language:disjoin-join-elim}.
\end{definition}

\begin{figure}
\flushleft \shadebox{$s \desugarFwdR e$}
\begin{smathpar}
   \inferrule*[
      lab={\ruleName{$\desugarFwdR$-op}}
   ]
   {
      \strut
   }
   {
      \exOp{\primOp} \desugarFwdR \primOp
   }
   \and
   \inferrule*[
      lab={\ruleName{$\desugarFwdR$-binary-apply}}
   ]
   {
      s \desugarFwdR e
      \\
      s' \desugarFwdR e'
   }
   {
      \exBinaryApp{s}{\primOp}{s'} \desugarFwdR \exApp{\exApp{\primOp}{e}}{e'}
   }
   \and
   \inferrule*[
      lab={\ruleName{$\desugarFwdR$-nil}}
   ]
   {
      \strut
   }
   {
      \annNil{\alpha} \desugarFwdR \annNil{\alpha}
   }
   \and
   \inferrule*[
      lab={\ruleName{$\desugarFwdR$-non-empty-list}}
   ]
   {
      s \desugarFwdR e
      \\
      r \desugarFwdR e'
   }
   {
      \annList{s}{r}{\alpha} \desugarFwdR \annCons{e}{e'}{\alpha}
   }
   \and
   \inferrule*[
      lab={\ruleName{$\desugarFwdR$-cons}}
   ]
   {
      s \desugarFwdR e
      \\
      s' \desugarFwdR e'
   }
   {
      \annCons{s}{s'}{\alpha} \desugarFwdR \annCons{e}{e'}{\alpha}
   }
   \and
   \inferrule*[
      lab={\ruleName{$\desugarFwdR$-let-rec}}
   ]
   {
      \vec{c_i} \clausesFwdR \sigma_i
      \quad
      (\forall i \numleq j)
      \\
      s \desugarFwdR e
   }
   {
      \exLetRecMutual{\seqRange{\bind{x_1}{\vec{c_1}}}{\bind{x_j}{\vec{c_j}}}}{s}
      \desugarFwdR
      \exLetRecMutual{\vec{\bind{x}{\sigma}}}{e}
   }
   \and
   \inferrule*[
      lab={\ruleName{$\desugarFwdR$-apply}}
   ]
   {
      s \desugarFwdR e
      \\
      s' \desugarFwdR e'
   }
   {
      \exApp{s}{s'} \desugarFwdR \exApp{e}{e'}
   }
   \and
   \inferrule*[
      lab={\ruleName{$\desugarFwdR$-match}}
   ]
   {
      s \desugarFwdR e
      \\
      \vec{c} \clausesFwdR \sigma
   }
   {
      \exMatch{s}{\vec{c}}
      \desugarFwdR
      \exApp{\exLambda{\sigma}}{e}
   }
   \and
   \inferrule*[
      lab={\ruleName{$\desugarFwdR$-let}}
   ]
   {
      s \desugarFwdR e
      \\
      s' \desugarFwdR e'
      \\
      p, e' \clauseFwdR \sigma
   }
   {
      \exLet{p}{s}{s'}
      \desugarFwdR
      \exApp{\exLambda{\sigma}}{e}
   }
   \and
   \inferrule*[
      lab={\ruleName{$\desugarFwdR$-if}}
   ]
   {
      s \desugarFwdR e
      \\
      s_1 \desugarFwdR e_1
      \\
      s_2 \desugarFwdR e_2
   }
   {
      \exIfThenElse{s}{s_1}{s_2}
      \desugarFwdR
      \exApp{\exLambda{\elimBool{e_1}{e_2}}}{e}
   }
   \and
   \inferrule*[
      lab={\ruleName{$\desugarFwdR$-list-enum}},
   ]
   {
      s \desugarFwdR e
      \\
      s' \desugarFwdR e'
   }
   {
      \exListEnum{s}{s'}
      \desugarFwdR
      \exApp{\exApp{\funEnumFromTo}{e}}{e'}
   }
   \and
   \inferrule*[lab={\ruleName{$\desugarFwdR$-list-comp-done}}]
   {
      s \desugarFwdR e
   }
   {
      \annListComp{s}{\seqEmpty}{\alpha}
      \desugarFwdR
      \annCons{e}{\annNil{\alpha}}{\alpha}
   }
   \and
   \inferrule*[
      lab={\ruleName{$\desugarFwdR$-list-comp-guard}}
   ]
   {
      \annListComp{s}{\vec{q}}{\alpha} \desugarFwdR e
      \\
      s' \desugarFwdR e'
   }
   {
      \annListComp{s}{\qualGuard{s'} \concat \vec{q}}{\alpha}
      \desugarFwdR
      \exApp{\exLambda{\elimBool{e}{\annNil{\alpha}}}}{e'}
   }
   \and
   \inferrule*[
      lab={\ruleName{$\desugarFwdR$-list-comp-decl}}
   ]
   {
      \annListComp{s}{\vec{q}}{\alpha} \desugarFwdR e
      \\
      p, e \clauseFwdR \sigma
      \\
      s' \desugarFwdR e
   }
   {
      \annListComp{s}{\qualDeclaration{p}{s'} \concat \vec{q}}{\alpha}
      \desugarFwdR
      \exApp{\exLambda{\sigma}}{e}
   }
   \and
   \inferrule*[
      lab={\ruleName{$\desugarFwdR$-list-comp-gen}}
   ]
   {
      \annListComp{s}{\vec{q}}{\alpha} \desugarFwdR e
      \\
      p, e \clauseFwdR \sigma
      \\
      \totaliseFwd{\sigma}{\alpha}{p}{\sigma'}
      \\
      s' \desugarFwdR e'
   }
   {
      \annListComp{s}{\qualGenerator{p}{s'} \concat \vec{q}}{\alpha}
      \desugarFwdR
      \exApp{\exApp{\funConcatMap}{\exLambda{\sigma'}}}{e'}
   }
   \end{smathpar}

   \vspace{5pt}
   \flushleft \shadebox{$r \desugarFwdR e$}
   \begin{smathpar}
   \inferrule*[
      lab={\ruleName{$\desugarFwdR$-list-rest-end}}
   ]
   {
      \strut
   }
   {
      \annListEnd{\alpha} \desugarFwdR \annNil{\alpha}
   }
   \and
   \inferrule*[
      lab={\ruleName{$\desugarFwdR$-list-rest-cons}}
   ]
   {
      s \desugarFwdR e
      \\
      r \desugarFwdR e'
   }
   {
      (\annListNext{s}{r}{\alpha}) \desugarFwdR \annCons{e}{e'}{\alpha}
   }
\end{smathpar}
\caption{Desugaring -- forward slicing (selected rules)}
\label{fig:desugar-fwd}
\end{figure}

\begin{figure}
\flushleft\shadebox{$\vec{p}, \kappa \clauseFwdR \sigma$}
\begin{smathpar}
\inferrule*[
   lab={\ruleName{$\clauseFwdR$-var}}
]
{
   \strut
}
{
   x, \kappa
   \clauseFwdR
   \elimVar{x}{\kappa}
}
\and
\inferrule*[
   lab={\ruleName{$\clauseFwdR$-true}}
]
{
   \strut
}
{
   \pattTrue, \kappa
   \clauseFwdR
   \elimTrue{\kappa}
}
\and
\inferrule*[
   lab={\ruleName{$\clauseFwdR$-false}}
]
{
   \strut
}
{
   \pattFalse, \kappa
   \clauseFwdR
   \elimFalse{\kappa}
}
\and
\inferrule*[
   lab={\ruleName{$\clauseFwdR$-nil}}
]
{
   \strut
}
{
   \pattNil, \kappa
   \clauseFwdR
   \elimNil{\kappa}
}
\and
\inferrule*[
   lab={\ruleName{$\clauseFwdR$-cons}}
]
{
   p', \kappa \clauseFwdR \sigma
   \\
   p, \sigma \clauseFwdR \sigma'
}
{
   \pattCons{p}{p'}, \kappa
   \clauseFwdR
   \elimCons{\sigma'}
}
\and
\inferrule*[
   lab={\ruleName{$\clauseFwdR$-non-empty-list}}
]
{
   o, \kappa \clauseFwdR \sigma
   \\
   p, \sigma \clauseFwdR \sigma'
}
{
   \pattList{p}{o}, \kappa
   \clauseFwdR
   \elimCons{\sigma'}
}
\and
\inferrule*[
   lab={\ruleName{$\clauseFwdR$-unit}}
]
{
   \strut
}
{
   \pattRecEmpty, \kappa
   \clauseFwdR
   \elimRecEmpty{\kappa}
}
\and
\inferrule*[
   lab={\ruleName{$\clauseFwdR$-record}}
]
{
   p', \kappa \clauseFwdR \sigma
   \\
   \pattRec{\vec{\bind{x}{p}}}, \sigma \clauseFwdR \elimRec{\vec{x}}{\sigma'}
}
{
   \pattRec{\vec{\bind{x}{p}} \concat \bind{y}{p'}}, \kappa
   \clauseFwdR
   \elimRec{\vec{x} \concat y}{\sigma'}
}
\and
\inferrule*[
   lab={\ruleName{$\clauseFwdR$-seq}},
   right={$\vec{p} \neq \seqEmpty$}
]
{
   \vec{p}, e \clauseFwdR \sigma
   \\
   p, \exLambda{\sigma} \clauseFwdR \sigma'
}
{
   p \concat \vec{p}, e
   \clauseFwdR
   \sigma'
}
\end{smathpar}
\vspace{3mm}

\flushleft\shadebox{$o, \kappa \clauseFwdR \sigma$}
\begin{smathpar}
\inferrule*[
   lab={\ruleName{$\clauseFwdR$-list-rest-end}}
]
{
   \strut
}
{
   \pattListEnd, \kappa
   \clauseFwdR
   \elimNil{\kappa}
}
\and
\inferrule*[
   lab={\ruleName{$\clauseFwdR$-list-rest-cons}}
]
{
   o, \kappa \clauseFwdR \sigma
   \\
   p, \sigma \clauseFwdR \sigma'
}
{
   (\pattListNext{p}{o}), \kappa
   \clauseFwdR
   \elimCons{\sigma'}
}
\end{smathpar}
\vspace{3mm}

\flushleft\shadebox{$\vec{c} \clausesFwdR \sigma$}
\begin{smathpar}
\inferrule*[
   lab={\ruleName{$\clausesFwdR$-clause}}
]
{
   s \desugarFwdR e
   \\
   \vec{p}, e \clauseFwdR \sigma
}
{
   \clause{\vec{p}}{s}
   \clausesFwdR
   \sigma
}
\and
\inferrule*[
   lab={\ruleName{$\clausesFwdR$-clause-seq}},
   right={$\vec{c} \neq \seqEmpty$}
]
{
   c \clausesFwdR \sigma
   \\
   \vec{c} \clausesFwdR \sigma'
   \\
   \sigma \disjjoin \sigma' = \tau
}
{
   c \concat \vec{c}
   \clausesFwdR
   \tau
}
\end{smathpar}
\caption{Desugaring clauses -- forward slicing}
\label{fig:surface-language:clause-fwd}
\end{figure}

\begin{figure}
\flushleft \shadebox{$\totaliseFwd{\kappa}{\alpha}{\vec{\pi}}{\kappa'}$}
\begin{smathpar}
\inferrule*[
   lab={\ruleName{$\totaliseFwdS$-empty}}
]
{
   \strut
}
{
   \totaliseFwd{e}{\alpha}{\seqEmpty}{e}
}
\and
\inferrule*[
   lab={\ruleName{$\totaliseFwdS$-var}}
]
{
   \totaliseFwd{\kappa}{\alpha}{\vec{\pi}}{\kappa'}
}
{
   \totaliseFwd{\elimVar{x}{\kappa}}
               {\alpha}
               {\pattVar{x} \concat \vec{\pi}}
               {\elimVar{x}{\kappa'}}
}
\and
\inferrule*[
   lab={\ruleName{$\totaliseFwdS$-true}}
]
{
   \totaliseFwd{\kappa}{\alpha}{\vec{\pi}}{\kappa'}
}
{
   \totaliseFwd{\elimTrue{\kappa}}
               {\alpha}
               {\pattTrue \concat \vec{\pi}}
               {\elimBool{\kappa'}{\annNil{\alpha}}}
}
\and
\inferrule*[
   lab={\ruleName{$\totaliseFwdS$-false}}
]
{
   \totaliseFwd{\kappa}{\alpha}{\vec{\pi}}{\kappa'}
}
{
   \totaliseFwd{\elimFalse{\kappa}}
               {\alpha}
               {\pattFalse \concat \vec{\pi}}
               {\elimBool{\annNil{\alpha}}{\kappa'}}
}
\and
\inferrule*[
   lab={\ruleName{$\totaliseFwdS$-unit}}
]
{
   \totaliseFwd{\kappa}{\alpha}{\vec{\pi}}{\kappa'}
}
{
   \totaliseFwd{\elimRecEmpty{\kappa}}
               {\alpha}
               {\pattRecEmpty \concat \vec{\pi}}
               {\elimRecEmpty{\kappa'}}
}
\and
\inferrule*[
   lab={\ruleName{$\totaliseFwdS$-record}}
]
{
   \totaliseFwd{\elimRec{\vec{x}}{\sigma}}
               {\alpha}
               {\pattRec{\vec{\bind{x}{p}}} \concat p' \concat \vec{\pi}}
               {\elimRec{\vec{x}}{\sigma'}}
}
{
   \totaliseFwd{\elimRec{\vec{x} \concat y}{\sigma}}
               {\alpha}
               {\pattRec{\vec{\bind{x}{p}} \concat \bind{y}{p'}} \concat \vec{\pi}}
               {\elimRec{\vec{x} \concat y}{\sigma'}}
}
\and
\inferrule*[
   lab={\ruleName{$\totaliseFwdS$-nil}}
]
{
   \totaliseFwd{\kappa}{\alpha}{\vec{\pi}}{\kappa'}
}
{
   \totaliseFwd{\elimNil{\kappa}}
               {\alpha}
               {\pattNil \concat \vec{\pi}}
               {\elimList{\kappa'}{\elimVar{x}{\elimVar{y}{\annNil{\alpha}}}}}
}
\and
\inferrule*[
   lab={\ruleName{$\totaliseFwdS$-cons}}
]
{
   \totaliseFwd{\sigma}{\alpha}{p \concat p' \concat \vec{\pi}}{\sigma'}
}
{
   \totaliseFwd{\elimCons{\sigma}}
               {\alpha}
               {(\pattCons{p}{p'}) \concat \vec{\pi}}
               {\elimList{\annNil{\alpha}}{\sigma'}}
}
\and
\inferrule*[
   lab={\ruleName{$\totaliseFwdS$-non-empty-list}}
]
{
   \totaliseFwd{\sigma}{\alpha}{p \concat o \concat \vec{\pi}}{\sigma'}
}
{
   \totaliseFwd{\elimCons{\sigma}}
               {\alpha}
               {(\pattList{p}{o}) \concat \vec{\pi}}
               {\elimList{\annNil{\alpha}}{\sigma'}}
}
\and
\inferrule*[
   lab={\ruleName{$\totaliseFwdS$-list-rest-end}}
]
{
   \totaliseFwd{\kappa}{\alpha}{\vec{\pi}}{\kappa'}
}
{
   \totaliseFwd{\elimNil{\kappa}}
               {\alpha}
               {\pattListEnd \concat \vec{\pi}}
               {\elimList{\kappa'}{\elimVar{x}{\elimVar{y}{\annot{\exNil}{\alpha}}}}}
}
\and
\inferrule*[
   lab={\ruleName{$\totaliseFwdS$-list-rest-cons}}
]
{
   \totaliseFwd{\sigma}{\alpha}{p \concat o \concat \vec{\pi}}{\sigma'}
}
{
   \totaliseFwd{\elimCons{\sigma}}
               {\alpha}
               {(\pattListNext{p}{o}) \concat \vec{\pi}}
               {\elimList{\annNil{\alpha}}{\sigma'}}
}
\end{smathpar}
\caption{Totalise -- forward slicing}
\label{fig:surface-language:totalise-fwd}
\end{figure}

\begin{figure}
\flushleft \shadebox{$e \desugarBwdR{t} s$}
\begin{smathpar}
   \inferrule*[
      lab={\ruleName{$\desugarBwdR{}$-eq}}
   ]
   {
      e \mathrel{\eq\desugarBwdR{t}} s
   }
   {
      e \desugarBwdR{t} s
   }
   \and
   \inferrule*[
      lab={\ruleName{$\desugarBwdR{}$-binary-apply}}
   ]
   {
      e \desugarBwdR{t} s
      \\
      e' \desugarBwdR{t'} s'
   }
   {
      \exApp{\exApp{\exOp{\primOp}}{e}}{e'}
      \desugarBwdR{\exBinaryApp{t}{\primOp}{t'}}
      \exBinaryApp{s}{\primOp}{s'}
   }
   \and
   \inferrule*[
      lab={\ruleName{$\desugarBwdR{}$-nil}}
   ]
   {
      \strut
   }
   {
      \annNil{\alpha} \desugarBwdR{\exNil} \annNil{\alpha}
   }
   \and
   \inferrule*[
      lab={\ruleName{$\desugarBwdR{}$-non-empty-list}}
   ]
   {
      e \desugarBwdR{t} s
      \\
      e' \desugarBwdR{r} r'
   }
   {
      \annCons{e}{e'}{\alpha}
      \desugarBwdR{\exList{t}{r}}
      \annList{s}{r'}{\alpha}
   }
   \and
   \inferrule*[
      lab={\ruleName{$\desugarBwdR{}$-cons}}
   ]
   {
      e \desugarBwdR{t} s
      \\
      e' \desugarBwdR{t'} s'
   }
   {
      \annCons{e}{e'}{\alpha}
      \desugarBwdR{\exCons{t}{t'}}
      \annCons{s}{s'}{\alpha}
   }
   \and
   \inferrule*[
      lab={\ruleName{$\desugarBwdR{}$-let-rec}}
   ]
   {
      \sigma_i \clausesBwdR{\vec{c_i}} \vec{c_i}'
      \quad
      (\forall i \numleq j)
      \\
      e \desugarBwdR{t} s
   }
   {
      \exLetRecMutual{\vec{\bind{x}{\sigma}}}{e}
      \desugarBwdR{\exLetRecMutual{\seqRange{\bind{x_1}{\vec{c_1}}}{\bind{x_j}{\vec{c_j}}}}{t}}
      \exLetRecMutual{\seqRange{\bind{x_1}{\vec{c_1}'}}{\bind{x_j}{\vec{c_j}'}}}{s}
   }
   \and
   \inferrule*[
      lab={\ruleName{$\desugarBwdR{}$-apply}}
   ]
   {
      e \desugarBwdR{t} s
      \\
      e' \desugarBwdR{t'} s'
   }
   {
      \exApp{e}{e'}
      \desugarBwdR{\exApp{t}{t'}}
      \exApp{s}{s'}
   }
   \and
   \inferrule*[
      lab={\ruleName{$\desugarBwdR{}$-if}}
   ]
   {
      e \desugarBwdR{t} s
      \\
      e_1 \desugarBwdR{t_1} s_1
      \\
      e_2 \desugarBwdR{t_2} s_2
   }
   {
      \exApp{\exLambda{\elimBool{e_1}{e_2}}}{e}
      \desugarBwdR{\exIfThenElse{t}{t_1}{t_2}}
      \exIfThenElse{s}{s_1}{s_2}
   }
   \and
   \inferrule*[
      lab={\ruleName{$\desugarBwdR{}$-match}}
   ]
   {
      \sigma \clausesBwdR{\vec{c}} \vec{c}'
      \\
      e \desugarBwdR{t} s
   }
   {
      \exApp{\exLambda{\sigma}}{e}
      \desugarBwdR{\exMatch{t}{\vec{c}}}
      \exMatch{s}{\vec{c}'}
   }
   \and
   \inferrule*[
      lab={\ruleName{$\desugarBwdR{}$-let}}
   ]
   {
      \sigma \clauseBwdR{p} e'
      \\
      e' \desugarBwdR{t'} s'
      \\
      e \desugarBwdR{t} s
   }
   {
      \exApp{\exLambda{\sigma}}{e}
      \desugarBwdR{\exLet{p}{t}{t'}}
      \exLet{p}{s}{s'}
   }
   \and
   \inferrule*[
      lab={\ruleName{$\desugarBwdR{}$-list-enum}},
   ]
   {
      e \desugarBwdR{t} s
      \\
      e' \desugarBwdR{t'} s'
   }
   {
      \exApp{\exApp{\funEnumFromTo}{e}}{e'}
      \desugarBwdR{\exListEnum{t}{t'}}
      \exListEnum{s}{s'}
   }
   \and
   \inferrule*[lab={\ruleName{$\desugarBwdR{}$-list-comp-done}}]
   {
      e \desugarBwdR{t} s
   }
   {
      \annCons{e}{\annNil{\alpha}}{\alpha'}
      \desugarBwdR{\exListComp{t}{\seqEmpty}}
      \annListComp{s}{\seqEmpty}{\alpha \join \alpha'}
   }
   \and
   \inferrule*[
      lab={\ruleName{$\desugarBwdR{}$-list-comp-guard}},
   ]
   {
      e' \desugarBwdR{t'} s'
      \\
      e \desugarBwdR{\exListComp{t}{\vec{q}}} \annListComp{s}{\vec{q}'}{\beta}
   }
   {
      \exApp{\exLambda{\elimBool{e}{\annNil{\alpha}}}}{e'}
      \desugarBwdR{\exListComp{t}{\qualGuard{t'} \concat \vec{q}}}
      \annListComp{s}{\qualGuard{s'} \concat \vec{q}'}{\alpha \join \beta}
   }
   \and
   \inferrule*[
      lab={\ruleName{$\desugarBwdR{}$-list-comp-decl}},
   ]
   {
      \sigma \clauseBwdR{p} e'
      \\
      e'
      \desugarBwdR{\exListComp{t'}{\vec{q}}}
      \annListComp{s'}{\vec{q}'}{\beta}
      \\
      e \desugarBwdR{t} s
   }
   {
      \exApp{\exLambda{\sigma}}{e}
      \desugarBwdR{\exListComp{t'}{\qualDeclaration{p}{t} \concat \vec{q}}}
      \annListComp{s'}{\qualDeclaration{p}{s} \concat \vec{q}'}{\beta}
   }
   \and
   \inferrule*[
      lab={\ruleName{$\desugarBwdR{}$-list-comp-gen}},
   ]
   {
      e \desugarBwdR{t} s
      \\
      \totaliseBwd{\sigma}{p}{\sigma'}{\beta}
      \\
      \sigma'
      \clauseBwdR{p}
      e'
      \\
      e'
      \desugarBwdR{\exListComp{t'}{\vec{q}}}
      \annListComp{s'}{\vec{q}'}{\beta'}
   }
   {
      \exApp{\exApp{\funConcatMap}{\exLambda{\sigma}}}{e}
      \desugarBwdR{\exListComp{t'}{\qualGenerator{p}{t} \concat \vec{q}}}
      \annListComp{s'}{\qualGenerator{p}{s} \concat \vec{q}'}{\beta \join \beta'}
   }
\end{smathpar}

\vspace{3mm}
\flushleft \shadebox{$e \desugarBwdR{r} r'$}
\begin{smathpar}
   \inferrule*[
      lab={\ruleName{$\desugarBwdR{}$-list-rest-end}}
   ]
   {
      \strut
   }
   {
      \annot{\exNil}{\alpha}
      \desugarBwdR{\exListEnd}
      \annot{\exListEnd}{\alpha}
   }
   \and
   \inferrule*[
      lab={\ruleName{$\desugarBwdR{}$-list-rest-cons}}
   ]
   {
      e \desugarBwdR{t} s
      \\
      e' \desugarBwdR{r} r'
   }
   {
      \annCons{e}{e'}{\alpha}
      \desugarBwdR{(\exListNext{t}{r})}
      (\annListNext{s}{r'}{\alpha})
   }
\end{smathpar}
\caption{Desugaring -- backward slicing (selected rules)}
\label{fig:desugar-bwd}
\end{figure}

\begin{figure}
\flushleft\shadebox{$\sigma \clauseBwdR{\vec{p}} \kappa$}
\begin{smathpar}
\inferrule*[
   lab={\ruleName{$\clauseBwdS$-eq}}
]
{
   \sigma \mathrel{\eq\clauseBwdR{\vec{p}}} \kappa
}
{
   \sigma \clauseBwdR{\vec{p}} \kappa
}
\and
\inferrule*[
   lab={\ruleName{$\clauseBwdS$-var}}
]
{
   \strut
}
{
   \elimVar{x}{\kappa}
   \clauseBwdR{\pattVar{x}}
   \kappa
}
\and
\inferrule*[
   lab={\ruleName{$\clauseBwdS$-true}}
]
{
   \strut
}
{
   \elimTrue{\kappa}
   \clauseBwdR{\pattTrue}
   \kappa
}
\and
\inferrule*[
   lab={\ruleName{$\clauseBwdS$-false}}
]
{
   \strut
}
{
   \elimFalse{\kappa}
   \clauseBwdR{\pattFalse}
   \kappa
}
\and
\inferrule*[
   lab={\ruleName{$\clauseBwdS$-nil}}
]
{
   \strut
}
{
   \elimNil{\kappa}
   \clauseBwdR{\pattNil}
   \kappa
}
\and
\inferrule*[
   lab={\ruleName{$\clauseBwdS$-cons}}
]
{
   \sigma \clauseBwdR{p} \tau
   \\
   \tau \clauseBwdR{p'} \kappa
}
{
   \elimCons{\sigma}
   \clauseBwdR{\pattCons{p}{p'}}
   \kappa
}
\and
\inferrule*[
   lab={\ruleName{$\clauseBwdS$-non-empty-list}}
]
{
   \sigma \clauseBwdR{p} \tau
   \\
   \tau \clauseBwdR{o} \kappa
}
{
   \elimCons{\sigma}
   \clauseBwdR{\pattList{p}{o}}
   \kappa
}
\and
\inferrule*[
   lab={\ruleName{$\clauseBwdS$-unit}}
]
{
   \strut
}
{
   \elimRecEmpty{\kappa}
   \clauseBwdR{\pattRecEmpty}
   \kappa
}
\and
\inferrule*[
   lab={\ruleName{$\clauseBwdS$-record}}
]
{
   \elimRec{\vec{x}}{\sigma} \clauseBwdR{\pattRec{\vec{\bind{x}{p}}}} \tau
   \\
   \tau \clauseBwdR{p'} \kappa
}
{
   \elimRec{\vec{x} \concat y}{\sigma}
   \clauseBwdR{\pattRec{\vec{\bind{x}{p}} \concat \bind{y}{p'}}}
   \kappa
}
\and
\inferrule*[
   lab={\ruleName{$\clauseBwdS$-seq}},
   right={$\vec{p} \neq \seqEmpty$}
]
{
   \sigma \clauseBwdR{p}\eq \exLambda{\sigma'}
   \\
   \sigma' \clauseBwdR{\vec{p}} e
}
{
   \sigma
   \clauseBwdR{p \concat \vec{p}}
   e
}
\end{smathpar}

\vspace{5pt}
\flushleft\shadebox{$\sigma \clauseBwdR{o} \kappa$}
\begin{smathpar}
\inferrule*[
   lab={\ruleName{$\clauseBwdS$-list-rest-eq}}
]
{
   \sigma \mathrel{\eq\clauseBwdR{o}} \kappa
}
{
   \sigma \clauseBwdR{o} \kappa
}
\and
\inferrule*[
   lab={\ruleName{$\clauseBwdS$-list-rest-end}}
]
{
   \strut
}
{
   \elimNil{\kappa}
   \clauseBwdR{\pattListEnd}
   \kappa
}
\and
\inferrule*[
   lab={\ruleName{$\clauseBwdS$-list-rest-cons}}
]
{
   \sigma \clauseBwdR{p} \tau
   \\
   \tau \clauseBwdR{o} \kappa
}
{
   \elimCons{\sigma}
   \clauseBwdR{(\pattListNext{p}{o})}
   \kappa
}
\end{smathpar}

\vspace{5pt}
\flushleft\shadebox{$\sigma \clausesBwdR{\vec{c}} \vec{c}'$}
\begin{smathpar}
\inferrule*[
   lab={\ruleName{$\clausesBwdS$-clause}}
]
{
   \sigma \clauseBwdR{\vec{p}} e
   \\
   e \desugarBwdR{s} s'
}
{
   \sigma
   \clausesBwdR{\clause{\vec{p}}{s}}
   \clause{\vec{p}}{s'}
}
\and
\inferrule*[
   lab={\ruleName{$\clausesBwdS$-clause-seq}},
   right={$\vec{c} \neq \seqEmpty$}
]
{
   \sigma \eq \sigma' \disjjoin \tau
   \\
   \sigma' \clausesBwdR{c} c'
   \\
   \tau \clausesBwdR{\vec{c}} \vec{c}'
}
{
   \sigma
   \clauseBwdR{c \concat \vec{c}}
   c' \concat \vec{c}'
}
\end{smathpar}
\caption{Desugaring clauses -- backward slicing}
\label{fig:surface-language:clause-bwd}
\end{figure}

\begin{figure}
\flushleft \shadebox{$\totaliseBwd{\kappa}{\vec{\pi}}{\kappa'}{\alpha}$}
\begin{smathpar}
\inferrule*[
   lab={\ruleName{$\totaliseBwdS$-empty}}
]
{
   \strut
}
{
   \totaliseBwd{e}{\seqEmpty}{e}{\FF}
}
\and
\inferrule*[
   lab={\ruleName{$\totaliseBwdS$-var}}
]
{
   \sigma \eq \elimVar{x}{\kappa}
   \\
   \totaliseBwd{\kappa}{\vec{\pi}}{\kappa'}{\alpha}
}
{
   \totaliseBwd{\sigma}{x \concat \vec{\pi}}{\elimVar{x}{\kappa'}}{\alpha}
}
\\
\and
\inferrule*[
   lab={\ruleName{$\totaliseBwdS$-true}}
]
{
   \sigma \eq \elimBool{\kappa}{\annot{\exNil}{\alpha}}
   \\
   \totaliseBwd{\kappa}{\vec{\pi}}{\kappa'}{\beta}
}
{
   \totaliseBwd{\sigma}
               {\pattTrue \concat \vec{\pi}}
               {\elimTrue{\kappa'}}
               {\alpha \join \beta}
}
\and
\inferrule*[
   lab={\ruleName{$\totaliseBwdS$-false}}
]
{
   \sigma \eq \elimBool{\annot{\exNil}{\alpha}}{\kappa}
   \\
   \totaliseBwd{\kappa}{\vec{\pi}}{\kappa'}{\beta}
}
{
   \totaliseBwd{\sigma}
               {\pattFalse \concat \vec{\pi}}
               {\elimFalse{\kappa'}}
               {\alpha \join \beta}
}
\and
\inferrule*[
   lab={\ruleName{$\totaliseBwdS$-unit}}
]
{
   \sigma \eq \elimRecEmpty{\kappa}
   \\
   \totaliseBwd{\kappa}{\vec{\pi}}{\kappa'}{\alpha}
}
{
   \totaliseBwd{\sigma}
               {\pattRecEmpty \concat \vec{\pi}}
               {\elimRecEmpty{\kappa'}}
               {\alpha}
}
\and
\inferrule*[
   lab={\ruleName{$\totaliseBwdS$-record}}
]
{
   \sigma \eq \elimRec{\vec{x} \concat y}{\sigma'}
   \\
   \totaliseBwd{\sigma'}{\pattRec{\vec{\bind{x}{p}}} \concat p' \concat \vec{\pi}}{\tau}{\beta}
}
{
   \totaliseBwd{\sigma}
               {\pattRec{\vec{\bind{x}{p}} \concat \bind{y}{p'}} \concat \vec{\pi}}
               {\elimRec{\vec{x} \concat y}{\tau}}
               {\beta}
}
\and
\inferrule*[
   lab={\ruleName{$\totaliseBwdS$-nil}}
]
{
   \sigma \eq \elimList{\kappa}{\elimVar{x}{\elimVar{y}{\annot{\exNil}{\alpha}}}}
   \\
   \totaliseBwd{\kappa}{\vec{\pi}}{\kappa'}{\beta}
}
{
   \totaliseBwd{\sigma}
               {\pattNil \concat \vec{\pi}}
               {\elimNil{\kappa'}}
               {\alpha \join \beta}
}
\and
\inferrule*[
   lab={\ruleName{$\totaliseBwdS$-cons}}
]
{
   \sigma \eq \elimList{\annot{\exNil}{\alpha}}{\sigma'}
   \\
   \totaliseBwd{\sigma'}{p \concat p' \concat \vec{\pi}}{\tau}{\beta}
}
{
   \totaliseBwd{\sigma}
               {(\pattCons{p}{p'}) \concat \vec{\pi}}
               {\elimCons{\tau}}
               {\alpha \join \beta}
}
\and
\inferrule*[
   lab={\ruleName{$\totaliseBwdS$-non-empty-list}}
]
{
   \sigma \eq \elimList{\annot{\exNil}{\alpha}}{\sigma'}
   \\
   \totaliseBwd{\sigma'}{p \concat o \concat \vec{\pi}}{\tau}{\beta}
}
{
   \totaliseBwd{\sigma}
               {(\pattList{p}{o}) \concat \vec{\pi}}
               {\elimCons{\tau}}
               {\alpha \join \beta}
}
\and
\inferrule*[
   lab={\ruleName{$\totaliseBwdS$-list-rest-end}}
]
{
   \sigma \eq \elimList{\kappa}{\elimVar{x}{\elimVar{y}{\annot{\exNil}{\alpha}}}}
   \\
   \totaliseBwd{\kappa}{\vec{\pi}}{\kappa'}{\beta}
}
{
   \totaliseBwd{\sigma}
               {\pattListEnd \concat \vec{\pi}}
               {\elimNil{\kappa'}}
               {\alpha \join \beta}
}
\and
\inferrule*[
   lab={\ruleName{$\totaliseBwdS$-list-rest-cons}}
]
{
   \sigma \eq \elimList{\annot{\exNil}{\alpha}}{\sigma'}
   \\
   \totaliseBwd{\sigma'}{p \concat o \concat \vec{\pi}}{\tau}{\beta}
}
{
   \totaliseBwd{\sigma}
               {(\pattListNext{p}{o}) \concat \vec{\pi}}
               {\elimCons{\tau}}
               {\alpha \join \beta}
}
\end{smathpar}
\caption{Totalise --- backward slicing}
\label{fig:surface-language:totalise-bwd}
\end{figure}

   \clearpage
   \section{Proofs: Core language}
\label{app:proofs}

\subsection{Conventions}

The symbol ${\qedLocal}$ indicates that a proof obligation is being discharged, and IH
stands for ``inductive hypothesis''. We make free use of the totality of the forward and backward slicing functions, and the fact that all term constructors preserve and reflect $\leq$, for example that $\exCons{u}{u'} \leq
\exCons{v}{v'}$ if and only if $u \leq v$ and $u' \leq v'$.

\subsection{\thmref{core-language:match:gc}}
\label{app:proofs:match:gc}
\setcounter{equation}{0}
\proofContext{match-fwd-after-bwd}
\subsubsection{Forward after backward direction.}
Induction on the $\matchBwdS$ derivation.
\begin{proof}

\small
\begin{flalign}
   \intertext{\crossrule}
   &
   \caseDerivation{\derivationWidth}{
   \begin{smathpar}
      \inferrule*[lab={\ruleName{$\matchBwdS$-var}}]
      {
         \strut
      }
      {
         \matchBwd{\bind{x}{v}}{\kappa}{\alpha}{\matchVar{x}}{v}{\elimVar{x}{\kappa}}
      }
   \end{smathpar}
   }
   &
   \notag
   \\
   &
   \qedLocal
   \derivation{\derivationWidth}{
   \begin{smathpar}
      \inferrule*[
         lab={\ruleName{$\matchFwdS$-var}}
      ]
      {
         \strut
      }
      {
         \matchFwd{v}{\elimVar{x}{\kappa}}{\matchVar{x}}{\bind{x}{v}}{\kappa}{\TT}
      }
   \end{smathpar}
   }
   &
   \notag
   \\
   &
   \qedLocal
   \TT
   \geq
   \alpha
   \notag
   \\
   \intertext{\crossrule}
   &
   \caseDerivation{\derivationWidth}{
   \begin{smathpar}
      \inferrule*[lab={\ruleName{$\matchBwdS$-true}}]
      {
         \strut
      }
      {
         \matchBwd{\seqEmpty}{\kappa}{\alpha}{\matchTrue}{\annTrue{\alpha}}{\elimBool{\kappa}{\hole}}
      }
   \end{smathpar}
   }
   &
   \notag
   \\
   &
   \qedLocal
   \derivation{\derivationWidth}{
   \begin{smathpar}
      \inferrule*[
         lab={\ruleName{$\matchFwdS$-true}}
      ]
      {
         \strut
      }
      {
         \matchFwd{\annTrue{\alpha}}{\elimBool{\kappa}{\hole}}{\matchTrue}{\seqEmpty}{\kappa}{\alpha}
      }
   \end{smathpar}
   }
   &
   \notag
   \\
   \intertext{\crossrule}
   &
   \caseDerivation{\derivationWidth}{
   \begin{smathpar}
      \inferrule*[lab={\ruleName{$\matchBwdS$-false}}]
      {
         \strut
      }
      {
         \matchBwd{\seqEmpty}{\kappa}{\alpha}{\matchFalse}{\annFalse{\alpha}}{\elimBool{\hole}{\kappa}}
      }
   \end{smathpar}
   }
   &
   \notag
   \\
   &
   \qedLocal
   \derivation{\derivationWidth}{
   \begin{smathpar}
      \inferrule*[
         lab={\ruleName{$\matchFwdS$-false}}
      ]
      {
         \strut
      }
      {
         \matchFwd{\annFalse{\alpha}}{\elimBool{\hole}{\kappa}}{\matchFalse}{\seqEmpty}{\kappa}{\alpha}
      }
   \end{smathpar}
   }
   &
   \notag
   \\
   \intertext{\crossrule}
   &
   \caseDerivation{\derivationWidth}{
   \begin{smathpar}
      \inferrule*[lab={\ruleName{$\matchBwdS$-pair}}]
      {
         \matchBwd{\rho_2}{\kappa}{\alpha}{w'}{v'}{\sigma}
         \\
         \matchBwd{\rho_1}{\sigma}{\alpha}{w}{v}{\tau}
      }
      {
         \matchBwd{\rho_1 \concat \rho_2}{\kappa}{\alpha}{\matchPair{w}{w'}}{\annPair{v}{v'}{\alpha}}{\elimProd{\tau}}
      }
   \end{smathpar}
   }
   &
   \notag
   \\
   &
   \matchFwd{v}{\tau}{w}{\rho_1'}{\tau'}{\beta}
   \geq
   \rho_1, \sigma, \alpha
   \quad
   (\exists \rho_1', \tau', \beta)
   &
   \text{IH}
   \locallabel{pair-premise-one}
   \\
   &
   \matchFwdGeq{v'}{\sigma}{w'}{\rho_2}{\kappa}{\alpha}
   &
   \text{IH}
   \notag
   \\
   &
   \matchFwd{v'}{\tau'}{w'}{\rho_2'}{\kappa'}{\beta'}
   \geq
   \rho_2, \kappa, \alpha
   \quad
   (\exists \rho_2', \kappa', \beta)
   &
   \text{monotonicity}
   \locallabel{pair-premise-two}
   \\
   &
   \qedLocal
   \derivation{\derivationWidth}{
   \begin{smathpar}
      \inferrule*[
         lab={\ruleName{$\matchFwdS$-pair}}
      ]
      {
         \matchFwd{v}{\tau}{w}{\rho_1'}{\tau'}{\beta}
         \\
         \matchFwd{v'}{\tau'}{w'}{\rho_2'}{\kappa'}{\beta'}
      }
      {
         \matchFwd{\annPair{v}{v'}{\alpha}}
                  {\elimProd{\tau}}{\matchPair{w}{w'}}
                  {\rho_1' \concat \rho_2'}{\kappa'}{\alpha \meet \beta \meet \beta'}
      }
   \end{smathpar}
   }
   &
   \text{(\localref{pair-premise-one}, \localref{pair-premise-two})}
   \notag
   \\
   &
   \qedLocal
   (\rho_1' \concat \rho_2', \kappa', \alpha \meet \beta \meet \beta')
   \geq
   (\rho_1 \concat \rho_2, \kappa, \alpha)
   &
   \notag
   \\
   \intertext{\crossrule}
   &
   \caseDerivation{\derivationWidth}{
   \begin{smathpar}
      \inferrule*[lab={\ruleName{$\matchBwdS$-nil}}]
      {
         \strut
      }
      {
         \matchBwd{\seqEmpty}{\kappa}{\alpha}{\matchNil}{\annNil{\alpha}}{\elimList{\kappa}{\hole}}
      }
   \end{smathpar}
   }
   &
   \notag
   \\
   &
   \qedLocal
   \derivation{\derivationWidth}{
   \begin{smathpar}
      \inferrule*[
         lab={\ruleName{$\matchFwdS$-nil}}
      ]
      {
         \strut
      }
      {
         \matchFwd{\annNil{\alpha}}{\elimList{\kappa}{\hole}}{\matchNil}{\seqEmpty}{\kappa}{\alpha}
      }
   \end{smathpar}
   }
   &
   \notag
   \\
   \intertext{\crossrule}
   &
   \caseDerivation{\derivationWidth}{
   \begin{smathpar}
      \inferrule*[lab={\ruleName{$\matchBwdS$-cons}}]
      {
         \matchBwd{\rho_2}{\kappa}{\alpha}{w'}{v'}{\sigma}
         \\
         \matchBwd{\rho_1}{\sigma}{\alpha}{w}{v}{\tau}
      }
      {
         \matchBwd{\rho_1 \concat \rho_2}
                  {\kappa}{\alpha}{\matchCons{w}{w'}}{\annCons{v}{v'}{\alpha}}{\elimList{\hole}{\tau}}
      }
   \end{smathpar}
   }
   &
   \notag
   \\
   &
   \matchFwd{v}{\tau}{w}{\rho_1'}{\tau'}{\beta}
   \geq
   \rho_1, \sigma, \alpha
   \quad
   (\exists \rho_1', \tau', \beta)
   &
   \text{IH}
   \locallabel{cons-premise-one}
   \\
   &
   \matchFwdGeq{v'}{\sigma}{w'}{\rho_2}{\kappa}{\alpha}
   &
   \text{IH}
   \notag
   \\
   &
   \matchFwd{v'}{\tau'}{w'}{\rho_2'}{\kappa'}{\beta'}
   \geq
   \rho_2, \kappa, \alpha
   \quad
   (\exists \rho_2', \kappa', \beta')
   &
   \text{monotonicity}
   \locallabel{cons-premise-two}
   \\
   &
   \qedLocal
   \derivation{\derivationWidth}{
   \begin{smathpar}
      \inferrule*[
         lab={\ruleName{$\matchFwdS$-cons}}
      ]
      {
         \matchFwd{v}{\tau}{w}{\rho_1'}{\tau'}{\beta}
         \\
         \matchFwd{v'}{\tau'}{w'}{\rho_2'}{\kappa'}{\beta'}
      }
      {
         \matchFwd{\annCons{v}{v'}{\alpha}}
                  {\elimList{\hole}{\tau}}
                  {\matchCons{w}{w'}}
                  {\rho_1' \concat \rho_2'}{\kappa'}{\alpha \meet \beta \meet \beta'}
      }
   \end{smathpar}
   }
   &
   \text{(\localref{cons-premise-one}, \localref{cons-premise-two})}
   \notag
   \\
   &
   \qedLocal
   (\rho_1' \concat \rho_2', \kappa', \alpha \meet \beta \meet \beta')
   \geq
   (\rho_1 \concat \rho_2, \kappa, \alpha)
   &
   \notag
\end{flalign}
\end{proof}

\setcounter{equation}{0}
\proofContext{match-bwd-after-fwd}
\subsubsection{Backward after forward direction.}
Induction on the $\matchFwdS$ derivation.
\begin{proof}

\small
\begin{flalign}
   \intertext{\crossrule}
   &
   \caseDerivation{\derivationWidth}{
   \begin{smathpar}
      \inferrule*[
         lab={\ruleName{$\matchFwdS$-var}}
      ]
      {
         \sigma \eq \elimVar{x}{\kappa}
      }
      {
         \matchFwd{v}{\sigma}{\matchVar{x}}{\bind{x}{v}}{\kappa}{\TT}
      }
   \end{smathpar}
   }
   &
   \notag
   \\
   &
   \qedLocal
   \derivation{\derivationWidth}{
   \begin{smathpar}
      \inferrule*[lab={\ruleName{$\matchBwdS$-var}}]
      {
         \strut
      }
      {
         \matchBwd{\bind{x}{v}}{\kappa}{\TT}{\matchVar{x}}{v}{\elimVar{x}{\kappa}}
      }
   \end{smathpar}
   }
   &
   \notag
   \\
   &
   \qedLocal
   (v, \elimVar{x}{\kappa})
   \eq
   (v, \sigma)
   \notag
   \\
   \intertext{\crossrule}
   &
   \caseDerivation{\derivationWidth}{
   \begin{smathpar}
      \inferrule*[
         lab={\ruleName{$\matchFwdS$-true}}
      ]
      {
         v \eq \annTrue{\alpha}
         \\
         \sigma \eq \elimBool{\kappa}{\kappa'}
      }
      {
         \matchFwd{v}{\sigma}{\matchTrue}{\seqEmpty}{\kappa}{\alpha}
      }
   \end{smathpar}
   }
   &
   \notag
   \\
   &
   \qedLocal
   \derivation{\derivationWidth}{
   \begin{smathpar}
      \inferrule*[lab={\ruleName{$\matchBwdS$-true}}]
      {
         \strut
      }
      {
         \matchBwd{\seqEmpty}{\kappa}{\alpha}{\matchTrue}{\annTrue{\alpha}}{\elimBool{\kappa}{\hole}}
      }
   \end{smathpar}
   }
   &
   \notag
   \\
   &
   \qedLocal
   (\annTrue{\alpha}, \elimBool{\kappa}{\hole})
   \leq
   (\annTrue{\alpha}, \elimBool{\kappa}{\kappa'})
   \eq
   (v, \sigma)
   \notag
   \\
   \intertext{\crossrule}
   &
   \caseDerivation{\derivationWidth}{
   \begin{smathpar}
      \inferrule*[
         lab={\ruleName{$\matchFwdS$-false}}
      ]
      {
         v \eq \annFalse{\alpha}
         \\
         \sigma \eq \elimBool{\kappa}{\kappa'}
      }
      {
         \matchFwd{v}{\sigma}{\matchFalse}{\seqEmpty}{\kappa'}{\alpha}
      }
   \end{smathpar}
   }
   &
   \notag
   \\
   &
   \qedLocal
   \derivation{\derivationWidth}{
   \begin{smathpar}
      \inferrule*[lab={\ruleName{$\matchBwdS$-false}}]
      {
         \strut
      }
      {
         \matchBwd{\seqEmpty}{\kappa'}{\alpha}{\matchFalse}{\annFalse{\alpha}}{\elimBool{\hole}{\kappa'}}
      }
   \end{smathpar}
   }
   &
   \notag
   \\
   &
   \qedLocal
   (\annFalse{\alpha}, \elimBool{\hole}{\kappa'})
   \leq
   (\annFalse{\alpha},  \elimBool{\kappa}{\kappa'})
   \eq
   (v, \sigma)
   \notag
   \\
   \intertext{\crossrule}
   &
   \caseDerivation{\derivationWidth}{
   \begin{smathpar}
      \inferrule*[
         lab={\ruleName{$\matchFwdS$-pair}}
      ]
      {
         v \eq \annPair{v_1}{v_2}{\alpha}
         \\
         \sigma \eq \elimProd{\sigma'}
         \\
         \matchFwd{v_1}{\sigma'}{w_1}{\rho_1}{\tau}{\beta}
         \\
         \matchFwd{v_2}{\tau}{w_2}{\rho_2}{\kappa}{\beta'}
      }
      {
         \matchFwd{v}{\sigma'}{\matchPair{w_1}{w_2}}{\rho_1 \concat \rho_2}{\kappa}{\alpha \meet \beta \meet \beta'}
      }
   \end{smathpar}
   }
   &
   \notag
   \\
   &
   \matchBwdLeq{\rho_2}{\kappa}{\beta'}{w_2}{v_2}{\tau}
   &
   \text{IH}
   \notag
   \\
   &
   \matchBwd{\rho_2}{\kappa}{\alpha \meet \beta \meet \beta'}{w_2}{u_2}{\tau'}
   \leq
   v_2, \tau
   \quad
   (\exists u_2, \tau')
   &
   \text{monotonicity}
   \locallabel{pair-premise-one}
   \\
   &
   \matchBwdLeq{\rho_1}{\tau}{\beta}{w_1}{v_1}{\sigma'}
   &
   \text{IH}
   \notag
   \\
   &
   \matchBwd{\rho_1}{\tau'}{\alpha \meet \beta \meet \beta'}{w_1}{u_1}{\tau^\twoPrime}
   \leq
   v_1, \sigma'
   \quad
   (\exists u_1, \tau^\twoPrime)
   &
   \text{monotonicity}
   \locallabel{pair-premise-two}
   \\
   &
   \qedLocal
   \derivation{\derivationWidth}{
   \begin{smathpar}
      \inferrule*[lab={\ruleName{$\matchBwdS$-pair}}]
      {
         \matchBwd{\rho_2}{\kappa}{\alpha \meet \beta \meet \beta'}{w_2}{u_2}{\tau'}
         \\
         \matchBwd{\rho_1}{\tau'}{\alpha \meet \beta \meet \beta'}{w_1}{u_1}{\tau^\twoPrime}
      }
      {
         \matchBwd{\rho_1 \concat \rho_2}
                  {\kappa}
                  {\alpha \meet \beta \meet \beta'}
                  {\matchPair{w_1}{w_2}}
                  {\annPair{u_1}{u_2}{\alpha \meet \beta \meet \beta'}}{\elimProd{\tau^\twoPrime}}
      }
   \end{smathpar}
   }
   &
   \text{
      (\localref{pair-premise-one},
       \localref{pair-premise-two})
   }
   \notag
   \\
   &
   \qedLocal
   (\annPair{u_1}{u_2}{\alpha \meet \beta \meet \beta'}, \elimProd{\tau^\twoPrime})
   \leq
   (\annPair{v_1}{v_2}{\alpha}, \elimProd{\sigma'})
   \eq
   (v, \sigma)
   &
   \notag
   \\
   \intertext{\crossrule}
   &
   \caseDerivation{\derivationWidth}{
   \begin{smathpar}
      \inferrule*[
         lab={\ruleName{$\matchFwdS$-nil}}
      ]
      {
         v \eq \annNil{\alpha}
         \\
         \sigma \eq \elimList{\kappa}{\sigma'}
      }
      {
         \matchFwd{v}{\sigma}{\matchNil}{\seqEmpty}{\kappa}{\alpha}
      }
   \end{smathpar}
   }
   &
   \notag
   \\
   &
   \qedLocal
   \derivation{\derivationWidth}{
   \begin{smathpar}
      \inferrule*[lab={\ruleName{$\matchBwdS$-nil}}]
      {
         \strut
      }
      {
         \matchBwd{\seqEmpty}{\kappa}{\alpha}{\matchNil}{\annNil{\alpha}}{\elimList{\kappa}{\hole}}
      }
   \end{smathpar}
   }
   &
   \notag
   \\
   &
   \qedLocal
   (\annNil{\alpha}, \elimList{\kappa}{\hole})
   \leq
   (\annNil{\alpha}, \elimList{\kappa}{\sigma'})
   \eq
   (v, \sigma)
   \notag
   \\
   \intertext{\crossrule}
   &
   \caseDerivation{\derivationWidth}{
   \begin{smathpar}
      \inferrule*[
         lab={\ruleName{$\matchFwdS$-cons}}
      ]
      {
         v \eq \annCons{v_1}{v_2}{\alpha}
         \\
         \sigma \eq \elimList{\kappa}{\sigma'}
         \\
         \matchFwd{v_1}{\sigma'}{w_1}{\rho_1}{\tau}{\beta}
         \\
         \matchFwd{v_2}{\tau}{w_2}{\rho_2}{\kappa}{\beta'}
      }
      {
         \matchFwd{v}{\sigma}{\matchCons{w_1}{w_2}}{\rho_1 \concat \rho_2}{\kappa}{\alpha \meet \beta \meet \beta'}
      }
   \end{smathpar}
   }
   &
   \notag
   \\
   &
   \matchBwdLeq{\rho_2}{\kappa}{\beta'}{w_2}{v_2}{\tau}
   &
   \text{IH}
   \notag
   \\
   &
   \matchBwd{\rho_2}{\kappa}{\alpha \meet \beta \meet \beta'}{w_2}{u_2}{\tau'}
   \leq
   v_2, \tau
   \quad
   (\exists u_2, \tau')
   &
   \text{monotonicity}
   \locallabel{cons-premise-one}
   \\
   &
   \matchBwdLeq{\rho_1}{\tau}{\beta}{w_1}{v_1}{\sigma'}
   &
   \text{IH}
   \notag
   \\
   &
   \matchBwd{\rho_1}{\tau'}{\alpha \meet \beta \meet \beta'}{w_1}{u_1}{\tau^\twoPrime}
   \leq
   v_1, \sigma'
   \quad
   (\exists u_1, \tau^\twoPrime)
   &
   \text{monotonicity}
   \locallabel{cons-premise-two}
   \\
   &
   \qedLocal
   \derivation{\derivationWidth}{
   \begin{smathpar}
      \inferrule*[lab={\ruleName{$\matchBwdS$-cons}}]
      {
         \matchBwd{\rho_2}{\kappa}{\alpha \meet \beta \meet \beta'}{w_2}{u_2}{\tau'}
         \\
         \matchBwd{\rho_1}{\tau'}{\alpha \meet \beta \meet \beta'}{w_1}{u_1}{\tau^\twoPrime}
      }
      {
         \matchBwd{\rho_1 \concat \rho_2}
                  {\kappa}
                  {\alpha \meet \beta \meet \beta'}
                  {\matchCons{w_1}{w_2}}
                  {\annCons{u_1}{u_2}{\alpha \meet \beta \meet \beta'}}
                  {\elimList{\hole}{\tau^\twoPrime}}
      }
   \end{smathpar}
   }
   &
   \text{(\localref{cons-premise-one}, \localref{cons-premise-two})}
   \notag
   \\
   &
   \qedLocal
   (\annCons{u_1}{u_2}{\alpha \meet \beta \meet \beta'}, \elimList{\hole}{\tau^\twoPrime})
   \leq
   (\annCons{v_1}{v_2}{\alpha}, \elimList{\kappa}{\sigma'})
   \eq
   (v, \sigma)
   &
   \notag
\end{flalign}
\end{proof}

\subsection{\lemref{core-language:env-get-put}}
\label{app:proofs:lookup:gc}
\setcounter{equation}{0}
\proofContext{env-get-put}
If $\envLookupBwd{\rho}{\rho'}{\bind{x}{v}}$ then $\envLookup{\rho}{x}{v}$.

\begin{proof}
   By induction on the proof that $\envLookupBwd{\rho}{\rho'}{\bind{x}{v}}$:
\small
\begin{flalign}
   \intertext{\crossrule}
   &
   \caseDerivation{\derivationWidth}{
   \begin{smathpar}
      \inferrule*[
         lab={\ruleName{$\envLookupBwdS$-head}}
      ]
      {
        \strut
      }
      {
        \envLookupBwd{(\hole_{\rho'} \concat \bind{x}{v})}{\rho' \concat \bind{x}{u}}{\bind{x}{v}}
      }
   \end{smathpar}
   }
   &
   \notag
   \\
   &
   \qedLocal
   \derivation{\derivationWidth}{
   \begin{smathpar}
      \inferrule*[
         lab={\ruleName{$\envLookupS$-head}}
      ]
      {
        \strut
      }
      {
        \envLookup{\hole_{\rho'} \concat \bind{x}{v}}{x}{v}
      }
   \end{smathpar}
   }
   &
   \notag
   \\
   \intertext{\crossrule}
   &
   \caseDerivation{\derivationWidth}{
      \begin{smathpar}
         \inferrule*[
            lab={\ruleName{$\envLookupBwdS$-tail}},
            right={$x \neq x'$}
         ]
         {
            \envLookupBwd{\rho}{\rho'}{\bind{x}{v}}
         }
         {
            \envLookupBwd{(\rho \concat \bind{x'}{\hole})}{\rho' \concat \bind{x'}{u}}{\bind{x}{v}}
         }
      \end{smathpar}
   }
   &
   \notag
   \\
   &
   \envLookup{\rho}{x}{v}
   &
   \text{IH}
   \locallabel{lookup-ih}
   \\
   &
   \qedLocal
   \derivation{\derivationWidth}{
      \begin{smathpar}
         \inferrule*[
            lab={\ruleName{$\envLookupS$-tail}},
            right={$x' \neq x$}
         ]
         {
            \envLookup{\rho}{x}{v}
         }
         {
            \envLookup{(\rho \concat \bind{x'}{\hole})}{x}{v}
         }
      \end{smathpar}
   }
   &
   (\localref{lookup-ih})
   \notag
\end{flalign}
\end{proof}

\noindent
If $\envLookup{\rho}{x}{v}$ then $\exists \rho'$ such that $\envLookupBwd{\rho'}{\rho}{\bind{x}{v}}$ and $\rho' \leq \rho$.
\begin{proof}
   By induction on the proof that $\envLookup{\rho}{x}{v}$.
\small
\begin{flalign}
   \intertext{\crossrule}
   &
   \caseDerivation{\derivationWidth}{
   \begin{smathpar}
      \inferrule*[
         lab={\ruleName{$\envLookupS$-head}}
      ]
      {
         \strut
      }
      {
         \envLookup{(\rho \concat \bind{x}{v})}{x}{v}
      }
   \end{smathpar}
   }
   &
   \notag
   \\
   &
   \qedLocal
   \derivation{\derivationWidth}{
   \begin{smathpar}
      \inferrule*[
         lab={\ruleName{$\envLookupBwdS$-head}}
      ]
      {
         \strut
      }
      {
         \envLookupBwd{(\hole_{\raw{\rho}} \concat \bind{x}{v})}
                      {\rho \concat \bind{x}{u}}
                      {\bind{x}{v}}
      }
   \end{smathpar}
   }
   &
   \notag
   \\
   &
   \qedLocal
   (\hole_{\raw{\rho}} \concat \bind{x}{v}) \leq (\rho \concat \bind{x}{v})
   &
   \notag
   \\
   \intertext{\crossrule}
   &
   \caseDerivation{\derivationWidth}{
      \begin{smathpar}
         \inferrule*[
            lab={\ruleName{$\envLookupS$-tail}},
            right={$x' \neq x$}
         ]
         {
            \envLookup{\rho}{x}{v}
         }
         {
            \envLookup{(\rho \concat \bind{x'}{u})}{x}{v}
         }
      \end{smathpar}
   }
   &
   \notag
   \\
   &
   \rho \geq \envLookupBwd{\rho'}{\rho}{\bind{x}{v}} \quad (\exists \rho')
   &
   \text{(IH)}
   \notag
   \\
   &
   \qedLocal
   \derivation{\derivationWidth}{
   \begin{smathpar}
      \inferrule*[
         lab={\ruleName{$\envLookupBwdS$-tail}},
         right={$x' \neq x$}
      ]
      {
         \envLookupBwd{\rho'}{\rho}{\bind{x}{v}}
      }
      {
         \envLookupBwd{(\rho' \concat \bind{x'}{\hole})}{\rho \concat \bind{x'}{u}}{\bind{x}{v}}
      }
   \end{smathpar}
   }
   &
   \notag
   \\
   &
   \qedLocal
   (\rho' \concat \bind{x'}{\hole}) \leq (\rho \concat \bind{x'}{u})
   &
   \notag
\end{flalign}
\end{proof}

\setcounter{equation}{0}

\subsection{\thmref{core-language:closeDefs:gc}}
\label{app:proofs:closeDefs:gc}
\proofContext{closeDefs}
\subsubsection{Forward afer backward direction}
Suppose $\rho, h \closeDefsR \rho'$.  Then $\closeDefsFwdF{\rho,h}(\closeDefsBwdF{\rho,h}(\rho')) \geq \rho'$.
\begin{proof}
\small
\begin{flalign}
   \intertext{\crossrule}
   &
   \vec{\bind{x}{\exClosure{\rho}{h}{\sigma}}}
   \closeDefsBwdR
   (\bigjoin\vec{\rho}, \vec{\bind{x}{\sigma}} \join {\bigjoin\vec{h}})
   \quad
   \text{where} \vec{h} = \seqRange{\vec{\bind{x_1}{\sigma_1'}}}{\vec{\bind{x_j}{\sigma_j'}}}
   &
   \text{(Suppose)}
   \notag
   \\
   &
   \vec{\bind{x}{\tau}} = \vec{\bind{x}{\sigma}} \join {\bigjoin\vec{h}}
   \quad
   (\exists \vec{\tau})
   &
   \notag
   \\
   &
   (\bigjoin\vec{\rho},  \vec{\bind{x}{\tau}})
   \closeDefsR
   \seqRange{\bind{x_1}{\exClosure{\bigjoin\vec{\rho}}{ \vec{\bind{x}{\tau}} }{\tau_1}}}
            {\bind{x_j}{\exClosure{\bigjoin\vec{\rho}}{ \vec{\bind{x}{\tau}} }{\tau_j}}}
   &
   \text{(Def.~$\closeDefsR$, \figref{core-language:semantics})}
   \notag
   \\
   &
   \bigjoin\vec{\rho} \geq \rho_i
   \quad
   (\forall i \numleq j)
   &
   \locallabel{recbindbtf:1}
   \\
   &
   \vec{\bind{x}{\tau}} \geq h_i
   \quad
   (\forall i \numleq j)
   &
   \locallabel{recbindbtf:2}
   \\
   &
   \tau_i = \sigma_i \join \sigma_{1, i}' \join .. \join \sigma_{j, i}'
   \quad
   (\forall i \numleq j)
   &
   \notag
   \\
   &
   \tau_i \geq \sigma_i
   &
   \locallabel{recbindbtf:3}
   \\
   &
   \qedLocal
   \seqRange{\bind{x_1}{\exClosure{\bigjoin\vec{\rho}}{\vec{\bind{x}{\tau}}}{\tau_1}}}
   {\bind{x_j}{\exClosure{\bigjoin\vec{\rho}}{\vec{\bind{x}{\tau}}}{\tau_j}}}
   \geq
   \vec{\bind{x}{\exClosure{\rho}{h}{\sigma}}}
   &
   \text{(\localref{recbindbtf:1}, \localref{recbindbtf:2}, \localref{recbindbtf:3})}
   \notag
\end{flalign}
\end{proof}

\subsubsection{Backward after forward direction}
Suppose $\rho, h \closeDefsR \rho'$.  Then $\closeDefsBwdF{\rho,h}(\closeDefsFwdF{\rho,h}(\rho, h)) \leq (\rho, h)$.
\begin{proof}
\small
\begin{flalign}
   \intertext{\crossrule}
   &
   \rho, h
   \closeDefsR
   \seqRange{\bind{x_1}{\exClosure{\rho}{h}{\sigma_1}}}{\bind{x_j}{\exClosure{\rho}{h}{\sigma_j}}}
   \hspace{5pt}
   \text{where $h = \vec{\bind{x}{\sigma}}$}
   &
   \text{(Suppose)}
   \notag
   \\
   &
   \seqRange{\bind{x_1}{\exClosure{\rho}{h}{\sigma_1}}}{\bind{x_j}{\exClosure{\rho}{h}{\sigma_j}}}
   \closeDefsBwdR
   (\rho, \vec{\bind{x}{\sigma}} \join h) = (\rho, h)
   &
   \text{(Def.~$\closeDefsBwdR$, \figref{data-dependencies:bwd})}
   \notag
   \\
   &
   \qedLocal
   (\rho, h) \leq (\rho, h)
   \notag
\end{flalign}
\end{proof}

\subsection{\thmref{core-language:eval:gc}}
\label{app:proofs:eval:gc}
\setcounter{equation}{0}
\proofContext{eval-fwd-after-bwd}
\subsubsection{Forward after backward direction.}
Induction on the $\evalBwdS$ derivation.
\begin{proof}

\small

\end{proof}

\setcounter{equation}{0}
\proofContext{eval-bwd-after-fwd}
\subsubsection{Backward after forward direction.}
Induction on the $\evalFwdS$ derivation.
\begin{proof}

\small
%
\end{proof}

   \section{Proofs: Surface language}
\label{app:proofs-surface}

\subsection{Auxiliary lemmas}

\begin{definition}
   Suppose $r \desugarFwdR e$. Define $\listRestFwdF{r}: \Below{r} \to \Below{e}$ and $\listRestBwdF{r}: \Below{e} \to \Below{r}$ to be $\desugarFwdR$ domain-restricted to $\Below{r}$ and $\desugarBwdR{r}$ domain-restricted to $\Below{e}$ respectively.
\end{definition}

\begin{lemma}[Galois connection for desugaring list rest]
  \label{lem:aux:desugarlistrest:gc}
  Suppose $r \desugarFwdR e$. Then:
  \begin{enumerate}
     \item \label{lem:aux:desugarlistrest:gc:1} $\listRestFwdF{r}$ and $\listRestBwdF{r}$ are monotonic.
     \item \label{lem:aux:desugarlistrest:gc:2} $\listRestFwdF{r}(\listRestBwdF{r}(e')) \geq e'$.
     \item \label{lem:aux:desugarlistrest:gc:3} $\listRestBwdF{r}(\listRestFwdF{r}(r')) \leq r'$.
  \end{enumerate}
\end{lemma}

\begin{definition}
   Suppose $\vec{p}, e \clauseFwdR \sigma$. Define $\clauseFwdF{\clause{\vec{p}}{e}}: \Below{e} \to \Below{\sigma}$ so that $\clauseFwdF{\clause{p}{e}}(e') = \sigma'$ iff $p, e' \clauseFwdR \sigma'$ and $\clauseBwdF{\clause{\vec{p}}{e}}: \Below{\sigma} \to \Below{e}$ so that $\clauseBwdF{\clause{\vec{p}}{e}}(\sigma') = e'$ iff $\sigma' \clauseBwdR{p} e'$.
\end{definition}

\begin{lemma}[Galois connection for clause]
  \label{lem:aux:clause:gc}
  Suppose $\vec{p}, e \clauseFwdR \sigma$. Then:
  \begin{enumerate}
     \item \label{lem:aux:clause:gc:1} $\clauseFwdF{\clause{\vec{p}}{e}}$ and $\clauseBwdF{\clause{\vec{p}}{e}}$ are monotonic.
     \item \label{lem:aux:clause:gc:2} $\clauseFwdF{\clause{\vec{p}}{e}}(\clauseBwdF{\clause{\vec{p}}{e}}(\sigma')) \geq \sigma'$.
     \item \label{lem:aux:clause:gc:3} $\clauseBwdF{\clause{\vec{p}}{e}}(\clauseFwdF{\clause{\vec{p}}{e}}(e')) \leq (e')$.
  \end{enumerate}
\end{lemma}

\begin{definition}
   Suppose $\vec{c} \clauseFwdR \sigma$. Define $\clausesFwdF{\vec{c}}: \Below{\vec{c}} \to \Below{\sigma}$ and $\clausesBwdF{\vec{c}}: \Below{\sigma} \to \Below{\vec{c}}$ to be $\clauseFwdR$ domain-restricted to $\Below{\vec{c}}$ and $\clauseBwdR{\vec{c}}$ domain-restricted to $\Below{\sigma}$ respectively.
\end{definition}

\begin{lemma}[Galois connection for clauses]
  \label{lem:aux:clauses:gc}
  Suppose $\vec{c} \clausesFwdR \sigma$. Then:
  \begin{enumerate}
     \item $\clausesFwdF{\vec{c}}$ and $\clausesBwdF{\vec{c}}$ are monotonic.
     \item $\clausesFwdF{\vec{c}}(\clausesBwdF{\vec{c}}(\sigma')) \geq \sigma'$.
     \item $\clausesBwdF{\vec{c}}(\clausesFwdF{\vec{c}}(\vec{c})) \leq \vec{c}'$.
  \end{enumerate}
\end{lemma}

\begin{definition}
   Suppose $\totaliseFwd{\kappa}{\alpha}{\vec{\pi}}{\kappa'}$. Define $\totaliseFwdF{\kappa,\alpha,\vec{\pi}}: \Below{(\kappa,\alpha)} \to \Below{\kappa'}$ and $\totaliseBwdF{\kappa,\alpha,\vec{\pi}}: \Below{\kappa'} \to \Below{(\kappa,\alpha)}$ to be $\totaliseFwdR{\vec{\pi}}$ domain-restricted to $\Below{(\kappa,\alpha)}$ and $\totaliseBwdR{\vec{\pi}}$ domain-restricted to $\Below{\kappa'}$ respectively.
\end{definition}

\begin{lemma}[Galois connection for totalise]
  \label{lem:aux:list-gen:gc}
  \item
  Suppose $\totaliseFwd{\kappa}{\alpha}{\vec{\pi}}{\kappa'}$. Then:
  \begin{enumerate}
     \item \label{lem:aux:totalise:gc:1} $\totaliseFwdF{\kappa,\alpha,\vec{\pi}}$ and $\totaliseBwdF{\kappa,\alpha,\vec{\pi}}$ are monotonic.
     \item \label{lem:aux:totalise:gc:2} $\totaliseFwdF{\kappa,\alpha,\vec{\pi}}(\totaliseBwdF{\kappa,\alpha,\vec{\pi}}(\smash{\kappa^\dagger})) \geq \smash{\kappa^\dagger}$.
     \item \label{lem:aux:totalise:gc:3} $\totaliseBwdF{\kappa,\alpha,\vec{\pi}}(\totaliseFwdF{\kappa,\alpha,\vec{\pi}}(\smash{\kappa^\dagger}, \alpha')) \leq (\smash{\kappa^\dagger}, \alpha')$.
  \end{enumerate}
\end{lemma}

\subsection{\thmref{surface-language:desugar:gc}}
\label{app:proofs-surface:desugar:gc}
\setcounter{equation}{0}
\proofContext{desugar-fwd-after-bwd}
\subsubsection{Forwards after backwards direction.}
Induction on the $\desugarBwdR{}$ derivation.
\begin{proof}
\small

\end{proof}

\setcounter{equation}{0}
\proofContext{desugar-bwd-after-fwd}
\subsubsection{Backwards after forwards direction.}
\begin{proof}
   Induction on the $\desugarFwdR$ derivation.
\small
%
\end{proof}

\setcounter{equation}{0}
\proofContext{list-rest}
\subsection{\lemref{aux:desugarlistrest:gc}}
Suppose $r \desugarFwdR e$. Then $\listRestFwdF{r}(\listRestBwdF{r}(e')) \geq e'$.
\begin{proof}
  \ref{lem:aux:desugarlistrest:gc:2}
\small
\begin{flalign}
   \intertext{\crossrule}
   &
   \caseDerivation{\derivationWidth}{
      \begin{smathpar}
         \inferrule*[
            lab={\ruleName{$\desugarBwdR{}$-list-rest-end}}
         ]
         {
            \strut
         }
         {
            \annot{\exNil}{\alpha}
            \desugarBwdR{\exListEnd}
            \annot{\exListEnd}{\alpha}
         }
      \end{smathpar}
   }
   &
   \notag
   \\
   &
   \qedLocal
   \derivation{\derivationWidth}{
      \begin{smathpar}
         \inferrule*[
            lab={\ruleName{$\desugarFwdR$-list-rest-end}}
         ]
         {
            \strut
         }
         {
            \annListEnd{\alpha} \desugarFwdR \annNil{\alpha}
         }
      \end{smathpar}
   }
   &
   \notag
   \\
   &
   \qedLocal
   \annNil{\alpha} \geq \annNil{\alpha}
   &
   \notag
   \intertext{\crossrule}
   &
   \caseDerivation{\derivationWidth}{
      \begin{smathpar}
         \inferrule*[
            lab={\ruleName{$\desugarBwdR{}$-list-rest-cons}}
         ]
         {
            e_1 \desugarBwdR{t} s
            \\
            e_2 \desugarBwdR{r} r
         }
         {
            \annCons{e_1}{e_2}{\alpha}
            \desugarBwdR{(\exListNext{t}{r})}
            (\annListNext{s}{r}{\alpha})
         }
      \end{smathpar}
   }
   &
   \notag
   \\
   &
   s \desugarFwdR e_1' \geq e_1 \quad (\exists e_1')
   &
   \text{(IH)}
   \notag
   \\
   &
   r \desugarFwdR e_2' \geq e_2 \quad (\exists e_2')
   &
   \text{(IH)}
   \notag
   \\
   &
   \qedLocal
   \derivation{\derivationWidth}{
      \begin{smathpar}
         \inferrule*[
            lab={\ruleName{$\desugarFwdR$-list-rest-cons}}
         ]
         {
            s \desugarFwdR e_1'
            \\
            r \desugarFwdR e_2'
         }
         {
            (\annListNext{s}{r}{\alpha}) \desugarFwdR \annCons{e_1'}{e_2'}{\alpha}
         }
      \end{smathpar}
   }
   &
   \notag
   \\
   &
   \qedLocal
   \annCons{e_1'}{e_2'}{\alpha} \geq \annCons{e_1}{e_2}{\alpha}
   &
   \notag
\end{flalign}
\end{proof}
\noindent
Suppose $r \desugarFwdR e$. Then $\listRestBwdF{r}(\listRestFwdF{r}(r')) \leq r'$.
\begin{proof}
  \ref{lem:aux:desugarlistrest:gc:3}
\small
\begin{flalign}
   \intertext{\crossrule}
   &
   \caseDerivation{\derivationWidth}{
      \begin{smathpar}
         \inferrule*[
            lab={\ruleName{$\desugarFwdR$-list-rest-end}}
         ]
         {
            \strut
         }
         {
            \annListEnd{\alpha} \desugarFwdR \annNil{\alpha}
         }
      \end{smathpar}
   }
   &
   \notag
   \\
   &
   \qedLocal
   \derivation{\derivationWidth}{
      \begin{smathpar}
         \inferrule*[
            lab={\ruleName{$\desugarBwdR{}$-list-rest-end}}
         ]
         {
            \strut
         }
         {
            \annot{\exNil}{\alpha}
            \desugarBwdR{\exListEnd}
            \annot{\exListEnd}{\alpha}
         }
      \end{smathpar}
   }
   &
   \notag
   \\
   &
   \qedLocal
   \annot{\exListEnd}{\alpha} \leq \annot{\exListEnd}{\alpha}
   &
   \notag
   \intertext{\crossrule}
   &
   \caseDerivation{\derivationWidth}{
      \begin{smathpar}
         \inferrule*[
            lab={\ruleName{$\desugarFwdR$-list-rest-cons}}
         ]
         {
            s \desugarFwdR e_1
            \\
            r \desugarFwdR e_2
         }
         {
            (\annListNext{s}{r}{\alpha}) \desugarFwdR \annCons{e_1}{e_2}{\alpha}
         }
      \end{smathpar}
   }
   &
   \notag
   \\
   &
   e_1 \desugarBwdR{t} s' \leq s \quad (\exists s')
   &
   \text{(IH)}
   \notag
   \\
   &
   e_2 \desugarBwdR{r} r' \leq r \quad (\exists r')
   &
   \text{(IH)}
   \notag
   \\
   &
   \qedLocal
   \derivation{\derivationWidth}{
      \begin{smathpar}
         \inferrule*[
            lab={\ruleName{$\desugarBwdR{}$-list-rest-cons}}
         ]
         {
            e_1 \desugarBwdR{t} s'
            \\
            e_2 \desugarBwdR{r} r'
         }
         {
            \annCons{e_1}{e_2}{\alpha}
            \desugarBwdR{(\exListNext{t}{r})}
            (\annListNext{s'}{r'}{\alpha})
         }
      \end{smathpar}
   }
   &
   \notag
   \\
   &
   \qedLocal
   (\annListNext{s'}{r'}{\alpha}) \leq (\annListNext{s}{r}{\alpha})
   &
   \notag
\end{flalign}
\end{proof}

\setcounter{equation}{0}
\proofContext{clause}
\subsection{\lemref{aux:clause:gc}, part (i)}
Suppose $\sigma \clauseBwdR{\vec{p}} e$. We show $\vec{p},e \clauseFwdR \sigma' \geq \sigma$.
\begin{proof}
\small
\begin{flalign}
   \intertext{\crossrule}
   &
   \caseDerivation{\derivationWidth}{
      \begin{smathpar}
         \inferrule*[
            lab={\ruleName{$\clauseBwdS$-var}}
         ]
         {
            \strut
         }
         {
            \elimVar{x}{\kappa}
            \clauseBwdR{x}
            \kappa
         }
      \end{smathpar}
   }
   &
   \notag
   \\
   &
   \qedLocal
   \derivation{\derivationWidth}{
      \begin{smathpar}
         \inferrule*[
            lab={\ruleName{$\clauseFwdR$-var}}
         ]
         {
            \strut
         }
         {
            x, \kappa
            \clauseFwdR
            \elimVar{x}{\kappa}
         }
      \end{smathpar}
   }
   &
   \notag
   \\
   &
   \qedLocal
   \elimVar{x}{\kappa} \geq \elimVar{x}{\kappa}
   &
   \notag
   \\
   \intertext{\crossrule}
   &
   \caseDerivation{\derivationWidth}{
      \begin{smathpar}
         \inferrule*[
            lab={\ruleName{$\clauseBwdS$-nil}}
         ]
         {
            \strut
         }
         {
            \elimNil{\kappa}
            \clauseBwdR{\exNil}
            \kappa
         }
      \end{smathpar}
   }
   &
   \notag
   \\
   &
   \qedLocal
   \derivation{\derivationWidth}{
      \begin{smathpar}
         \inferrule*[
            lab={\ruleName{$\clauseFwdR$-nil}}
         ]
         {
            \strut
         }
         {
            \pattNil, \kappa
            \clauseFwdR
            \elimNil{\kappa}
         }
      \end{smathpar}
   }
   &
   \notag
   \\
   &
   \qedLocal
   \elimNil{\kappa} \geq \elimNil{\kappa}
   &
   \notag
   \\
   \intertext{\crossrule}
   &
   \caseDerivation{\derivationWidth}{
      \begin{smathpar}
         \inferrule*[
            lab={\ruleName{$\clauseBwdS$-cons}}
         ]
         {
            \sigma \clauseBwdR{p} \tau
            \\
            \tau \clauseBwdR{p'} \kappa
         }
         {
            \elimCons{\sigma}
            \clauseBwdR{\pattCons{p}{p'}}
            \kappa
         }
      \end{smathpar}
   }
   &
   \notag
   \\
   &
   p', \kappa \clauseFwdR \tau' \geq \tau \quad (\exists \tau')
   &
   \text{(IH)}
   \notag
   \\
   &
   p, \tau \clauseFwdR \geq \sigma
   &
   \text{(IH)}
   \notag
   \\
   &
   p, \tau' \clauseFwdR \sigma' \geq \sigma \quad (\exists \sigma')
   &
   \text{(Monotonicity)}
   \notag
   \\
   &
   \qedLocal
   \derivation{\derivationWidth}{
      \begin{smathpar}
         \inferrule*[
            lab={\ruleName{$\clauseFwdR$-cons}}
         ]
         {
            p', \kappa \clauseFwdR \tau'
            \\
            p, \tau' \clauseFwdR \sigma'
         }
         {
            \pattCons{p}{p'}, \kappa
            \clauseFwdR
            \elimCons{\sigma'}
         }
      \end{smathpar}
   }
   &
   \notag
   \\
   &
   \elimCons{\sigma'} \geq \elimCons{\sigma}
   &
   \notag
   \\
   \intertext{\crossrule}
   &
   \caseDerivation{\derivationWidth}{
      \begin{smathpar}
         \inferrule*[
            lab={\ruleName{$\clauseBwdS$-non-empty-list}}
         ]
         {
            \sigma \clauseBwdR{p} \tau
            \\
            \tau \clauseBwdR{o} \kappa
         }
         {
            \elimCons{\sigma}
            \clauseBwdR{\pattList{p}{o}}
            \kappa
         }
      \end{smathpar}
   }
   &
   \notag
   \\
   &
   o, \kappa \clauseFwdR \tau' \geq \tau \quad (\exists \tau')
   &
   \text{(IH)}
   \notag
   \\
   &
   p, \tau \clauseFwdR \geq \sigma
   &
   \text{(IH)}
   \notag
   \\
   &
   p, \tau' \clauseFwdR \sigma' \geq \sigma \quad (\exists \sigma')
   &
   \text{(Monotonicity)}
   \notag
   \\
   &
   \qedLocal
   \derivation{\derivationWidth}{
      \begin{smathpar}
         \inferrule*[
            lab={\ruleName{$\clauseFwdR$-non-empty-list}}
         ]
         {
            o, \kappa \clauseFwdR \tau'
            \\
            p, \tau' \clauseFwdR \sigma'
         }
         {
            \pattList{p}{o}, \kappa
            \clauseFwdR
            \elimCons{\sigma'}
         }
      \end{smathpar}
   }
   &
   \notag
   \\
   &
   \qedLocal
   \elimCons{\sigma'} \geq \elimCons{\sigma}
   &
   \notag
   \\
   \intertext{\crossrule}
   &
   \caseDerivation{\derivationWidth}{
      \begin{smathpar}
         \inferrule*[
            lab={\ruleName{$\clauseBwdS$-pair}}
         ]
         {
            \sigma \clauseBwdR{p} \tau
            \\
            \tau \clauseBwdR{p'} \kappa
         }
         {
            \elimProd{\sigma}
            \clauseBwdR{\pattPair{p}{p'}}
            \kappa
         }
      \end{smathpar}
   }
   &
   \notag
   \\
   &
   p', \kappa \clauseFwdR \tau' \geq \tau \quad (\exists \tau')
   &
   \text{(IH)}
   \notag
   \\
   &
   p, \tau \clauseFwdR \geq \sigma
   &
   \text{(IH)}
   \notag
   \\
   &
   p, \tau' \clauseFwdR \sigma' \geq \sigma \quad (\exists \sigma')
   &
   \text{(Monotonicity)}
   \notag
   \\
   &
   \qedLocal
   \derivation{\derivationWidth}{
      \begin{smathpar}
         \inferrule*[
            lab={\ruleName{$\clauseFwdR$-pair}}
         ]
         {
            p', \kappa \clauseFwdR \tau'
            \\
            p, \tau' \clauseFwdR \sigma'
         }
         {
            \pattPair{p}{p'}, \kappa
            \clauseFwdR
            \elimProd{\sigma'}
         }
      \end{smathpar}
   }
   &
   \notag
   \\
   &
   \qedLocal
   \elimProd{\sigma'} \geq \elimProd{\sigma}
   &
   \notag
   \\
   \intertext{\crossrule}
   &
   \caseDerivation{\derivationWidth}{
      \begin{smathpar}
         \inferrule*[
            lab={\ruleName{$\clauseBwdS$-seq}},
            right={$\vec{p} \neq \seqEmpty$}
         ]
         {
            \sigma \clauseBwdR{p}\eq \exLambda{\tau}
            \\
            \tau \clauseBwdR{\vec{p}} e
         }
         {
            \sigma
            \clauseBwdR{p \concat \vec{p}}
            e
         }
      \end{smathpar}
   }
   &
   \notag
   \\
   &
   \vec{p}, e \clauseFwdR \tau' \geq \tau \quad (\exists \tau')
   &
   \text{(IH)}
   \notag
   \\
   &
   p, \exLambda{\tau} \clauseFwdR \geq \sigma
   &
   \text{(IH)}
   \notag
   \\
   &
   p, \exLambda{\tau'} \clauseFwdR \sigma' \geq \sigma \quad (\exists \sigma')
   &
   \text{(Monotonicity)}
   \notag
   \\
   &
   \qedLocal
   \derivation{\derivationWidth}{
      \begin{smathpar}
         \inferrule*[
            lab={\ruleName{$\clauseFwdR$-seq}},
            right={$\vec{p} \neq \seqEmpty$}
         ]
         {
            \vec{p}, e \clauseFwdR \tau'
            \\
            p, \exLambda{\tau'} \clauseFwdR \sigma'
         }
         {
            p \concat \vec{p}, e
            \clauseFwdR
            \sigma'
         }
      \end{smathpar}
   }
   &
   \notag
   \\
   &
   \qedLocal
   \sigma' \geq \sigma
   &
   \notag
   \\
   \intertext{\crossrule}
   &
   \caseDerivation{\derivationWidth}{
      \begin{smathpar}
         \inferrule*[
            lab={\ruleName{$\clauseBwdS$-list-rest-end}}
         ]
         {
            \strut
         }
         {
            \elimNil{\kappa}
            \clauseBwdR{\pattListEnd}
            \kappa
         }
      \end{smathpar}
   }
   &
   \notag
   \\
   &
   \qedLocal
   \derivation{\derivationWidth}{
      \begin{smathpar}
         \inferrule*[
            lab={\ruleName{$\clauseFwdR$-list-rest-end}}
         ]
         {
            \strut
         }
         {
            \pattListEnd, \kappa
            \clauseFwdR
            \elimNil{\kappa}
         }
      \end{smathpar}
   }
   &
   \notag
   \\
   &
   \qedLocal
   \elimNil{\kappa} \geq \elimNil{\kappa}
   &
   \notag
   \\
   \intertext{\crossrule}
   &
   \caseDerivation{\derivationWidth}{
      \begin{smathpar}
         \inferrule*[
            lab={\ruleName{$\clauseBwdS$-list-rest-cons}}
         ]
         {
            \sigma \clauseBwdR{p} \tau
            \\
            \tau \clauseBwdR{o} \kappa
         }
         {
            \elimCons{\sigma}
            \clauseBwdR{(\pattListNext{p}{o})}
            \kappa
         }
      \end{smathpar}
   }
   &
   \notag
   \\
   &
   o, \kappa \clauseFwdR \tau' \geq \tau \quad (\exists \tau')
   &
   \text{(IH)}
   \notag
   \\
   &
   p, \tau \clauseFwdR \geq \sigma
   &
   \text{(IH)}
   \notag
   \\
   &
   p, \tau' \clauseFwdR \sigma' \geq \sigma \quad (\exists \sigma')
   &
   \text{(Monotonicity)}
   \notag
   \\
   &
   \qedLocal
   \derivation{\derivationWidth}{
      \begin{smathpar}
         \inferrule*[
            lab={\ruleName{$\clauseFwdR$-list-rest-cons}}
         ]
         {
            o, \kappa \clauseFwdR \tau'
            \\
            p, \tau' \clauseFwdR \sigma'
         }
         {
            (\pattListNext{p}{o}), \kappa
            \clauseFwdR
            \elimCons{\sigma'}
         }
      \end{smathpar}
   }
   &
   \notag
   \\
   &
   \qedLocal
   \elimCons{\sigma'} \geq \elimCons{\sigma}
   &
   \notag
\end{flalign}
\end{proof}

\noindent
Suppose $\vec{p}, e \clauseFwdR \sigma$. We show $\sigma \clauseBwdR{\vec{p}} e' \leq e$.
\begin{proof}
\small
\begin{flalign}
   \intertext{\crossrule}
   &
   \caseDerivation{\derivationWidth}{
      \begin{smathpar}
         \inferrule*[
            lab={\ruleName{$\clauseFwdR$-var}}
         ]
         {
            \strut
         }
         {
            x, \kappa
            \clauseFwdR
            \elimVar{x}{\kappa}
         }
      \end{smathpar}
   }
   &
   \notag
   \\
   &
   \qedLocal
   \derivation{\derivationWidth}{
      \begin{smathpar}
         \inferrule*[
            lab={\ruleName{$\clauseBwdS$-var}}
         ]
         {
            \strut
         }
         {
            \elimVar{x}{\kappa}
            \clauseBwdR{x}
            \kappa
         }
      \end{smathpar}
   }
   &
   \notag
   \\
   &
   \qedLocal
   (x, \kappa) \leq (x, \kappa)
   &
   \notag
   \\
   \intertext{\crossrule}
   &
   \caseDerivation{\derivationWidth}{
      \begin{smathpar}
         \inferrule*[
            lab={\ruleName{$\clauseFwdR$-nil}}
         ]
         {
            \strut
         }
         {
            \pattNil, \kappa
            \clauseFwdR
            \elimNil{\kappa}
         }
      \end{smathpar}
   }
   &
   \notag
   \\
   &
   \qedLocal
   \derivation{\derivationWidth}{
      \begin{smathpar}
         \inferrule*[
            lab={\ruleName{$\clauseBwdS$-nil}}
         ]
         {
            \strut
         }
         {
            \elimNil{\kappa}
            \clauseBwdR{\exNil}
            \kappa
         }
      \end{smathpar}
   }
   &
   \notag
   \\
   &
   (\pattNil, \kappa) \leq (\pattNil, \kappa)
   &
   \notag
   \\
   \intertext{\crossrule}
   &
   \caseDerivation{\derivationWidth}{
      \begin{smathpar}
         \inferrule*[
            lab={\ruleName{$\clauseFwdR$-cons}}
         ]
         {
            p', \kappa \clauseFwdR \tau
            \\
            p, \tau \clauseFwdR \sigma
         }
         {
            \pattCons{p}{p'}, \kappa
            \clauseFwdR
            \elimCons{\sigma}
         }
      \end{smathpar}
   }
   &
   \notag
   \\
   &
   \sigma \clauseBwdR{p} \tau' \leq \tau \quad (\exists \tau')
   &
   \text{(IH)}
   \notag
   \\
   &
   \tau \clauseBwdR{p'} \leq \kappa
   &
   \text{(IH)}
   \notag
   \\
   &
   \tau' \clauseBwdR{p'} \kappa' \leq \kappa \quad (\exists \kappa')
   &
   \text{(Monotonicity)}
   \notag
   \\
   &
   \qedLocal
   \derivation{\derivationWidth}{
      \begin{smathpar}
         \inferrule*[
            lab={\ruleName{$\clauseBwdS$-cons}}
         ]
         {
            \sigma \clauseBwdR{p} \tau'
            \\
            \tau' \clauseBwdR{p'} \kappa'
         }
         {
            \elimCons{\sigma}
            \clauseBwdR{\pattCons{p}{p'}}
            \kappa'
         }
      \end{smathpar}
   }
   &
   \notag
   \\
   &
   \qedLocal
   (\pattCons{p}{p'}, \kappa') \leq (\pattCons{p}{p'}, \kappa)
   &
   \notag
   \\
   \intertext{\crossrule}
   &
   \caseDerivation{\derivationWidth}{
      \begin{smathpar}
         \inferrule*[
            lab={\ruleName{$\clauseFwdR$-non-empty-list}}
         ]
         {
            o, \kappa \clauseFwdR \tau
            \\
            p, \tau \clauseFwdR \sigma
         }
         {
            \pattList{p}{o}, \kappa
            \clauseFwdR
            \elimCons{\sigma}
         }
      \end{smathpar}
   }
   &
   \notag
   \\
   &
   \sigma \clauseBwdR{p} \tau' \leq \tau \quad (\exists \tau')
   &
   \text{(IH)}
   \notag
   \\
   &
   \tau \clauseBwdR{o} \leq \kappa
   &
   \text{(IH)}
   \notag
   \\
   &
   \tau' \clauseBwdR{o} \kappa' \leq \kappa \quad (\exists \kappa')
   &
   \text{(Monotonicity)}
   \notag
   \\
   &
   \qedLocal
   \derivation{\derivationWidth}{
      \begin{smathpar}
         \inferrule*[
            lab={\ruleName{$\clauseBwdS$-non-empty-list}}
         ]
         {
            \sigma \clauseBwdR{p} \tau'
            \\
            \tau' \clauseBwdR{o} \kappa'
         }
         {
            \elimCons{\sigma}
            \clauseBwdR{\pattList{p}{o}}
            \kappa'
         }
      \end{smathpar}
   }
   &
   \notag
   \\
   &
   \qedLocal
   (\pattList{p}{o}, \kappa') \leq (\pattList{p}{o}, \kappa)
   &
   \notag
   \\
   \intertext{\crossrule}
   &
   \caseDerivation{\derivationWidth}{
      \begin{smathpar}
         \inferrule*[
            lab={\ruleName{$\clauseFwdR$-pair}}
         ]
         {
            p', \kappa \clauseFwdR \tau
            \\
            p, \tau \clauseFwdR \sigma
         }
         {
            \pattPair{p}{p'}, \kappa
            \clauseFwdR
            \elimProd{\sigma}
         }
      \end{smathpar}
   }
   &
   \notag
   \\
   &
   \sigma \clauseBwdR{p} \tau' \leq \tau \quad (\exists \tau')
   &
   \text{(IH)}
   \notag
   \\
   &
   \tau \clauseBwdR{p'} \leq \kappa
   &
   \text{(IH)}
   \notag
   \\
   &
   \tau' \clauseBwdR{p'} \kappa' \leq \kappa \quad (\exists \kappa')
   &
   \text{(Monotonicity)}
   \notag
   \\
   &
   \qedLocal
   \derivation{\derivationWidth}{
      \begin{smathpar}
         \inferrule*[
            lab={\ruleName{$\clauseBwdS$-pair}}
         ]
         {
            \sigma \clauseBwdR{p} \tau'
            \\
            \tau' \clauseBwdR{p'} \kappa'
         }
         {
            \elimProd{\sigma}
            \clauseBwdR{\pattPair{p}{p'}}
            \kappa'
         }
      \end{smathpar}
   }
   &
   \notag
   \\
   &
   \qedLocal
   (\pattPair{p}{p'}, \kappa') \leq (\pattPair{p}{p'}, \kappa)
   &
   \notag
   \\
   \intertext{\crossrule}
   &
   \caseDerivation{\derivationWidth}{
      \begin{smathpar}
         \inferrule*[
            lab={\ruleName{$\clauseFwdR$-seq}},
            right={$\vec{p} \neq \seqEmpty$}
         ]
         {
            \vec{p}, e \clauseFwdR \tau
            \\
            p, \exLambda{\tau} \clauseFwdR \sigma
         }
         {
            p \concat \vec{p}, e
            \clauseFwdR
            \sigma
         }
      \end{smathpar}
   }
   &
   \notag
   \\
   &
   \sigma \clauseBwdR{p}\eq \exLambda{\tau'} \leq \exLambda{\tau} \quad (\exists \tau')
   &
   \text{(IH)}
   \notag
   \\
   &
   \tau \clauseBwdR{\vec{p}} \leq e
   &
   \text{(IH)}
   \notag
   \\
   &
   \tau' \clauseBwdR{\vec{p}} e' \leq e \quad (\exists e')
   &
   \text{(Monotonicity)}
   \notag
   \\
   &
   \qedLocal
   \derivation{\derivationWidth}{
      \begin{smathpar}
         \inferrule*[
            lab={\ruleName{$\clauseBwdS$-seq}},
            right={$\vec{p} \neq \seqEmpty$}
         ]
         {
            \sigma \clauseBwdR{p}\eq \exLambda{\tau'}
            \\
            \tau' \clauseBwdR{\vec{p}} e'
         }
         {
            \sigma
            \clauseBwdR{p \concat \vec{p}}
            e'
         }
      \end{smathpar}
   }
   &
   \notag
   \\
   &
   \qedLocal
   (p \concat \vec{p}, e') \leq (p \concat \vec{p}, e)
   &
   \notag
   \\
   \intertext{\crossrule}
   &
   \caseDerivation{\derivationWidth}{
      \begin{smathpar}
         \inferrule*[
            lab={\ruleName{$\clauseFwdR$-list-rest-end}}
         ]
         {
            \strut
         }
         {
            \pattListEnd, \kappa
            \clauseFwdR
            \elimNil{\kappa}
         }
      \end{smathpar}
   }
   &
   \notag
   \\
   &
   \qedLocal
   \derivation{\derivationWidth}{
      \begin{smathpar}
         \inferrule*[
            lab={\ruleName{$\clauseBwdS$-list-rest-end}}
         ]
         {
            \strut
         }
         {
            \elimNil{\kappa}
            \clauseBwdR{\pattListEnd}
            \kappa
         }
      \end{smathpar}
   }
   &
   \notag
   \\
   &
   \qedLocal
   (\pattListEnd, \kappa) \leq (\pattListEnd, \kappa)
   &
   \notag
   \\
   \intertext{\crossrule}
   &
   \caseDerivation{\derivationWidth}{
      \begin{smathpar}
         \inferrule*[
            lab={\ruleName{$\clauseFwdR$-list-rest-cons}}
         ]
         {
            o, \kappa \clauseFwdR \tau
            \\
            p, \tau \clauseFwdR \sigma
         }
         {
            (\pattListNext{p}{o}), \kappa
            \clauseFwdR
            \elimCons{\sigma}
         }
      \end{smathpar}
   }
   &
   \notag
   \\
   &
   \sigma \clauseBwdR{p} \tau' \leq \tau \quad (\exists \tau')
   &
   \text{(IH)}
   \notag
   \\
   &
   \tau \clauseBwdR{o} \leq \kappa
   &
   \text{(IH)}
   \notag
   \\
   &
   \tau' \clauseBwdR{o} \kappa' \leq \kappa \quad (\exists \kappa')
   &
   \text{(Monotonicity)}
   \notag
   \\
   &
   \qedLocal
   \derivation{\derivationWidth}{
      \begin{smathpar}
         \inferrule*[
            lab={\ruleName{$\clauseBwdS$-list-rest-cons}}
         ]
         {
            \sigma \clauseBwdR{p} \tau'
            \\
            \tau' \clauseBwdR{o} \kappa'
         }
         {
            \elimCons{\sigma}
            \clauseBwdR{(\pattListNext{p}{o})}
            \kappa'
         }
      \end{smathpar}
   }
   &
   \notag
   \\
   &
   \qedLocal
   (\pattListNext{p}{o}, \kappa') \leq (\pattListNext{p}{o}, \kappa)
   &
   \notag
\end{flalign}
\end{proof}

\setcounter{equation}{0}
\proofContext{clauses}
\subsection{\lemref{aux:clauses:gc}}
Suppose $\sigma \clausesBwdR{\vec{c}} \vec{c}'$. We show $\vec{c}' \clausesFwdR\geq \sigma$.
\begin{proof}
\small
\begin{flalign}
   \intertext{\crossrule}
   &
   \caseDerivation{\derivationWidth}{
      \begin{smathpar}
         \inferrule*[
            lab={\ruleName{$\clausesBwdS$-clause}}
         ]
         {
            \sigma \clauseBwdR{\vec{p}} e
            \\
            e \desugarBwdR{s} s'
         }
         {
            \sigma
            \clausesBwdR{\clause{\vec{p}}{s}}
            \clause{\vec{p}}{s'}
         }
      \end{smathpar}
   }
   &
   \notag
   \\
   &
   s' \desugarFwdR e' \geq e
   \quad
   (\exists e')
   &
   \text{\thmref{surface-language:desugar:gc}}
   \locallabel{fwd-bwd:clause:one}
   \\
   &
   \vec{p}, e \clauseFwdR\geq \sigma
   &
   \text{\lemref{aux:clause:gc}}
   \notag
   \\
   &
   \vec{p}, e' \clauseFwdR \sigma' \geq \sigma
   \quad
   (\exists \sigma')
   &
   \text{monotonicity}
   \locallabel{fwd-bwd:clause:two}
   \\
   &
   \qedLocal
   \derivation{\derivationWidth}{
      \begin{smathpar}
         \inferrule*[
            lab={\ruleName{$\clausesFwdR$-clause}}
         ]
         {
            s' \desugarFwdR e'
            \\
            \vec{p}, e' \clauseFwdR \sigma'
         }
         {
            \clause{\vec{p}}{s'}
            \clausesFwdR
            \sigma'
         }
      \end{smathpar}
   }
   &
   (\localref{fwd-bwd:clause:one}, \localref{fwd-bwd:clause:two})
   \notag
   \\
   &
   \qedLocal
   \sigma' \geq \sigma
   &
   \notag
   \\
   \intertext{\crossrule}
   &
   \caseDerivation{\derivationWidth}{
      \begin{smathpar}
         \inferrule*[
            lab={\ruleName{$\clausesBwdS$-clause-seq}},
            right={$\vec{c} \neq \seqEmpty$}
         ]
         {
            \sigma \eq \sigma' \disjjoin \tau
            \\
            \sigma' \clausesBwdR{c} c'
            \\
            \tau \clausesBwdR{\vec{c}} \vec{c}'
         }
         {
            \sigma
            \clauseBwdR{c \concat \vec{c}}
            c' \concat \vec{c}'
         }
      \end{smathpar}
   }
   &
   \notag
   \\
   &
   c' \clausesFwdR \sigma^\dagger \geq \sigma'
   \quad
   (\exists \sigma^\dagger)
   &
   \text{IH}
   \locallabel{fwd-bwd:clause-seq:one}
   \\
   &
   \vec{c}' \clausesFwdR \tau' \geq \tau
   \quad
   (\exists \tau')
   &
   \text{IH}
   \locallabel{fwd-bwd:clause-seq:two}
   \\
   &
   \sigma^\dagger \disjjoin \tau' = \sigma^\ddagger
   \quad
   (\exists \sigma^\ddagger)
   &
   \text{
      $\sigma' \disjjoin \tau$ defined;
      (\localref{fwd-bwd:clause-seq:one}, \localref{fwd-bwd:clause-seq:two})
   }
   \notag
   \\
   &
   \qedLocal
   \derivation{\derivationWidth}{
      \begin{smathpar}
         \inferrule*[
            lab={\ruleName{$\clausesFwdR$-clause-seq}},
            right={$\vec{c}' \neq \seqEmpty$}
         ]
         {
            c' \clausesFwdR \sigma^\dagger
            \\
            \vec{c}' \clausesFwdR \tau'
            \\
            \sigma^\dagger \disjjoin \tau' = \sigma^\ddagger
         }
         {
            c' \concat \vec{c}'
            \clausesFwdR
            \sigma^\ddagger
         }
      \end{smathpar}
   }
   &
   \text{(\localref{fwd-bwd:clause-seq:one}, \localref{fwd-bwd:clause-seq:two})}
   \notag
   \\
   &
   \qedLocal
   \sigma^\ddagger \geq \sigma
   &
   \text{monotonicity of $\disjjoin$}
   \notag
\end{flalign}
\end{proof}
Suppose $\vec{c} \clausesFwdR \sigma$. We show $\sigma \clausesBwdR{\vec{c}}\leq \vec{c}$.
\begin{proof}
\small
\begin{flalign}
   \intertext{\crossrule}
   &
   \caseDerivation{\derivationWidth}{
      \begin{smathpar}
         \inferrule*[
            lab={\ruleName{$\clausesFwdR$-clause}}
         ]
         {
            s \desugarFwdR e
            \\
            \vec{p}, e \clauseFwdR \sigma
         }
         {
            \clause{\vec{p}}{s}
            \clausesFwdR
            \sigma
         }
      \end{smathpar}
   }
   &
   \notag
   \\
   &
   \sigma \clauseBwdR{\vec{p}} e' \leq e
   \quad
   (\exists e')
   &
   \text{\lemref{aux:clause:gc}}
   \locallabel{bwd-fwd:clause:one}
   \\
   &
   e \desugarBwdR{s}\leq s
   &
   \text{\thmref{surface-language:desugar:gc}}
   \notag
   \\
   &
   e' \desugarBwdR{s} s' \leq s
   &
   \text{monotonicity}
   \locallabel{bwd-fwd:clause:two}
   \\
   &
   \qedLocal
   \derivation{\derivationWidth}{
      \begin{smathpar}
         \inferrule*[
            lab={\ruleName{$\clausesBwdS$-clause}}
         ]
         {
            \sigma \clauseBwdR{\vec{p}} e'
            \\
            e' \desugarBwdR{s} s'
         }
         {
            \sigma
            \clausesBwdR{\clause{\vec{p}}{s}}
            \clause{\vec{p}}{s'}
         }
      \end{smathpar}
   }
   &
   \text{
      (\localref{bwd-fwd:clause:one}, \localref{bwd-fwd:clause:two})
   }
   \notag
   \\
   &
   \qedLocal
   s' \leq s
   &
   \notag
   \\
   \intertext{\crossrule}
   &
   \caseDerivation{\derivationWidth}{
      \begin{smathpar}
         \inferrule*[
            lab={\ruleName{$\clausesFwdR$-clause-seq}},
            right={$\vec{c} \neq \seqEmpty$}
         ]
         {
            c \clausesFwdR \sigma
            \\
            \vec{c} \clausesFwdR \sigma'
            \\
            \sigma \disjjoin \sigma' = \tau
         }
         {
            c \concat \vec{c}
            \clausesFwdR
            \tau
         }
      \end{smathpar}
   }
   &
   \notag
   \\
   &
   \sigma \clausesBwdR{c} c' \leq c
   \quad
   (\exists c')
   &
   \text{IH}
   \locallabel{bwd-fwd:clause-seq:one}
   \\
   &
   \sigma' \clausesBwdR{\vec{c}} \vec{c}' \leq \vec{c}
   \quad
   (\exists \vec{c}')
   &
   \text{IH}
   \locallabel{bwd-fwd:clause-seq:two}
   \\
   &
   \qedLocal
   \derivation{\derivationWidth}{
      \begin{smathpar}
         \inferrule*[
            lab={\ruleName{$\clausesBwdS$-clause-seq}},
            right={$\vec{c} \neq \seqEmpty$}
         ]
         {
            \tau \eq \sigma' \disjjoin \sigma
            \\
            \sigma \clausesBwdR{c} c'
            \\
            \sigma' \clausesBwdR{\vec{c}} \vec{c}'
         }
         {
            \tau
            \clauseBwdR{c \concat \vec{c}}
            c' \concat \vec{c}'
         }
      \end{smathpar}
   }
   &
   \text{(\localref{bwd-fwd:clause-seq:one}, \localref{bwd-fwd:clause-seq:two})}
   \notag
   \\
   &
   \qedLocal
   c' \concat \vec{c}' \leq c \concat \vec{c}
   &
   \notag
\end{flalign}
\end{proof}

\setcounter{equation}{0}
\proofContext{totalise}
\subsection{\lemref{aux:list-gen:gc}}
Suppose $\totaliseFwd{\kappa}{\alpha}{\vec{\pi}}{\kappa'}$. Then $\totaliseFwdF{\kappa,\alpha,\vec{\pi}}(\totaliseBwdF{\kappa,\alpha,\vec{\pi}}(\smash{\kappa^\dagger})) \geq \smash{\kappa^\dagger}$.
\begin{proof}
\ref{lem:aux:totalise:gc:2}
\small
\begin{flalign}
   \intertext{\crossrule}
   &
   \caseDerivation{\derivationWidth}{
    \begin{smathpar}
      \inferrule*[
        lab={\ruleName{$\totaliseBwdS$-var}}
      ]
      {
         \totaliseBwd{\kappa_1}{\vec{\pi}}{\kappa_2}{\alpha}
      }
      {
         \totaliseBwd{\elimVar{x}{\kappa_1}}{x \concat \vec{\pi}}{\elimVar{x}{\kappa_2}}{\alpha}
      }
    \end{smathpar}
   }
   &
   \notag
   \\
   &
   \totaliseFwd{\kappa_2}{\alpha}{\vec{\pi}}{\kappa_1'} \geq \kappa_1 \quad (\exists \kappa_1')
   &
   \text{(IH)}
   \notag
   \\
   &
   \qedLocal
   \derivation{\derivationWidth}{
      \begin{smathpar}
         \inferrule*[
            lab={\ruleName{$\totaliseFwdS$-elim-var}}
         ]
         {
            \totaliseFwd{\kappa_2}{\alpha}{\vec{\pi}}{\kappa_1'}
         }
         {
            \totaliseFwd{\elimVar{x}{\kappa_2}}
                        {\alpha}
                        {\pattVar{x} \concat \vec{\pi}}
                        {\elimVar{x}{\kappa_1'}}
         }
      \end{smathpar}
   }
   &
   \notag
   \\
   &
   \qedLocal
   {\elimVar{x}{\kappa_1'}} \geq {\elimVar{x}{\kappa_1}}
   &
   \notag
   \intertext{\crossrule}
   &
   \caseDerivation{\derivationWidth}{
    \begin{smathpar}
      \inferrule*[
         lab={\ruleName{$\totaliseBwdS$-true}}
      ]
      {
         \totaliseBwd{\kappa_1}{\vec{\pi}}{\kappa_2}{\beta}
      }
      {
         \totaliseBwd{\elimBool{\kappa_1}{\annot{\exNil}{\alpha}}}
                     {\pattTrue \concat \vec{\pi}}
                     {\elimTrue{\kappa_2}}
                     {\alpha \join \beta}
      }
    \end{smathpar}
   }
   &
   \notag
   \\
   &
   \totaliseFwdGeq{\kappa_2}{\beta}{\vec{\pi}}{\kappa_1}
   &
   \text{(IH)}
   \notag
   \\
   &
   \totaliseFwd{\kappa_2}{\alpha \join \beta}{\vec{\pi}}{\kappa_1'} \geq \kappa_1 \quad (\exists \kappa_1')
   &
   \text{(Monotonicity)}
   \notag
   \\
   &
   \qedLocal
   \derivation{\derivationWidth}{
      \begin{smathpar}
         \inferrule*[
            lab={\ruleName{$\totaliseFwdS$-elim-true}}
         ]
         {
            \totaliseFwd{\kappa_2}{\alpha \join \beta}{\vec{\pi}}{\kappa_1'}
         }
         {
            \totaliseFwd{\elimTrue{\kappa_2}}
                        {\alpha \join \beta}
                        {\pattTrue \concat \vec{\pi}}
                        {\elimBool{\kappa_1'}{\annNil{\alpha \join \beta}}}
         }
      \end{smathpar}
   }
   &
   \notag
   \\
   &
   \qedLocal
   \elimBool{\kappa_1'}{\annNil{\alpha \join \beta}} \geq \elimBool{\kappa_1}{\annot{\exNil}{\alpha}}
   &
   \notag
   \intertext{\crossrule}
   &
   \caseDerivation{\derivationWidth}{
    \begin{smathpar}
      \inferrule*[
         lab={\ruleName{$\totaliseBwdS$-false}}
      ]
      {
         \totaliseBwd{\kappa_1}{\vec{\pi}}{\kappa_2}{\beta}
      }
      {
         \totaliseBwd{\elimBool{\annot{\exNil}{\alpha}}{\kappa_1}}
                     {\pattFalse \concat \vec{\pi}}
                     {\elimFalse{\kappa_2}}
                     {\alpha \join \beta}
      }
    \end{smathpar}
   }
   &
   \notag
   \\
   &
   \totaliseFwdGeq{\kappa_2}{\beta}{\vec{\pi}}{\kappa_1}
   &
   \text{(IH)}
   \notag
   \\
   &
   \totaliseFwd{\kappa_2}{\alpha \join \beta}{\vec{\pi}}{\kappa_1'} \geq \kappa_1 \quad (\exists \kappa_1')
   &
   \text{(Monotonicity)}
   \notag
   \\
   &
   \qedLocal
   \derivation{\derivationWidth}{
      \begin{smathpar}
         \inferrule*[
            lab={\ruleName{$\totaliseFwdS$-elim-false}}
         ]
         {
            \totaliseFwd{\kappa_2}{\alpha \join \beta}{\vec{\pi}}{\kappa_1'}
         }
         {
            \totaliseFwd{\elimFalse{\kappa_2}}
                        {\alpha \join \beta}
                        {\pattFalse \concat \vec{\pi}}
                        {\elimBool{\annNil{\alpha \join \beta}}{\kappa_1'}}
         }
      \end{smathpar}
   }
   &
   \notag
   \\
   &
   \qedLocal
   \elimBool{\annNil{\alpha \join \beta}}{\kappa_1'} \geq \elimBool{\annot{\exNil}{\alpha}}{\kappa_1}
   &
   \notag
   \intertext{\crossrule}
   &
   \caseDerivation{\derivationWidth}{
    \begin{smathpar}
      \inferrule*[
         lab={\ruleName{$\totaliseBwdS$-prod}}
      ]
      {
         \totaliseBwd{\sigma_1}{p \concat p' \concat \vec{\pi}}{\sigma_2}{\alpha}
      }
      {
         \totaliseBwd{\elimProd{\sigma_1}}{\pattPair{p}{p'} \concat \vec{\pi}}{\elimProd{\sigma_2}}{\alpha}
      }
    \end{smathpar}
   }
   &
   \notag
   \\
   &
   \totaliseFwd{\sigma_2}{\alpha}{p \concat p' \concat \vec{\pi}}{\sigma_1'} \geq \sigma_1 \quad (\exists \sigma_1')
   &
   \text{(IH)}
   \notag
   \\
   &
   \qedLocal
   \derivation{\derivationWidth}{
      \begin{smathpar}
         \inferrule*[
            lab={\ruleName{$\totaliseFwdS$-elim-prod}}
         ]
         {
            \totaliseFwd{\sigma_2}{\alpha}{p \concat p' \concat \vec{\pi}}{\sigma_1'}
         }
         {
            \totaliseFwd{\elimProd{\sigma_2}}
                        {\alpha}
                        {\pattPair{p}{p'} \concat \vec{\pi}}
                        {\elimProd{\sigma_1'}}
         }
      \end{smathpar}
   }
   &
   \notag
   \\
   &
   \qedLocal
   \elimProd{\sigma_1'} \geq \elimProd{\sigma_1}
   &
   \notag
   \intertext{\crossrule}
   &
   \caseDerivation{\derivationWidth}{
    \begin{smathpar}
      \inferrule*[
         lab={\ruleName{$\totaliseBwdS$-nil}}
      ]
      {
         \totaliseBwd{\kappa_1}{\vec{\pi}}{\kappa_2}{\beta}
      }
      {
         \totaliseBwd{\elimList{\kappa_1}{\elimVar{x}{\elimVar{y}{\annot{\exNil}{\alpha}}}}}
                     {\pattNil \concat \vec{\pi}}
                     {\elimNil{\kappa_2}}
                     {\alpha \join \beta}
      }
    \end{smathpar}
   }
   &
   \notag
   \\
   &
   \totaliseFwdGeq{\kappa_2}{\beta}{\vec{\pi}}{\kappa_1}
   &
   \text{(IH)}
   \notag
   \\
   &
   \totaliseFwd{\kappa_2}{\alpha \join \beta}{\vec{\pi}}{\kappa_1'} \geq \kappa_1 \quad (\exists \kappa_1')
   &
   \text{(Monotonicity)}
   \notag
   \\
   &
   \qedLocal
   \derivation{\derivationWidth}{
      \begin{smathpar}
         \inferrule*[
            lab={\ruleName{$\totaliseFwdS$-elim-nil}}
         ]
         {
            \totaliseFwd{\kappa_2}{\alpha \join \beta}{\vec{\pi}}{\kappa_1'}
         }
         {
            \totaliseFwd{\elimNil{\kappa_2}}
                        {\alpha \join \beta}
                        {\pattNil \concat \vec{\pi}}
                        {\elimList{\kappa_1'}{\elimVar{x}{\elimVar{y}{\annNil{\alpha \join \beta}}}}}
         }
      \end{smathpar}
   }
   &
   \notag
   \\
   &
   \qedLocal
   \elimList{\kappa_1'}{\elimVar{x}{\elimVar{y}{\annNil{\alpha \join \beta}}}} \geq \elimList{\kappa_1}{\elimVar{x}{\elimVar{y}{\annot{\exNil}{\alpha}}}} \eq
   &
   \notag
   \intertext{\crossrule}
   &
   \caseDerivation{\derivationWidth}{
    \begin{smathpar}
      \inferrule*[
         lab={\ruleName{$\totaliseBwdS$-cons}}
      ]
      {
         \totaliseBwd{\sigma_1}{p \concat p' \concat \vec{\pi}}{\sigma_2}{\beta}
      }
      {
         \totaliseBwd{\elimList{\annot{\exNil}{\alpha}}{\sigma_1}}
                     {(\pattCons{p}{p'}) \concat \vec{\pi}}
                     {\elimCons{\sigma_2}}
                     {\alpha \join \beta}
      }
    \end{smathpar}
   }
   &
   \notag
   \\
   &
   \totaliseFwdGeq{\sigma_2}{\beta}{p \concat p' \concat \vec{\pi}}{\sigma_1}
   &
   \text{(IH)}
   \notag
   \\
   &
   \totaliseFwd{\sigma_2}{\alpha \join \beta}{p \concat p' \concat \vec{\pi}}{\sigma_1'} \geq \sigma_1 \quad (\exists \sigma_1')
   &
   \text{(Monotonicity)}
   \notag
   \\
   &
   \qedLocal
   \derivation{\derivationWidth}{
      \begin{smathpar}
         \inferrule*[
            lab={\ruleName{$\totaliseFwdS$-elim-cons}}
         ]
         {
            \totaliseFwd{\sigma_2}{\alpha \join \beta}{p \concat p' \concat \vec{\pi}}{\sigma_1'}
         }
         {
            \totaliseFwd{\elimCons{\sigma_2}}
                        {\alpha \join \beta}
                        {(\pattCons{p}{p'}) \concat \vec{\pi}}
                        {\elimList{\annNil{\alpha \join \beta}}{\sigma_1'}}
         }
      \end{smathpar}
   }
   &
   \notag
   \\
   &
   \qedLocal
   \elimList{\annNil{\alpha \join \beta}}{\sigma_1'} \geq \elimList{\annot{\exNil}{\alpha}}{\sigma_1}
   &
   \notag
   \intertext{\crossrule}
   &
   \caseDerivation{\derivationWidth}{
    \begin{smathpar}
      \inferrule*[
         lab={\ruleName{$\totaliseBwdS$-non-empty-list}}
      ]
      {
         \totaliseBwd{\sigma_1}{p \concat o \concat \vec{\pi}}{\sigma_2}{\beta}
      }
      {
         \totaliseBwd{\elimList{\annot{\exNil}{\alpha}}{\sigma_1}}
                     {(\pattList{p}{o}) \concat \vec{\pi}}
                     {\elimCons{\sigma_2}}
                     {\alpha \join \beta}
      }
    \end{smathpar}
   }
   &
   \notag
   \\
   &
   \totaliseFwdGeq{\sigma_2}{\beta}{p \concat o \concat \vec{\pi}}{\sigma_1}
   &
   \text{(IH)}
   \notag
   \\
   &
   \totaliseFwd{\sigma_2}{\alpha \join \beta}{p \concat o \concat \vec{\pi}}{\sigma_1'} \geq \sigma_1 \quad (\exists \sigma_1')
   &
   \text{(Monotonicity)}
   \notag
   \\
   &
   \qedLocal
   \derivation{\derivationWidth}{
      \begin{smathpar}
         \inferrule*[
            lab={\ruleName{$\totaliseFwdS$-elim-non-empty-list}}
         ]
         {
            \totaliseFwd{\sigma_2}{\alpha \join \beta}{p \concat o \concat \vec{\pi}}{\sigma_1'}
         }
         {
            \totaliseFwd{\elimCons{\sigma_2}}
                        {\alpha \join \beta}
                        {(\pattList{p}{o}) \concat \vec{\pi}}
                        {\elimList{\annNil{\alpha \join \beta}}{\sigma_1'}}
         }
      \end{smathpar}
   }
   &
   \notag
   \\
   &
   \qedLocal
   \elimList{\annNil{\alpha \join \beta}}{\sigma_1'} \geq \elimList{\annot{\exNil}{\alpha}}{\sigma_1}
   &
   \notag
   \intertext{\crossrule}
   &
   \caseDerivation{\derivationWidth}{
    \begin{smathpar}
      \inferrule*[
         lab={\ruleName{$\totaliseBwdS$-list-rest-end}}
      ]
      {
         \totaliseBwd{\kappa_1}{\vec{\pi}}{\kappa_2}{\beta}
      }
      {
         \totaliseBwd{\elimList{\kappa_1}{\elimVar{x}{\elimVar{y}{\annot{\exNil}{\alpha}}}}}
                     {\pattListEnd \concat \vec{\pi}}
                     {\elimNil{\kappa_2}}
                     {\alpha \join \beta}
      }
    \end{smathpar}
   }
   &
   \notag
   \\
   &
   \totaliseFwdGeq{\kappa_2}{\beta}{\vec{\pi}}{\kappa_1}
   &
   \text{(IH)}
   \notag
   \\
   &
   \totaliseFwd{\kappa_2}{\alpha \join \beta}{\vec{\pi}}{\kappa_1'} \geq \kappa_1 \quad (\exists \kappa_1')
   &
   \text{(Monotonicity)}
   \notag
   \\
   &
   \qedLocal
   \derivation{\derivationWidth}{
      \begin{smathpar}
         \inferrule*[
            lab={\ruleName{$\totaliseFwdS$-elim-list-rest-end}}
         ]
         {
            \totaliseFwd{\kappa_2}{\alpha \join \beta}{\vec{\pi}}{\kappa_1'}
         }
         {
            \totaliseFwd{\elimNil{\kappa_2}}
                        {\alpha \join \beta}
                        {\pattListEnd \concat \vec{\pi}}
                        {\elimList{\kappa_1'}{\elimVar{x}{\elimVar{y}{\annot{\exNil}{\alpha \join \beta}}}}}
         }
      \end{smathpar}
   }
   &
   \notag
   \\
   &
   \qedLocal
   \elimList{\kappa_1'}{\elimVar{x}{\elimVar{y}{\annot{\exNil}{\alpha \join \beta}}}} \geq \elimList{\kappa_1}{\elimVar{x}{\elimVar{y}{\annot{\exNil}{\alpha}}}}
   &
   \notag
   \intertext{\crossrule}
   &
   \caseDerivation{\derivationWidth}{
    \begin{smathpar}
      \inferrule*[
         lab={\ruleName{$\totaliseBwdS$-list-rest-cons}}
      ]
      {
         \totaliseBwd{\sigma_1}{p \concat o \concat \vec{\pi}}{\sigma_2}{\beta}
      }
      {
         \totaliseBwd{\elimList{\annot{\exNil}{\alpha}}{\sigma_1}}
                     {(\pattListNext{p}{o}) \concat \vec{\pi}}
                     {\elimCons{\sigma_2}}
                     {\alpha \join \beta}
      }
    \end{smathpar}
   }
   &
   \notag
   \\
   &
   \totaliseFwdGeq{\sigma_2}{\beta}{p \concat o \concat \vec{\pi}}{\sigma_1}
   &
   \text{(IH)}
   \notag
   \\
   &
   \totaliseFwd{\sigma_2}{\alpha \join \beta}{p \concat o \concat \vec{\pi}}{\sigma_1'} \geq \sigma_1 \quad (\exists \sigma_1')
   &
   \text{(Monotonicity)}
   \notag
   \\
   &
   \qedLocal
   \derivation{\derivationWidth}{
      \begin{smathpar}
         \inferrule*[
            lab={\ruleName{$\totaliseFwdS$-elim-list-rest-cons}}
         ]
         {
            \totaliseFwd{\sigma_2}{\alpha \join \beta}{p \concat o \concat \vec{\pi}}{\sigma_1'}
         }
         {
            \totaliseFwd{\elimCons{\sigma_2}}
                        {\alpha \join \beta}
                        {(\pattListNext{p}{o}) \concat \vec{\pi}}
                        {\elimList{\annNil{\alpha \join \beta}}{\sigma_1'}}
         }
      \end{smathpar}
   }
   &
   \notag
   \\
   &
   \qedLocal
   \elimList{\annNil{\alpha \join \beta}}{\sigma_1'} \geq \elimList{\annot{\exNil}{\alpha}}{\sigma_1}
   &
   \notag
\end{flalign}
\end{proof}

\noindent
Suppose $\totaliseFwd{\kappa}{\alpha}{\vec{\pi}}{\kappa'}$. Then $\totaliseBwdF{\kappa,\alpha,\vec{\pi}}(\totaliseFwdF{\kappa,\alpha,\vec{\pi}}(\smash{\kappa^\dagger}, \alpha')) \leq (\smash{\kappa^\dagger}, \alpha')$.
\begin{proof}
   \ref{lem:aux:totalise:gc:3}
   \small
   \begin{flalign}
      \intertext{\crossrule}
      &
      \caseDerivation{\derivationWidth}{
         \begin{smathpar}
            \inferrule*[
               lab={\ruleName{$\totaliseFwdS$-elim-var}}
            ]
            {
               \totaliseFwd{\kappa_1}{\alpha}{\vec{\pi}}{\kappa_2}
            }
            {
               \totaliseFwd{\elimVar{x}{\kappa_1}}
                           {\alpha}
                           {\pattVar{x} \concat \vec{\pi}}
                           {\elimVar{x}{\kappa_2}}
            }
         \end{smathpar}
      }
      &
      \notag
      \\
      &
      \totaliseBwd{\kappa_2}{\vec{\pi}}{\kappa_1'}{\beta} \leq \kappa_1, \alpha \quad (\exists \kappa_1', \beta)
      &
      \text{(IH)}
      \notag
      \\
      &
      \qedLocal
      \derivation{\derivationWidth}{
         \begin{smathpar}
            \inferrule*[
               lab={\ruleName{$\totaliseBwdS$-var}}
            ]
            {
               \totaliseBwd{\kappa_2}{\vec{\pi}}{\kappa_1'}{\beta}
            }
            {
               \totaliseBwd{\elimVar{x}{\kappa_2}}{x \concat \vec{\pi}}{\elimVar{x}{\kappa_1'}}{\beta}
            }
         \end{smathpar}
      }
      &
      \notag
      \\
      &
      \qedLocal
      (\elimVar{x}{\kappa_1'}, \beta) \leq (\elimVar{x}{\kappa_1}, \alpha)
      &
      \notag
      \intertext{\crossrule}
      &
      \caseDerivation{\derivationWidth}{
         \begin{smathpar}
            \inferrule*[
               lab={\ruleName{$\totaliseFwdS$-elim-true}}
            ]
            {
               \totaliseFwd{\kappa_1}{\alpha}{\vec{\pi}}{\kappa_2}
            }
            {
               \totaliseFwd{\elimTrue{\kappa_1}}
                           {\alpha}
                           {\pattTrue \concat \vec{\pi}}
                           {\elimBool{\kappa_2}{\annNil{\alpha}}}
            }
         \end{smathpar}
      }
      &
      \notag
      \\
      &
      \totaliseBwd{\kappa_2}{\vec{\pi}}{\kappa_1'}{\beta} \leq \kappa_1, \alpha \quad (\exists \kappa_1', \beta)
      &
      \text{(IH)}
      \notag
      \\
      &
      \qedLocal
      \derivation{\derivationWidth}{
         \begin{smathpar}
            \inferrule*[
               lab={\ruleName{$\totaliseBwdS$-true}}
            ]
            {
               \totaliseBwd{\kappa_2}{\vec{\pi}}{\kappa_1'}{\beta}
            }
            {
               \totaliseBwd{\elimBool{\kappa_2}{\annot{\exNil}{\alpha}}}
                           {\pattTrue \concat \vec{\pi}}
                           {\elimTrue{\kappa_1'}}
                           {\alpha \join \beta}
            }
         \end{smathpar}
      }
      &
      \notag
      \\
      &
      \qedLocal
      (\elimTrue{\kappa_1'}, \alpha \join \beta) \leq (\elimTrue{\kappa_1}, \alpha)
      &
      \notag
      \intertext{\crossrule}
      &
      \caseDerivation{\derivationWidth}{
         \begin{smathpar}
            \inferrule*[
               lab={\ruleName{$\totaliseFwdS$-elim-false}}
            ]
            {
               \totaliseFwd{\kappa_1}{\alpha}{\vec{\pi}}{\kappa_2}
            }
            {
               \totaliseFwd{\elimFalse{\kappa_1}}
                           {\alpha}
                           {\pattFalse \concat \vec{\pi}}
                           {\elimBool{\annNil{\alpha}}{\kappa_2}}
            }
         \end{smathpar}
      }
      &
      \notag
      \\
      &
      \totaliseBwd{\kappa_2}{\vec{\pi}}{\kappa_1'}{\beta} \leq \kappa_1, \alpha \quad (\exists \kappa_1', \beta)
      &
      \text{(IH)}
      \notag
      \\
      &
      \qedLocal
      \derivation{\derivationWidth}{
         \begin{smathpar}
            \inferrule*[
               lab={\ruleName{$\totaliseBwdS$-false}}
            ]
            {
               \totaliseBwd{\kappa_2}{\vec{\pi}}{\kappa_1'}{\beta}
            }
            {
               \totaliseBwd{\elimBool{\annot{\exNil}{\alpha}}{\kappa_2}}
                           {\pattFalse \concat \vec{\pi}}
                           {\elimFalse{\kappa_1'}}
                           {\alpha \join \beta}
            }
         \end{smathpar}
      }
      &
      \notag
      \\
      &
      \qedLocal
      (\elimFalse{\kappa_1'}, \alpha \join \beta) \leq (\elimFalse{\kappa_1}, \alpha)
      &
      \notag
      \intertext{\crossrule}
      &
      \caseDerivation{\derivationWidth}{
         \begin{smathpar}
            \inferrule*[
               lab={\ruleName{$\totaliseFwdS$-elim-prod}}
            ]
            {
               \totaliseFwd{\sigma_1}{\alpha}{p \concat p' \concat \vec{\pi}}{\sigma_2}
            }
            {
               \totaliseFwd{\elimProd{\sigma_1}}
                           {\alpha}
                           {\pattPair{p}{p'} \concat \vec{\pi}}
                           {\elimProd{\sigma_2}}
            }
         \end{smathpar}
      }
      &
      \notag
      \\
      &
      \totaliseBwd{\sigma_2}{p \concat p' \concat \vec{\pi}}{\sigma_1'}{\beta} \leq \sigma_1, \alpha \quad (\exists \sigma_1', \beta)
      &
      \text{(IH)}
      \notag
      \\
      &
      \qedLocal
      \derivation{\derivationWidth}{
         \begin{smathpar}
            \inferrule*[
               lab={\ruleName{$\totaliseBwdS$-prod}}
            ]
            {
               \totaliseBwd{\sigma_2}{p \concat p' \concat \vec{\pi}}{\sigma_1'}{\beta}
            }
            {
               \totaliseBwd{\elimProd{\sigma_2}}{\pattPair{p}{p'} \concat \vec{\pi}}{\elimProd{\sigma_1'}}{\beta}
            }
         \end{smathpar}
      }
      &
      \notag
      \\
      &
      \qedLocal
      (\elimProd{\sigma_1'}, \beta) \leq (\elimProd{\sigma_1}, \alpha)
      &
      \notag
      \intertext{\crossrule}
      &
      \caseDerivation{\derivationWidth}{
         \begin{smathpar}
            \inferrule*[
               lab={\ruleName{$\totaliseFwdS$-elim-nil}}
            ]
            {
               \totaliseFwd{\kappa_1}{\alpha}{\vec{\pi}}{\kappa_2}
            }
            {
               \totaliseFwd{\elimNil{\kappa_1}}
                           {\alpha}
                           {\pattNil \concat \vec{\pi}}
                           {\elimList{\kappa_2}{\elimVar{x}{\elimVar{y}{\annNil{\alpha}}}}}
            }
         \end{smathpar}
      }
      &
      \notag
      \\
      &
      \totaliseBwd{\kappa_2}{\vec{\pi}}{\kappa_1'}{\beta} \leq \kappa_1, \alpha \quad (\exists \kappa_1', \beta)
      &
      \text{(IH)}
      \notag
      \\
      &
      \qedLocal
      \derivation{\derivationWidth}{
         \begin{smathpar}
            \inferrule*[
               lab={\ruleName{$\totaliseBwdS$-nil}}
            ]
            {
               \totaliseBwd{\kappa_2}{\vec{\pi}}{\kappa_1'}{\beta}
            }
            {
               \totaliseBwd{\elimList{\kappa_2}{\elimVar{x}{\elimVar{y}{\annot{\exNil}{\alpha}}}}}
                           {\pattNil \concat \vec{\pi}}
                           {\elimNil{\kappa_1'}}
                           {\alpha \join \beta}
            }
         \end{smathpar}
      }
      &
      \notag
      \\
      &
      \qedLocal
      (\elimNil{\kappa_1'}, \alpha \join \beta) \leq (\elimNil{\kappa_1}, \alpha)
      &
      \notag
      \intertext{\crossrule}
      &
      \caseDerivation{\derivationWidth}{
         \begin{smathpar}
            \inferrule*[
               lab={\ruleName{$\totaliseFwdS$-elim-cons}}
            ]
            {
               \totaliseFwd{\sigma_1}{\alpha}{p \concat p' \concat \vec{\pi}}{\sigma_2}
            }
            {
               \totaliseFwd{\elimCons{\sigma_1}}
                           {\alpha}
                           {(\pattCons{p}{p'}) \concat \vec{\pi}}
                           {\elimList{\annNil{\alpha}}{\sigma_2}}
            }
         \end{smathpar}
      }
      &
      \notag
      \\
      &
      \totaliseBwd{\sigma_2}{p \concat p' \concat \vec{\pi}}{\sigma_1'}{\beta} \leq \sigma_1, \alpha \quad (\exists \sigma_1', \beta)
      &
      \text{(IH)}
      \notag
      \\
      &
      \qedLocal
      \derivation{\derivationWidth}{
         \begin{smathpar}
            \inferrule*[
               lab={\ruleName{$\totaliseBwdS$-cons}}
            ]
            {
               \totaliseBwd{\sigma_2}{p \concat p' \concat \vec{\pi}}{\sigma_1'}{\beta}
            }
            {
               \totaliseBwd{\elimList{\annot{\exNil}{\alpha}}{\sigma_2}}
                           {(\pattCons{p}{p'}) \concat \vec{\pi}}
                           {\elimCons{\sigma_1'}}
                           {\alpha \join \beta}
            }
         \end{smathpar}
      }
      &
      \notag
      \\
      &
      \qedLocal
      (\elimCons{\sigma_1'}, \alpha \join \beta) \leq (\elimCons{\sigma_1}, \alpha)
      &
      \notag
      \intertext{\crossrule}
      &
      \caseDerivation{\derivationWidth}{
         \begin{smathpar}
            \inferrule*[
               lab={\ruleName{$\totaliseFwdS$-elim-non-empty-list}}
            ]
            {
               \totaliseFwd{\sigma_1}{\alpha}{p \concat o \concat \vec{\pi}}{\sigma_2}
            }
            {
               \totaliseFwd{\elimCons{\sigma_1}}
                           {\alpha}
                           {(\pattList{p}{o}) \concat \vec{\pi}}
                           {\elimList{\annNil{\alpha}}{\sigma_2}}
            }
         \end{smathpar}
      }
      &
      \notag
      \\
      &
      \totaliseBwd{\sigma_2}{p \concat o \concat \vec{\pi}}{\sigma_1'}{\beta}\leq \sigma_1, \alpha \quad (\exists \sigma_1', \beta)
      &
      \text{(IH)}
      \notag
      \\
      &
      \qedLocal
      \derivation{\derivationWidth}{
         \begin{smathpar}
            \inferrule*[
               lab={\ruleName{$\totaliseBwdS$-non-empty-list}}
            ]
            {
               \totaliseBwd{\sigma_2}{p \concat o \concat \vec{\pi}}{\sigma_1'}{\beta}
            }
            {
               \totaliseBwd{\elimList{\annot{\exNil}{\alpha}}{\sigma_2}}
                           {(\pattList{p}{o}) \concat \vec{\pi}}
                           {\elimCons{\sigma_1'}}
                           {\alpha \join \beta}
            }
         \end{smathpar}
      }
      &
      \notag
      \\
      &
      \qedLocal
      (\elimCons{\sigma_1'}, \alpha \join \beta) \leq (\elimCons{\sigma_1}, \alpha)
      &
      \notag
      \intertext{\crossrule}
      &
      \caseDerivation{\derivationWidth}{
         \begin{smathpar}
            \inferrule*[
               lab={\ruleName{$\totaliseFwdS$-elim-list-rest-end}}
            ]
            {
               \totaliseFwd{\kappa_1}{\alpha}{\vec{\pi}}{\kappa_2}
            }
            {
               \totaliseFwd{\elimNil{\kappa_1}}
                           {\alpha}
                           {\pattListEnd \concat \vec{\pi}}
                           {\elimList{\kappa_2}{\elimVar{x}{\elimVar{y}{\annot{\exNil}{\alpha}}}}}
            }
         \end{smathpar}
      }
      &
      \notag
      \\
      &
      \totaliseBwd{\kappa_2}{\vec{\pi}}{\kappa_1'}{\beta} \leq \kappa_1, \alpha \quad (\exists \kappa_1', \beta)
      &
      \text{(IH)}
      \notag
      \\
      &
      \qedLocal
      \derivation{\derivationWidth}{
         \begin{smathpar}
            \inferrule*[
               lab={\ruleName{$\totaliseBwdS$-list-rest-end}}
            ]
            {
               \totaliseBwd{\kappa_2}{\vec{\pi}}{\kappa_1'}{\beta}
            }
            {
               \totaliseBwd{\elimList{\kappa_2}{\elimVar{x}{\elimVar{y}{\annot{\exNil}{\alpha}}}}}
                           {\pattListEnd \concat \vec{\pi}}
                           {\elimNil{\kappa_1'}}
                           {\alpha \join \beta}
            }
         \end{smathpar}
      }
      &
      \notag
      \\
      &
      \qedLocal
      (\elimNil{\kappa_1'}, \alpha \join \beta) \leq (\elimNil{\kappa_1}, \alpha)
      &
      \notag
      \intertext{\crossrule}
      &
      \caseDerivation{\derivationWidth}{
         \begin{smathpar}
            \inferrule*[
               lab={\ruleName{$\totaliseFwdS$-elim-list-rest-cons}}
            ]
            {
               \totaliseFwd{\sigma_1}{\alpha}{p \concat o \concat \vec{\pi}}{\sigma_2}
            }
            {
               \totaliseFwd{\elimCons{\sigma_1}}
                           {\alpha}
                           {(\pattListNext{p}{o}) \concat \vec{\pi}}
                           {\elimList{\annNil{\alpha}}{\sigma_2}}
            }
         \end{smathpar}
      }
      &
      \notag
      \\
      &
      \totaliseBwd{\sigma_2}{p \concat o \concat \vec{\pi}}{\sigma_1'}{\beta} \leq \sigma_1, \alpha \quad (\exists \sigma_1', \beta)
      &
      \text{(IH)}
      \notag
      \\
      &
      \qedLocal
      \derivation{\derivationWidth}{
         \begin{smathpar}
            \inferrule*[
               lab={\ruleName{$\totaliseBwdS$-list-rest-cons}}
            ]
            {
               \totaliseBwd{\sigma_2}{p \concat o \concat \vec{\pi}}{\sigma_1'}{\beta}
            }
            {
               \totaliseBwd{\elimList{\annot{\exNil}{\alpha}}{\sigma_2}}
                           {(\pattListNext{p}{o}) \concat \vec{\pi}}
                           {\elimCons{\sigma_1'}}
                           {\alpha \join \beta}
            }
         \end{smathpar}
      }
      &
      \notag
      \\
      &
      \qedLocal
      (\elimCons{\sigma_1'}, \alpha \join \beta) \leq (\elimCons{\sigma_1}, \alpha)
      &
      \notag
   \end{flalign}
\end{proof}

\fi

\end{document}